\documentclass[a4paper,11pt,fleqn]{article}
\pdfoutput=1
\usepackage{jheppub}

\usepackage{graphicx}
\usepackage[figuresright]{rotating}
\usepackage{bm,amsmath,amssymb}
\usepackage[mathscr]{eucal}
\usepackage{color}
\usepackage{multirow}
\usepackage{subcaption}

\long\def\comment#1{ }
\newcommand{\eqn}[1]{Eq.~\eqref{#1}}
\newcommand{\beq}{\begin{equation}}
\newcommand{\eeq}{\end{equation}}
\newcommand{\nn}{\nonumber\\}

\newcommand{\rmd}{{\rm{d }}}

\newcommand{\ombr}{\omega_{\textrm{br}}}

\newcommand{\abar}{\bar{\alpha}_s}

\newcommand{\rme}{{\rm e}}

\newcommand{\order}[1]{\mathcal{O}{(#1)}}

\newcommand{\bk}{\bm{k}_\perp}
\newcommand{\bq}{\bm{q}_\perp}
\newcommand{\bp}{\bm{p}_\perp}
\newcommand{\bx}{\bm{x}_\perp}
\newcommand{\by}{\bm{y}_\perp}
\newcommand{\bu}{\bm{u}}
\newcommand{\bv}{\bm{v}}

\newcommand{\br}{\bm{r}}

\usepackage{color}
\definecolor{darkgreen}{rgb}{0,0.5,0}
\definecolor{darkblue}{rgb}{0,0,0.7}
\definecolor{darkred}{rgb}{0.5,0,0.0}
\definecolor{darkorange}{rgb}{0.8,0.4,0.0}

\makeatletter
\g@addto@macro\bfseries{\boldmath}
\makeatother

\title{Jet polarisation in an anisotropic medium}

\author{S.~Hauksson}
\author{and E.~Iancu,}
\affiliation{Institut de Physique Th\'{e}orique, Universit\'{e} Paris-Saclay, CNRS, CEA, F-91191, Gif-sur-Yvette, France}

\emailAdd{sigtryggur.hauksson@ipht.fr}
\emailAdd{edmond.iancu@ipht.fr}
\abstract{We study the evolution of an energetic jet  which propagates in an anisotropic quark-gluon plasma, as created in the intermediate
stages of ultrarelativistic heavy-ion collisions. 
We argue that the partons of the jet should acquire a non-zero average polarisation proportional to the medium anisotropy.
We first observe that the
medium anisotropy introduces a difference between the rates for transverse momentum broadening along the two directions perpendicular
to the jet axis. In turn, this difference leads to a polarisation-dependent bias in the BDMPS-Z rates for medium-induced gluon branching.
Accordingly, the daughter gluons in a branching process can carry net polarisation even if their parent gluon was unpolarised.
Using these splitting rates, we construct kinetic equations which describe the production and transmission of polarisation
via multiple branching in an anisotropic medium. The solutions to these equations show that polarisation is efficiently
produced via quasi-democratic branchings, but then it is rapidly washed out by the subsequent branchings, 
due to the inability of soft gluons to keep trace of the polarisation of their parents.  
Based on that, we conclude that a net polarisation for the jet should survive in the final state if and only if
the medium anisotropy is sizeable as the jet escapes the medium.}
\begin{document}
\maketitle

\section{Introduction}


The physics of ``jet quenching'' --- a rather general concept which encompasses the ensemble of the modifications suffered
by an energetic jet or hadron which propagates through a quark-gluon plasma --- represents one of our
main sources of information about the properties of the dense QCD medium created in the intermediate stages of
ultrarelativistic heavy-ion collisions at RHIC and the LHC \cite{Casalderrey-Solana:2007knd,Qin:2015srf,Blaizot:2015lma}. 
The theory of jet quenching was originally developed for a limited set of phenomena
--- collisional transverse momentum broadening and medium-induced radiative energy loss ---,
for a relatively simple ``hard probe'' (an energetic parton), and for a weakly-coupled plasma in thermal equilibrium. More recently, 
the theory and phenomenology of jet quenching have progressively been extended to genuine jets (including
vacuum-like and medium-induced parton cascades), to more complex aspects of the jet-medium interactions,
like colour decoherence or medium back-reaction, and to a plasma which is far away from thermal equilibrium and
might be even strongly coupled. At the same time, the spectrum of associated observables has extended from 
inclusive quantities like the nuclear suppression of hadron and jet production (largely controlled by the in-medium
energy loss), to extremely complex, global, phenomena like the dijet asymmetry  (which likely probes all the stages of the in-medium
evolution, as well as the fine structure of the medium), and also to fine probes of the jet substructure 
(which arguably lies within the realm of QCD perturbation theory).

In this paper, we propose another effect that should emerge from the jet interactions in a dense QCD medium:
if the medium is anisotropic, then the jet partons should acquire net polarisation. There are several mechanisms for generating such an
anisotropy for the quark-gluon plasma created in the intermediate stages of a nucleus-nucleus collision. 

The high energy of experiments
naturally leads to a rapid longitudinal expansion of the quark-gluon plasma (QGP) medium created in the wake of a collision \cite{Bjorken:1982qr}.
In turn, the expansion of the QGP medium
leads to a pronounced anisotropy in the momentum distribution of the medium constituents: their momentum component $p_z$ along 
the collision axis is typically much smaller than the transverse components $p_x$ and $p_y$  \cite{Baier:2000sb}.
This has interesting consequences for various probes of the medium,  such as stronger binding of quarkonia and preferred orientation of the quarks comprising a quarkonium \cite{Dumitru:2009ni, Burnier:2009yu, Thakur:2012eb,Dong:2022mbo}, angular dependence in energy loss and momentum broadening of heavy quarks \cite{Prakash:2021lwt, Song:2019cqz, Romatschke:2004au} and a modification of the spectrum of dileptons radiated by the plasma \cite{Ryblewski:2015hea, Churchill:2020uvk, Coquet:2021lca}. 
An anisotropy in the transverse plane is possible as well, especially in not so central collisions where it leads to the elliptic flow of hadrons. This transverse anisotropy will be ignored in our subsequent study.

Another scenario leading to medium anisotropy is inherent in the glasma picture for the early stages
\cite{Lappi:2006fp}. The ``glasma''
(the precursor of the quark-gluon plasma) is a form of gluonic matter with large occupation numbers, that can be conveniently 
described in terms of  strong, classical, colour fields. The underlying theory (the ``colour glass condensate'', or CGC
\cite{Iancu:2003xm,Gelis:2010nm,Gelis:2012ri,Gelis:2015gza})  predicts that, right after the
collision, the distribution of these fields should be strongly anisotropic: their energy density is concentrated within longitudinal
(chromo-electric and chromo-magnetic) flux tubes which extend between the recessing nuclei.  This anisotropy has been argued to have observable consequences, such as inducing spin polarization of heavy quarks in the glasma \cite{Kumar:2022ylt}.

In both the case of the QGP medium and of the glasma, the anisotropy is expected to decrease with time. The longitudinal expansion of the QGP medium should compete with elastic collisions
among the plasma constituents which redistribute energy and momentum, and thus broaden the originally anisotropic momentum distribution.
In the glasma scenario, the longitudinal flux tubes are unstable and should eventually break and transmit their energy to gluons with transverse 
polarisations. Yet, explicit calculations in both scenarios --- lattice calculations for the glasma \cite{Berges:2013eia,Berges:2013fga,Ipp:2020mjc,Ipp:2020nfu} 
(that is, classical Yang-Mills theory with initial conditions from the CGC) and, respectively, numerical solutions to kinetic theory for a 
quark-gluon plasma undergoing boost-invariant longitudinal expansion \cite{Kurkela:2015qoa,Kurkela:2018vqr,Du:2020dvp}
and with initial conditions inspired by the ``bottom-up'' scenario \cite{Baier:2000sb} --- demonstrate that the process of isotropisation 
proceeds only slowly,
so that a sizeable anisotropy persists at all the times that should be relevant for the phenomenology of heavy ion collisions.

Furthermore, both scenarios predict that the medium anisotropy should affect the transverse momentum broadening of 
an energetic probe (parton or jet) which propagates through the plasma: if the hard probe propagates at central rapidities, 
say along the $x$ axis, then one should observe an asymmetry between its momentum broadening along the two directions orthogonal 
to the jet axis, that is, the $y$ and $z$ directions. For the glasma case,  lattice
calculations of the relevant chromo-electric and chromo-magnetic field correlators (2-point functions of the non-Abelian
Lorentz force) have found sizable anisotropy in momentum broadening \cite{Ipp:2020mjc,Ipp:2020nfu}.  These results are furthermore supported by analytic approximations \cite{Carrington:2021dvw,Carrington:2022bnv}. 

For a weakly-coupled plasma --- the case that we shall focus on in this paper ---, 
the transverse momentum broadening is a consequence of quasi-local elastic collisions and the respective rates
--- the jet quenching parameters $\widehat q_y $ and $\widehat q_z$ --- can in principle be computed within perturbation theory.
Yet, perturbative calculations for anisotropic plasmas appear to be unstable, due to the non-Abelian analog of the Weibel instabilities
(e.g., the polarisation effects introduce a pole at space-like momenta in the retarded propagator) \cite{Mrowczynski:1993qm,Mrowczynski:2016etf, Hauksson:2020wsm}. Whereas for electromagnetic
plasma such instabilities are physical and lead to charge filamentation, the fate of instabilities in a  non-Abelian plasma is less clear.
Classical Yang-Mills simulations on the lattice \cite{Berges:2013eia,Berges:2013fga}
point to the fact that plasma instabilities do not appear to play a dominant role in the non-equilibrium evolution of the glasma beyond very early times. 
So, in what follows we shall ignore this problem and assume that the physics of transverse momentum broadening in an
anisotropic plasma can be faithfully described in terms of two (generally different)  jet quenching parameters,
 $\widehat q_y $ and $\widehat q_z$, even though for the time being we still lack an explicit calculation of these parameters from
 first principles (see however \cite{Arnold:2002zm,Baier:2008js,Romatschke:2006bb,Hauksson:2021okc}).  

For simplicity we assume the medium to be homogeneous and static, although it
should be possible to incorporate the time-dependence of \(\widehat{q}\) due to longitudinal expansion of the medium along the collision axis
by following the discussions in \cite{Baier:1998yf,Zakharov:1998wq,Baier:2000sb,Arnold:2008iy,Iancu:2018trm,Adhya:2019qse,Caucal:2020uic}. Relaxing the assumption of a homogeneous medium introduces new effects in which gradients in temperature and density, as well as net flow of the medium, change the spectrum of radiated partons \cite{Barata:2022krd,Sadofyev:2021ohn,Andres:2022ndd}. These medium inhomogeneities arise naturally in the framework of ideal hydrodynamics where temperature and flow velocity vary in space and time but the medium is everywhere in local thermal equilibrium. Here our goal is to consider genuine non-equilibrium effects and how they modify the spectrum of radiated jet partons. In other words, we consider a QGP medium that is \emph{locally} out of equilibrium, as e.g. captured by pressure anisotropy and viscous corrections in second order hydrodynamics or non-equilibrium momentum distributions in kinetic theory. Such a non-equilibrium description of the QGP medium is known to be necessary to describe measurements of soft hadrons, see e.g. \cite{Romatschke:2007mq}. Conveniently, for jet partons at sufficiently high energy where the so-called ``harmonic approximation'' is valid, the effect of a non-equilibrium medium on the jet is completely captured by the two jet quenching parameters \(\widehat{q}_y\) and \(\widehat{q}_z\) such that a detailed microscopic description of the medium is not needed.  Our work could be extended to include medium inhomogeneities, in addition to the  homogeneous local deviations from equilibrium considered here.

 Within this framework, our main observation is that the medium anisotropy  should also introduce a polarisation-dependent
  bias in the rates for medium-induced gluon branching ($g\to gg$). Thus the two daughter gluons can carry non-zero net polarisation even when the
parent gluon is unpolarised. We demonstrate this effect within the BDMPS-Z approach, where medium-induced radiation 
is linked to transverse momentum broadening: parton branching is triggered by multiple soft scattering, leading to a loss
of coherence between the daughter partons and their parent. This approach has been originally developed~\cite{Baier:1996kr,Baier:1996sk,Zakharov:1996fv,Zakharov:1997uu,Baier:1998kq,Baier:1998yf,Zakharov:1998wq,Wiedemann:1999fq,Wiedemann:2000za}  for an isotropic plasma and for the branching of unpolarised partons, but here we shall provide its
generalisation to an anisotropic medium and to gluons with definite polarisation states. (We leave the inclusion of
quarks to a further study.)

By using the in-medium branching rates for polarised gluons, we construct kinetic equations which describe the evolution of the jet
energy and polarisation distributions via multiple branchings. The equation satisfied by the unpolarised distribution is
formally the same as for an isotropic medium and its solution is well understood  \cite{Blaizot:2013hx,Blaizot:2013vha}: it exhibits {\it wave turbulence},
that is, the soft gluons rapidly multiply via quasi-democratic branchings, thus allowing for an efficient transfer of energy from the leading parton
to low-energy gluons. By using both numerical methods and analytic approximations, we also solve the equation for the polarised distribution which quantifies the degree of polarisation of partons at a given energy. We do this for the case where the leading parton is originally unpolarised. We find that the democratic branchings play an important role for the polarised distribution,
with two opposite effects. On one hand, democratic branchings enhance the net polarisation via the branching of unpolarised partons
in the presence of the anisotropy. On the other hand, they rapidly randomise the parton polarisation, due to the inability of the soft
gluons to keep trace of the polarisation of their parents. The competition between these two tendencies implies that the parton
polarisation is created and washed out quasi-locally in energy and time, so that the polarised distribution closely follows the unpolarised one:
for sufficiently low energies, these two distributions are simply proportional to each other, with a proportionality coefficient which
would vanish for an isotropic medium. In particular, the polarised distribution shows the same characteristic enhancement at low-energies,
which is the hallmark of the BDMPS-Z spectrum and also a fixed point of the turbulent parton cascade.

The previous considerations imply that net  polarisation of jet partons is only present in the final state if the medium anisotropy
survives until the moment when the jet exits the medium. Indeed, at least for the ideal, turbulent cascade that we have
studied here, any net polarisation that might be acquired at early stages will be rapidly washed out (via multiple branching)
if the medium becomes fully isotropic. Conversely, an experimental observation that indicates a net polarisation 
of partons in a jet,  tells us that the medium anisotropy had persisted until relatively late times. It  furthermore gives an indication of the degree of anisotropy as the jet escapes the medium. 
Measurements of net jet polarisation could be signalled e.g. by a larger-than-expected abundance of non-zero spin hadrons in the jet
 fragmentation or indirectly through anisotropy in the distribution of hadrons in the jet cone.
Thinking further ahead, one could envisage correlating measurements of jet polarisation with jet energy loss, which probes the path length of the jet in medium and thus the time at which a jet escapes the medium. This might give the anisotropy of the medium at different times as the escape time for jets varies between collisions.


So far, we have considered the effect of the medium anisotropy on the polarisation of the partons in a jet, but one could
also imagine a scenario where this polarisation is transmitted back to the medium. This is related to the late stage of the bottom-up scenario,
which predicts that most of the energy of the medium comes from the quenching of mini-jets in the presence of a highly anisotropic background. The soft gluons created by the decay of the mini-jets are
expected to carry net polarisation and thus produce a polarised quark-gluon plasma. It would be interesting to identify observables for
such a polarised medium at early stages, like the spin distribution of heavy quarks.

This paper is organised as follows. Sect.~\ref{sec:ptbroad} presents general considerations about the collisional momentum broadening
of an energetic parton propagating through a non-equilibrium, weakly-coupled, quark gluon plasma, which is anisotropic. 
Our main purpose here is to motivate the relation between the anisotropy in the momentum distributions of the plasma constituents and that
in the  transverse momentum broadening of the hard probe. Then in Sect.~\ref{sec:1br} we investigate the consequences of this anisotropy on the 
branching rates for medium-induced  emissions of linearly polarised gluons.
After quickly re-deriving the polarised version of the DGLAP splitting functions, we proceed with extending the BDMPS-Z formalism to
an anisotropic medium and to polarised gluons. Our main result in that section is a set of polarised in-medium branching rates, shown in 
Eqs.~\eqref{Eq:Gammaxtox_full}--\eqref{Eq:Gammaytoy_full}, whose physical content is briefly discussed  in Sect.~\ref{sec:1split}. In
Sect.~\ref{sec:evol} we study the evolution of the jet distribution in energy and polarisation via multiple branching. We start by constructing
the respective evolution equations --- a set of coupled kinetic equations for the polarised and the unpolarised (or total) gluon distributions.
The equation for the unpolarised distribution is essentially the same as for an isotropic plasma (it differs only by a rescaling of the time
variable), so its solution is well known \cite{Blaizot:2013hx}. In Sects.~\ref{sec:Green} and~\ref{sec:sol}, we also solve the equation for the polarised
distribution, via the Green's function method. Our main results are encoded in Fig.~\ref{fig:Dtilde} together with the analytic approximation 
\eqref{polD}, valid for soft gluons. We summarise our conclusions together with some open problems in Sect.~\ref{sec:conc}.

\section{Transverse momentum broadening in an anisotropic plasma}
\label{sec:ptbroad}

For simplicity, we consider a jet made only of gluons. This is a good approximation at large number of colours \(N_c\). 
The jet is initiated by a ``leading gluon'' which propagates along the $x$ axis. The secondary gluons produced 
via successive gluon branchings are assumed to be quasi-collinear to the leading parton, hence to the $x$ axis.
In the BDMPSZ approach, the physics of medium-induced radiation is linked to that of transverse momentum broadening,
where by ``transverse'' we now mean the $(y,z)$ plane, which is perpendicular to the jet direction of motion. 
Since the jet dynamics is most naturally analysed w.r.t. the jet axis, we shall systematically use this convention from now
on: by ``transverse'' we shall always mean the $(y,z)$ plane, and not the plane $(x,y)$ which is orthogonal to the 
collision axis. Similarly, the $x$ axis will often be referred to as ``longitudinal''. 

The ``transverse momentum broadening'' in this context means that the partons from the jet 
suffer independent collisions with the plasma constituents, leading to a broadening of their
3-momentum distribution along the $y$ and $z$ directions: the transverse momenta $p_y$ and $p_z$,
as accumulated via collisions, are relatively small ($p_y^2,\,p_z^2\ll p_x^2$) and random, with dispersions which grow
linearly in time: $\langle p_y^2\rangle=\widehat q_y \Delta t$ and $\langle p_z^2\rangle=\widehat q_z \Delta t$.
The hallmark of an anisotropic plasma is the fact that the rates, $\widehat q_y $ and $\widehat q_z$, 
 for transverse momentum broadening along the two transverse directions are not the same:
\beq\label{hatqyz}
\widehat{q}_y\equiv  \frac{d \langle p^2_{y} \rangle}{dt}\, \ne\,
\widehat{q}_z \equiv \frac{d \langle p^2_{z} \rangle}{dt}
\,.
\eeq

%

Inspired by recent calculations of momentum broadening in the Glasma \cite{Ipp:2020nfu,Ipp:2020mjc,Carrington:2021dvw}, 
we shall sometimes assume that $\widehat{q}_z > \widehat{q}_y$ ---
the momentum broadening is stronger along the collision axis than perpendicular to it. That said,
our subsequent results do not crucially depend upon the sign of the difference $\widehat{q}_z - \widehat{q}_y$:
all that matters is the fact that these quantities are generally different. The limit of an isotropic plasma can be easily
obtained from our results by letting $\widehat{q}_z = \widehat{q}_y\equiv \hat q/2$.

Although the physical origin of the anisotropy is not essential for what follows,
it is still interesting to understand how the difference  between $\widehat{q}_z$ and $\widehat{q}_y$
may arise in the case of a weakly coupled quark-gluon plasma. To that aim, we shall briefly recall the
respective calculation of transverse momentum broadening, with emphasis on the possible sources of
anisotropy.

Before that, let us summarise our notations. The Minkowski coordinates of a generic space-time point $x$ 
will be written as $x^\mu=(t,x,y,z)=(t,\vec x) = (t,x, \bx)$, where the 3-vector $\vec x\equiv (x,y,z)$ encompasses
all the spatial coordinates, while the 2-vector $\bx\equiv (y,z)$ refers to the transverse plane alone. Similarly,
for the 4-momentum $p$ we will write $p^\mu=(p_0, p_x, p_y, p_z)=(p_0, \vec p) = (p_0, p_x, \bp)$. To study the
interactions between the jet and the medium, it will also be convenient to use light-cone vector notations
w.r.t. to the jet ($x$) axis,  defined as
\beq\label{LCdef}
x^+\equiv\frac{t+x}{\sqrt{2}},\quad x^-\equiv\frac{t-x}{\sqrt{2}},\quad p^+\equiv\frac{p_0+p_x}{\sqrt{2}},\quad
p^-\equiv\frac{p_0-p_x}{\sqrt{2}}\,.\eeq
The variable $x^+$ plays the role of the LC time, while $p^+$ is the LC longitudinal momentum.
To simplify wording and notations, we shall generally refer to $x^+$ simply as ``time'' (and use the simpler notation
$t\equiv x^+$), while $p^+$ will be the ``energy'' (sometimes denoted as $\omega$). Also, we shall ignore the ``perp'' subscript
$\perp$ on transverse vectors whenever there is no possible confusion.
 In LC notations, the 4-momentum of a parton reads 
$p^\mu=(p^+, p^-, \bp)$, where $\bp=(p_y, p_z)$ is the transverse momentum 
and $p^-=\bp^2/2p^+$ for an on-shell massless parton.


To study transverse momentum broadening, we consider the propagation of 
an energetic gluon (the ``hard probe'') through a weakly-coupled QGP.  The gluon longitudinal momentum $p^+$
is assumed to be much larger than both the (longitudinal and transverse) momenta of the plasma constituents
and the transverse momentum $\bp$ acquired by the probe via collisions. In this high-energy kinematics,
the hard gluon predominantly couples to the light-cone component $A^-\equiv (A_0-A_x)/\sqrt{2}$ 
of the gauge field generated by the plasma constituents. 
So, for the present purposes, the medium can be described 
as a random colour field $A^-_a$ with 2-point correlation function
\begin{align}\label{2pA}
\langle {A}^-_a(x^+, x^-, \bx){ A}^-_b(y^+,y^-,\by)\rangle
\,=\,\delta_{ab}
 \delta(x^+-y^+)\,\gamma(\bx-\by),
\end{align}
where the angular brackets denote the medium average. The restriction to the 2-point function 
is justified to leading order at weak coupling. The $\delta$--function in colour space, $\delta_{ab}$,
follows from gauge invariance, while that in LC time, $ \delta(x^+-y^+)$, from Lorentz time dilation (the energetic
parton has a poor resolution in $x^+$, hence it ``sees'' the medium correlations as quasi-local in time).
The 2-point function is independent of $x^-$ and $y^-$,
since the medium is probed near the trajectory of the energetic parton, at $x^-\simeq 0$. This is in agreement
with our assumption that the plasma fields $A^-_a$ carry relatively small longitudinal momenta $k^+\ll p^+$.
In turn, this implies that both the longitudinal momentum $p^+$ of the hard gluon and its polarisation are not
effected by the medium\footnote{These features are reminiscent of the eikonal approximation, but our
calculation is in fact more general --- we do not need to assume that the hard probe preserves a straight line
trajectory with a fixed transverse coordinate. Indeed, such an assumption would not be justified for the
gluons produced via medium-induced emissions (see below).}. Finally, $\gamma(\bx-\by)$ depends only upon
the transverse {\it separation} $\bx-\by$ by homogeneity, and is independent of $x^+$ because the plasma 
is assumed to be static. (More general situations, e.g. a plasma which expands along the collisional axis, can be
similarly considered.)

In this set-up, the medium anisotropy is encoded in the fact that the 2-point correlation $\gamma(\bx-\by)$ 
has no rotational symmetry in the transverse plane, that is, it  is {\it not} just a function of the distance 
$r\equiv |\bx-\by|$. When Fourier transformed to transverse momentum space, \eqn{2pA} becomes
(we suppress the irrelevant variables $x^-$ and $y^-$)
\beq\label{correlmed}
\left\langle A^-_a(x^+,\bk)A^{*-}_b(y^+,\bq)\right\rangle=
\delta_{ab}\delta(x^+-y^+)(2\pi)^2\delta^{(2)}(\bk-\bq)\,\tilde\gamma(\bk)\;.
\eeq
For an anisotropic case, the function $\tilde\gamma(\bk)$ (a.k.a. the ``collision kernel'')
depends upon the orientation of the 2-dimensional vector 
$\bk=(k_y, k_z)$, that is, it {\it separately} depends upon its 2 components.

To gain more insight in the structure of $\tilde\gamma(\bk)$ and thus
understand how an anisotropy might be generated, it is instructive to recall the relation between the
2-point function of the field and that of its colour sources (the quarks and gluons from the plasma).
Our discussion will be very schematic, since merely intended for illustration purposes. In particular,
we shall often ignore the Minkowski and colour indices, and write the gluon correlator simply as\footnote{In
general, this is a Minkowski tensor, $\langle {A}^\mu_a(x){ A}^\nu_b(y)\rangle=\delta_{ab} G^{\mu\nu}(x-y)$;
here, we only need its particular projection 
$G^{--}=n_\mu n_\nu  G^{\mu\nu}$, with $n^\mu=\delta^{\mu+}$ (hence, $n\cdot A= A^-$).
Also, the precise gauge choice is unimportant so long as we choose a gauge (such as the covariant
gauge, or the LC gauge $A^+=0$) in which the component $A^-$ is non-zero.}
$ G(x,y)\equiv \langle {A}(x){ A}(y)\rangle$. For a plasma which is static and homogeneous, 
this admits the Fourier representation
\begin{align}\label{G<}
G(x-y)&=\int\frac{\rmd k^-}{2\pi}\,\rme^{-ik^-(x^+-y^+)} \int\frac{\rmd k^+}{2\pi}\,
\rme^{-ik^+(x^--y^-)}
\int\frac{\rmd^2\bk}{(2\pi)^2}\, \rme^{i\bk\cdot(\bx-\by)}\,G(k^+,k^-,\bk)
\nonumber\\*[0.2cm]
&\simeq \int\frac{\rmd k^-}{2\pi}\,\rme^{-ik^-(x^+-y^+)}
 \int\frac{\rmd k^+}{2\pi}\int\frac{\rmd^2\bk}{(2\pi)^2}\, \rme^{i\bk\cdot(\bx-\by)}\,G(k^+,k^-=0,\bk),
\end{align}
where the second lines was obtained by using $x^-\simeq y^-\simeq 0$ (more precisely, 
$x^-\sim y^-\sim 1/p^+$ with $p^+\gg k^+$) and by approximating $k^-\simeq 0$ inside the correlator:
this is appropriate since the LC times $x^+$ and $y^+$ take relatively large values $x^+\sim y^+\sim 1/p^-$,
with $p^-=\bp^2/2p^+$ much smaller than the {\it typical} values of $k^-$. Hence, the integral is controlled
by unusually small values of $k^-$ (to avoid large oscillations of the phase $\rme^{-ik^-(x^+-y^+)}$,
so the function $G$ inside the integrand can be evaluated at $k^-=0$.
This condition $k^-=0$, or $k_0=k_x$, shows that we consider {\it space-like modes} of the
gluon correlator: $k^2=k_0^2-\vec k^2=-k_\perp^2 < 0$, where $\vec k=(k_x, k_y, k_z) =(k_x, \bk)$.

The integral over $k^-$ generates the $\delta$--function $\delta(x^+-y^+)$, so the final result in
 \eqn{G<} is indeed consistent with \eqn{correlmed}, with the following representation for the collision kernel:
 \beq\label{tildegamma}
\tilde\gamma(\bk)\simeq \int\frac{\rmd k^+}{2\pi}\,G^{--}(k^+,k^-=0,\bk).\eeq
For space-like modes, the statistical 2-point function of the gluon field emitted by the plasma
constituents has the general structure
\beq\label{GR}
G(k)\,=\,|G_R(k)|^2\,\Pi(k),
\eeq
where $G_R(k)$ is the retarded propagator and $\Pi(k)$ is the gluon polarisation tensor.
This structure can be understood as follows: introducing the colour charge density $\rho(k)$
of the plasma constituents, the (event-by-event) gauge field is schematically obtained as $A(k)=G_R(k)\rho(k)$,
hence its 2-point function has indeed the structure \eqref{GR} with $\Pi(k)=\langle \rho(k)\rho(-k)\rangle$
(the charge-charge correlator). The retarded propagator too is modified by polarisation effects, via the
retarded version of the polarisation tensor: $G_R^{-1}=G_{0, R}^{-1}+\Pi_R$, with $G_{0, R}$ the free retarded
propagator. 

Due to the mobility of the colour charges, the polarisation tensor is
{\it non-local}, i.e $\Pi(k)$ is a non-trivial function of the 4-momentum $k^\mu=(k_0, \vec k)$.
This non-locality reflects the momentum distribution of the plasma constituents: when this distribution
happens to be anisotropic, it implies a corresponding anisotropy in the ``collision kernel''
$\tilde\gamma(\bk)$. To be more specific, let us remind the reader of the expression of the polarisation tensor
in the Hard Thermal Loop (HTL) approximation. 

The HTLs describe the response of the medium
constituents to long-wavelength excitations, such as the soft gluons exchanged in elastic collisions
(see Refs. \cite{Blaizot:2001nr,Kapusta:2006pm}  for pedagogical discussions).
Originally introduced in the context of thermal equilibrium 
\cite{Braaten:1989mz,Frenkel:1989br,Blaizot:1993zk},  the HTLs have subsequently been extended to more general, 
non-equilibrium, situations, 
that can be still described in terms of quasi-particles and their occupation numbers. (See notably 
Refs.~\cite{Arnold:2002zm,Mrowczynski:2000ed,Romatschke:2003ms,Mrowczynski:2004kv,Hauksson:2021okc}
for HTL studies in the context of anisotropic plasmas.) For a static and homogeneous plasma, we shall denote
these occupation numbers as $f_g(\vec p)$ and $f_q(\vec p)$ for (on-shell) gluons and quarks, respectively.
In thermal equilibrium, the quasi-particles have typical energies and momenta of the order of the
temperature, $p_0=p\sim T$ (with $p=|\vec p|$), whereas the typical momenta exchanged via (small-angle) 
elastic collisions
are much softer: $k_0, k \sim gT\ll T$. Out of equilibrium, we shall assume that a similar hierarchy exists, between
the ``hard'' momenta of the medium constituents and the ``soft'' exchanges. 
Under these assumptions, the leading-order expression of the  polarisation tensor for soft gluons is
given by the gluon HTL, which reads (after restoring the Minkowski indices)
\beq\label{PiHTL}
\Pi^{\mu\nu}(k)\,=4\pi g^2 \int\frac{\rmd^3\vec p}{(2\pi)^3}\left[N_c f_g(\vec p)(1+f_g(\vec p))
+N_f f_q(\vec p)(1-f_q(\vec p))\right]
v^\mu v^\nu \,\delta(v\cdot k),
\eeq
where $v^\mu\equiv p^\mu/p = (1,\vec v)$ with $\vec v=\vec p/p$ (the particle velocity).
The piece proportional to $N_c$ ($N_f$) is the gluon (quark) contribution. The support of the $\delta$--function 
reflects the microscopic origin of the plasma polarisation ---
the absorption of the soft gluon by a plasma constituent  (``Landau damping''): for this process to be
kinematically allowed, the soft gluon must be space-like.
The argument of the  $\delta$--function can be rewritten as $v\cdot k=(p\cdot k)/p$ with 
\beq
p\cdot k 
=p^+k^-+p^-k^+-\bp\cdot\bk\simeq p^-k^+-\bp\cdot\bk,\eeq
where the last, approximate, equality, uses $k^-=0$, as in \eqn{G<}. 
So long as the transverse momentum $\bp$ of the hard parton
is kept fixed, the integrand of \eqn{PiHTL} is clearly anisotropic in the transverse plane: 
it depends upon the azimuthal angle $\phi$
between $\bk$ and $\bp$. 

In thermal equilibrium, this anisotropy is washed out by the integral over $\bp$, because
the thermal distributions  --- the Bose-Einstein distribution $f_g^{\rm eq}(p)$ for gluons and the Fermi-Dirac
distribution $f_q^{\rm eq}(p)$ for quarks --- are themselves isotropic.
It is then straightforward to perform the integral over $p$ and find 
\begin{align}\label{EqHTL}
\Pi^{\mu\nu}(k_0,\vec k)&\,=2\pi m_D^2 T\int\frac{\rmd\Omega}{4\pi}\,v^\mu v^\nu
\delta(k_0-\vec v\cdot\vec k), \end{align}
where the angular integral runs over the directions of the unit vector $\vec v$ and 
\beq
m_D^2
=\,(2N_c+N_f)\frac{g^2T^2}{6}\,,
\eeq
is the Debye mass. We also used the following integrals for the thermal distributions:
\begin{align}
\int\rmd p \,p^2 f_q(p)(1-f_q(p))=-T\int\rmd p\, p^2\frac{\rmd f_q}{\rmd p}=2T\int\rmd p\, pf_q(p)=\frac{\pi^2 T^3}{6}\,, 
\nonumber\\*[0.2cm]
\int\rmd p \,p^2 f_g(p)(1+f_g(p))=-T\int\rmd p\, p^2\frac{\rmd f_g}{\rmd p}=2T\int\rmd p\, pf_g(p)=\frac{\pi^2 T^3}{3}\,.
 \end{align}

Still in thermal equilibrium, the statistical self-energy $\Pi$ and the retarded one $\Pi_R$ are related 
via the KMS condition, which implies (for soft gluon modes with $k_0\ll T$)
\beq
\Pi^{\mu\nu}(k)\,=\,-\frac{T}{k_0}\,2\,{\rm Im}\, \Pi_R^{\mu\nu}(k)\,.
\eeq
Using this relation together with \eqn{EqHTL} one finds the expected result
for the imaginary part of the retarded HTL  \cite{Blaizot:2001nr,Kapusta:2006pm}.
The Debye mass $m_D^2$
acts as a screening mass in the propagator of the longitudinal gluon and also in the collision kernel.

Specifically, in thermal equilibrium and to leading-order in perturbative QCD, the collision kernel  \eqref{tildegamma} 
has been computed as\footnote{See e.g. Eqs.~(A.1)--(A.3)
in \cite{CaronHuot:2008ni}; as compared to those equations, our result includes an additional factor of $\sqrt{2}$
because the integration variable in \eqref{tildegamma} is $k^+=\sqrt{2}k^0$, rather than $k_0$.}
 \cite{Aurenche:2002pd,CaronHuot:2008ni}
\beq\label{gammaeq}
\tilde\gamma(k_\perp)
\,=\,\frac{1}{\sqrt{2}}\, \frac{Tm_D^2}{k_\perp^2(k_\perp^2+m_D^2)}
,\eeq
where the first (second) terms inside the parentheses refers to the exchange of a transverse (longitudinal)
gluon. For relatively large momenta $k_\perp\gg m_D$, we recognise the characteristic power tail 
$\sim 1/k_\perp^4$ of Rutherford scattering, but the singularity at low momenta $k_\perp\to 0$ gets milder
($1/k_\perp^2$ instead of $1/k_\perp^4$), due to Debye screening for $k_\perp\lesssim m_D$.

Out of equilibrium, the momentum distributions can be anisotropic for a variety of reasons.
One  natural mechanism in that respect is the expansion of the plasma along the collision ($z$) axis: 
the partons liberated in a heavy-ion collision
should follow straight-line trajectories and segregate themselves in $z$ according to their longitudinal velocity
$v_z = p_z/p$. This ``free-streaming'' scenario, expected to hold at very early stages, leads to an 
anisotropic momentum distribution, which is strongly oblate: $\langle p_z^2\rangle \ll \langle p^2\rangle $  in the local rest frame. 
With increasing time, the interactions are expected to become more important (notably, due to the emission of soft
gluons) and to reduce the anisotropy via elastic collisions.  In this ``bottom-up'' scenario for thermalisation \cite{Baier:2000sb},
that is supported by numerical solutions to kinetic theory \cite{Kurkela:2015qoa,Kurkela:2018vqr,Du:2020dvp},  the plasma is predicted
to slowly evolve towards isotropisation and eventually reach thermal equilibrium at very large times
--- much larger than the lifetime of the quark-gluon plasma produced in heavy ion collisions.

During this slow approach to isotropisation at late stages\footnote{We recall that our main goal in this paper is to
study medium-induced radiation, which is controlled by the behaviour at large times, of the order of the distance
$L$ travelled by the jet through the medium.}, it seems legitimate to neglect the time-dependence of the anisotropy.
In what follows, we shall make the stronger assumption that the medium is static as a whole --- that is, we also neglect its longitudinal expansion
along the collision axis. This assumption is only intended for simplicity and can be relaxed in further work: the effects of the medium expansion
can be included in an adiabatic approximation (e.g. by allowing the jet quenching parameters, $\widehat q_y $ and $\widehat q_z$, to
be time-dependent), as in  previous studies \cite{Baier:1998yf,Zakharov:1998wq,Baier:2000sb,Arnold:2008iy,Salgado:2002cd,Salgado:2003gb,Adhya:2019qse,Iancu:2018trm,Caucal:2020uic}
which assumed an isotropic medium.
For such a static but anisotropic medium, the momentum distributions of the plasma constituents take the generic form
\beq\label{Naniso}
f_\nu(\vec p) \,=\,F\left(\sqrt{p^2+\nu p_z^2}\right)
 ,\eeq
where the parameter $\nu$ characterises the strength of the anisotropy: a positive value $\nu > 0$ implies an oblate
distribution with $\langle p_z^2\rangle < \langle p^2_x\rangle =\langle p^2_y\rangle $. The precise form of the function $F$ is unimportant for what
follows (in some calculations, this is taken to be the equilibrium distribution \cite{Romatschke:2003ms,Hauksson:2021okc}). Using a momentum distribution such as Eq. \eqref{Naniso} in the general expressions in Eqs. \eqref{tildegamma}, \eqref{GR}, \eqref{EqHTL}, one would then expect to obtain an anisotropic collision kernel for momentum broadening \(\widetilde{\gamma}(\mathbf{k}_{\perp})\) from first principles. In practice this is hindered by e.g. plasma instabilities which are beyond the scope of this paper, see e.g. \cite{Hauksson:2021okc} for further discussion.

%


We now return to our energetic gluon probe and examine the effects of the collisions
on its transverse momentum distribution. This problem has been studied at length in the literature and here we shall
only focus on the new aspects which emerge when the medium is anisotropic. We would like to compute
compute the probability density ${\cal P}(\Delta\bp, \Delta t)$ for the gluon to acquire a transverse momentum 
$\Delta\bp$ after crossing the medium along a distance (or LC time) $\Delta t$. 
The calculation is most conveniently formulated in the transverse coordinate representation, which allows for an
efficient treatment of multiple scattering: the in-medium propagation of the test particle is governed 
by a 2-dimensional Schr\" odinger equation describing quantum diffusion in the random field $A^-$
(see Appendix~\ref{app:pt}). As well known, this  Schr\" odinger equation admits a formal solution
in terms of a path integral. By multiplying two such solutions --- for the gluon in the direct amplitude (DA) and in
the complex-conjugate amplitude (CCA), respectively -- and averaging over the random field $A^-$ according
to  \eqn{correlmed}, one effectively builds the $S$-matrix  $S( \br, \Delta t)$  
for the elastic scattering of a gluon-gluon dipole with transverse size $\br$. 
The probability density of interest is finally  obtained as the following Fourier transform  (see Appendix~\ref{app:pt} for details)
\beq\label{Pcoord}
{\cal P}(\Delta\bp, \Delta t)=\int \rmd^2\br \,\rme^{-i \Delta\bp\cdot\br}\,S( \br, \Delta t)\,,\qquad
S( \br, \Delta t)\,\equiv\,\rme^{-g^2{N_c}\Delta t [\gamma(0)-\gamma(\br)]}.
\eeq
The dipole $S$--matrix  obeys  $S(r\to 0)\to 1$ (``colour transparency''), which ensures the proper
normalisation for the probability density: 
\beq 
\int \frac{\rmd^2  \bp}{(2\pi)^2}\, {\cal P}(\bp, \Delta t)=1.\eeq
We are interested in the multiple-scattering regime where the distance $\Delta t $ travelled by the probe through the medium 
is much larger than the mean free path $\lambda_{\rm mfp}$ between two successive collisions. Accordingly, 
the momentum $\Delta p_\perp\equiv |\Delta \bp|$  accumulated during $\Delta t$
 is typically much larger than the Debye mass\footnote{For an anisotropic plasma, one can have different screening 
 masses along the $y$ and $z$ directions, but this difference should not matter to the leading-logarithmic accuracy
 of our calculation; see below.} $m_D$ (the typical momentum
 transfer in a single collision): $\Delta p_\perp \gg m_D$. In this regime, the function in the exponent of $S$, that is,
\beq\label{sigmadipole}
\gamma(0)-\gamma(\br)=\int\frac{\rmd^2\bk}{(2\pi)^2}\left(1-\rme^{i\bk\cdot\br}\right)\tilde\gamma(\bk),
\eeq
has a logarithmic domain of integration at $m_D\ll k_\perp\ll 1/r\sim \Delta p_\perp$.  Indeed, the 
polarisation effects (in or out of thermal equilibrium) cannot modify the fact that
 $\tilde\gamma(\bk)\propto 1/k_\perp^4$ for sufficiently large momenta $k_\perp\gg m_D$.
Hence, in the leading-logarithmic approximation which only keeps the contribution enhanced by the
large logarithm $\ln\frac{1}{r m_D}$, one can evaluate the integral by expanding
 the exponential $\rme^{i\bk\cdot\br}$ within the integrand.  It is natural to assume reflexion symmetry, e.g.
$\tilde\gamma(-k_y, k_z)=\tilde\gamma(k_y, k_z)$ --- that is, $\tilde\gamma(\bk)$ is truly a function of $|k_y|$ and $|k_z|$.
Then the linear terms and also the crossed quadratic terms in the expansion cancel after the integration and the leading logarithmic contribution 
comes from the diagonal quadratic terms:
\beq\label{sigmaexp}
\gamma(0)-\gamma(\br)\simeq \frac{1}{2}\int\frac{\rmd^2\bk}{(2\pi)^2}\left(k_y^2 r_y^2+k_z^2 r_z^2\right)
\tilde\gamma(\bk).
\eeq
This finally implies
\beq\label{SGauss}
S(\br, \Delta t)\simeq\exp\left\{-\frac{\Delta t}{2}\left(\widehat{q}_y r_y^2+ \widehat{q}_z r_z^2\right)\right\},
\eeq
where
\beq\label{qhaty}
\widehat{q}_y \equiv \frac{g^2 N_c}{2}\int\frac{\rmd^2\bk}{(2\pi)^2}\,k_y^2 \,\tilde\gamma(\bk)
,\eeq
together with a similar expression for $\widehat{q}_z$. As already mentioned, the above integral has a logarithmic
divergence at large values of $k_\perp$ which must be cut off at $k_\perp \sim 1/r$. For instance, for a plasma
in thermal equilibrium, we can use \eqref{gammaeq} to deduce  $\widehat{q}_y= \widehat{q}_z=\hat q/2$,
with
\beq\label{hatqeq}
\hat q\simeq
\frac{\alpha_s N_c T m_D^2}{\sqrt{2}} \int^{1/r^2}\frac{\rmd k_\perp^2}{k_\perp^2+m_D^2}\simeq
\sqrt{2} \alpha_s N_c T m_D^2 \ln\frac{1}{r m_D}\,.
\eeq



Similar  logarithmic dependences upon the dipole size $r$ are expected for both 
 $\widehat{q}_y$ and $\widehat{q}_z$ in the case of an anisotropic plasma. Such dependences
complicate the final Fourier transform  to transverse momentum space, cf. \eqn{Pcoord}.
A common approximation at this level, known as the ``harmonic approximation'', 
consists in ignoring this residual $r$-dependence
 \cite{Baier:1998yf,Zakharov:1998wq}.
 This is appropriate for describing the effects of multiple soft scattering. With this approximation,  the 
 Fourier transform  is easily computed as
\beq
{\cal P}(\Delta\bp, \Delta t)=\frac{2\pi}{\sqrt{\widehat{q}_y\widehat{q}_z} \Delta t}\,\exp\left\{
-\frac{(\Delta p_y)^2}{2\widehat{q}_y  \Delta t}
-\frac{(\Delta p_z)^2}{2\widehat{q}_z  \Delta t}
\right\}.
\eeq
This Gaussian probability distribution immediately implies the expected results for anisotropic
momentum broadening, namely $\langle \Delta p_y^2\rangle=\widehat q_y \Delta t$ and
$\langle \Delta p_z^2\rangle=\widehat q_z \Delta t$.

Incidentally, the above discussion shows that \eqn{SGauss} can be rewritten as 
\beq\label{SGaus2s}
S(\Delta t, \br)\simeq\exp\left\{-\frac{1}{2}\left(r_y^2
\langle \Delta p_y^2\rangle +  r_z^2\langle \Delta p_z^2\rangle\right)\right\}.
\eeq
This formula is more general than our present considerations at weak coupling: it is also found
when non-perturbatively evaluating the Wilson loop  which reduces to our gluon-gluon
dipole correlator in the LC gauge  $A^+=0$
(see e.g. Eq.~(31) in \cite{Ipp:2020mjc}). This formula was used in \cite{Ipp:2020mjc} to extract
$\langle\Delta p_y^2\rangle$ and $\langle\Delta p_z^2\rangle$ from real-time lattice simulations of
the Glasma. In our perturbative approach, this formula comes together with results like
\eqn{qhaty}, which relate $\widehat{q}_y$ and $\widehat{q}_z$ to the microscopic structure of the medium.
%
%

\section{Polarised gluon splitting in an anisotropic plasma}
\label{sec:1br}

Besides providing transverse momentum broadening, the transverse ``kicks'' received by a test parton via collisions in the
plasma also have the effect to trigger radiation.  
The medium-induced radiation in the kinematical range of interest is controlled by the BDMPS-Z  
mechanism~\cite{Baier:1996kr,Baier:1996sk,Zakharov:1996fv,Zakharov:1997uu,Baier:1998kq,Baier:1998yf,Zakharov:1998wq,Wiedemann:1999fq,Wiedemann:2000za}, which takes into account the coherence effects associated with 
multiple soft scattering during the quantum formation of an emission. This mechanism is effective so long as 
the characteristic formation time $t_f=\sqrt{2\omega/\hat q}$, with $\omega$ the energy of the emitted gluon and
$\hat q$ the jet quenching parameter, is much larger than the parton mean free path $\lambda_{\rm mfp}=
m_D^2/\hat q$ between two successive collisions, but smaller
than the medium size $L$ available to the parent parton. (Note that we ignore the medium anisotropy for these
physical  considerations: its effects, to be later discussed, do not modify the general picture.) These conditions
imply an energy window $\omega_{\rm BH} \ll \omega \lesssim \omega_c$. The lower limit
$\omega_{\rm BH}\equiv m_D^4/2\hat q$, corresponding to $t_f\simeq\lambda_{\rm mfp}$,
separates from the Bethe-Heitler regime, where emissions are triggered by a {\it single}, soft, scattering. The upper limit
$\omega_c\equiv \hat q L^2/2$ is the maximal energy of a gluon emitted via multiple soft scattering, for which $t_f\simeq L$.

In what follows, we shall focus on the  {\it typical} gluon emissions, those with energies much smaller than
$\omega_c$, but much larger than $\omega_{\rm BH}$. Such emissions have relatively short formation
times, $t_f(\omega)\ll L$, hence a large emission probability,
\beq\label{Pdeltat}
 \omega\frac{\rmd \mathcal{P}}{\rmd \omega}\,\simeq\,\frac{\alpha_s C_R}{\pi} \,\frac{L}{t_f(\omega)}
 \,=\,\frac{\alpha_s C_R}{\pi} \,\sqrt{\frac{\omega_c}{\omega}}
 \,,\eeq
so for them the effects of {\it multiple branchings} are expected to be important \cite{Blaizot:2012fh,Blaizot:2013hx,Blaizot:2013vha}.
On the other hand, such emissions occur quasi-instantaneously, hence multiple emissions can be simply resummed
by solving appropriate {\it rate equations} \cite{Blaizot:2013hx,Blaizot:2013vha}. The only non-trivial ingredient of these
equations is the {\it emission rate} (the emission probability per unit time), which in turn can be inferred from the BDMPS-Z 
spectrum for a single gluon emission with $\omega\ll \omega_c$. So, our main objective in this section is to
generalise the BDMPS-Z branching rate to the case of an anisotropic plasma and to gluons with fixed polarisations.
As we shall see, this generalisation brings no conceptual difficulties: the treatment of multiple soft scattering in
an anisotropic medium has already been discussed in the previous section and the polarisation dependence
of the branching rate is fully encoded in the (leading-order) DGLAP splitting functions, 
which are well known --- including for polarised partons (see e.g. \cite{Ellis:1996mzs} and Sect.~\ref{sec:DGLAP} below).


%
%
%
%


After this preparation, let us start our study of polarised medium-induced gluon branching ($g\to gg$) in an anisotropic plasma.
Consider the branching process $a\to b+c$, where the gluon labels $a, b$ and $c$ encompass all the
relevant ``quantum numbers'' (longitudinal and transverse momenta, polarisation, and colour). Both the parent gluon and
the two daughter ones are assumed to be on-shell, yet the branching is kinematically allowed due to the collisions in the medium.
As already argued, the parton longitudinal momenta (``energies'') are not significantly changed by the interactions with the medium, 
so we can assume  energy conservation: $p^+_a=p^+_b+p^+_c$. It is then convenient to use the simpler notation
 $\omega\equiv p^+_a$ and introduce the splitting fractions  $\zeta\equiv p^+_b/\omega$ and $1-\zeta= p^+_c/\omega$ 
 of the daughter gluons.  
The transverse momentum balance is more subtle: this is conserved at the QCD splitting vertex,
which is local in time, but not also for the overall branching process, which is {\it non-local}. Physically,
this non-conservation is associated with the collisional broadening during the formation time, estimated as 
$\Delta k_\perp^2 \simeq \hat q t_f$ with $t_f=\sqrt{2\zeta(1-\zeta)\omega/\hat q}$. Mathematically, this is expressed
by the fact that the QCD splitting vertices occur at different times, $t_1$ and $t_2$, in the DA
 and, respectively, the CCA, so they generally involve different transverse
momenta on their external legs: this difference accounts for the additional momentum broadening
during the interval $t_2-t_1$.

Before we discuss the full process including medium effects, let us consider the QCD splitting vertex
for gluons with linear polarisations (this will be an ingredient of the complete calculation).


\subsection{DGLAP splitting functions for linearly polarised gluons}
\label{sec:DGLAP}

As announced, the special geometry of the medium, which distinguishes between momentum broadening along the
 $y$ and the $z$ directions, makes it useful to work with linearly polarised states which are aligned along
 these 2 directions. To specify these states, we first need to fix the gauge. The natural gauge for describing quantum
 evolution in LC time $x^+$ is the LC gauge $A^+=0$ (with $A^+\equiv (A_0+A_x)/\sqrt{2}$ of course).  In this gauge and in
 LC notations, $\epsilon^\mu=(\epsilon^+, \epsilon^-, \bm{\epsilon}_\perp)$,
the polarisation vectors describing linear polarisations along the two transverse directions $y$ and $z$ read as follows
(for a gluon with transverse momentum $\bk$ and longitudinal momentum $k^+$):
\beq\label{pol}
\epsilon^\mu(k, \lambda)=\left(0,\,\frac{\bm{\epsilon}_\lambda\cdot\bk}{k^+}\,,\,\bm{\epsilon}_\lambda\right),\quad
\bm{\epsilon}_y=(1,0),\ \ \bm{\epsilon}_z=(0,1)\,,
\eeq
where $\lambda$ is the polarisation index: $\lambda=y$ or $z$.
These vectors obey $k\cdot \epsilon(k, \lambda)=0$.

The QCD vertex for the $g\to gg$ branching process has the familiar structure
\beq\label{VQCD}
V_{abc}^{\mu\nu\rho} (p_a,p_b,p_c)=
 ig f^{abc} \left[g^{\mu\nu} (p_a + p_b)^{\rho} + g^{\nu \rho} (-p_b + p_c)^{\mu} + g^{\rho\mu} (-p_c - p_a)^{\nu}\right],
\eeq
where we use the convention that the momentum \(p_a\) flows towards the vertex, while \(p_b\) and \(p_c\) flow away from the vertex;
hence, momentum conservation reads $p_a=p_b+p_c$. Notice that,  with a slight abuse of notations, the indices $a, b$ and $c$
are used both as labels for the 3 gluons, and as the respective colour indices in the adjoint representation.
When computing the branching amplitude, this vertex is projected onto the polarisation vectors for the three external gluons, that is,
$\epsilon^\mu_a\equiv \epsilon^\mu(p_a, \lambda_a)$
 for the parent gluon and similarly for the two daughter gluons. After using momentum conservation 
($p_a=p_b+p_c$) and the fact that \(\epsilon_a \cdot p_a = \epsilon_b \cdot p_b = \epsilon_c \cdot p_c =0 \), the result of
this projection can be written as \cite{Ellis:1996mzs}
\beq \epsilon_a^{\mu} \epsilon_b^{\nu} \epsilon_c^{\rho} \,
V^{abc}_{\mu\nu\rho} (p_a,p_b,p_c)=ig f^{abc} \Gamma_{a\rightarrow bc}\,,\eeq
with the new vertex $\Gamma_{a\rightarrow bc}$ defined as
\beq
\label{Eq:QCD_vertex}
\Gamma_{a\rightarrow bc} \equiv 2 \left[ \left(\epsilon_a \cdot \epsilon_b \right)\left( \epsilon_c \cdot p_b \right) - \left(\epsilon_b \cdot \epsilon_c \right)\left( \epsilon_a \cdot p_b \right) - \left(\epsilon_c \cdot \epsilon_a \right)\left( \epsilon_b \cdot p_c \right)\right].
\eeq
Notice that the ``minus'' components (e.g. $p_a^-$) of the parton 4-momenta do not
contribute to this vertex, due to our use of the LC gauge.

From \eqn{pol}, one easily finds\footnote{We here use the subscripts $\lambda=y$ or $z$
in the sense of discrete values, e.g. $\delta_{yy}=\delta_{zz}=1$ and $\delta_{yz}=0$.}
\beq
\epsilon_a \cdot \epsilon_b=-\bm{\epsilon}_{\lambda_a}\cdot \bm{\epsilon}_{\lambda_b}=-\delta_{\lambda_a \lambda_b}\,,
\eeq
(and similarly for the other pairs of gluons); this is simply the statement that 
two different polarisation states (e.g., $\lambda_a=y$ and $\lambda_b=z$) are orthogonal to each other.

To compute dot products like $ \epsilon_a \cdot p_b $, it is useful to introduce the {\it relative} transverse momentum 
between the 2 daughter partons (or between one daughter parton and its parent), defined as\footnote{We remind the reader
that  $\zeta\equiv p^+_b/\omega$ and $1-\zeta= p^+_c/\omega$ (with  $\omega\equiv p^+_a$)
are  the splitting fractions of the daughter gluons.}
\beq\label{Pdef}
\bm{P}\,\equiv \,(1-\zeta)\bm{p}_b-\zeta\bm{p}_c \,=\,\bm{p}_b-\zeta\bm{p}_a \,=\,(1-\zeta)\bm{p}_a-\bm{p}_c\,,\eeq
where the second and third equalities follow from transverse momentum conservation 
($\bm{p}_a= \bm{p}_b+\bm{p}_c$). One then easily finds
\begin{align}
 \epsilon_a \cdot p_b &\, = \zeta \bm{p}_a\cdot \bm{\epsilon}_{\lambda_a} - \bm{p}_b\cdot \bm{\epsilon}_{\lambda_a}=-\bm{P}\cdot \bm{\epsilon}_{\lambda_a}\,,\nonumber\\*[0.2cm]
  \epsilon_c \cdot p_b &\, =\frac{\zeta}{1-\zeta}\, \bm{p}_c\cdot \bm{\epsilon}_{\lambda_c} - \bm{p}_b\cdot \bm{\epsilon}_{\lambda_c}=-
  \frac{\bm{P}\cdot \bm{\epsilon}_{\lambda_c}}{1-\zeta}\,,
 \nonumber \\*[0.2cm]
  \epsilon_b \cdot p_c &\, =\frac{1-\zeta}{\zeta} \,\bm{p}_b\cdot \bm{\epsilon}_{\lambda_b} - \bm{p}_c\cdot \bm{\epsilon}_{\lambda_b}=
  \frac{\bm{P}\cdot \bm{\epsilon}_{\lambda_b}}{\zeta}\,.
 \end{align}
The final result of these manipulations can be compactly written us
\beq
\label{QCD_vertex}
\Gamma_{a\rightarrow bc} (\bm{P},\zeta )= 2\left[ \delta_{\lambda_a \lambda_b}\,\frac{P_c}{1-\zeta} - \delta_{\lambda_b \lambda_c} P_a + 
\delta_{\lambda_a \lambda_c}\,\frac{P_b}{\zeta}
\right].\eeq
As manifest from this result, the QCD vertex projected  onto the gluon polarisation states
 depends upon the gluon transverse momenta only via the relative transverse momentum $\bm{P}=(P_y, P_z)$.

The (medium-induced) branching probability to be presented in the next subsection will involve the product of
the vertex \eqref{QCD_vertex} in the DA times a similar vertex, but evaluated at a different value, $\overline{\bm{P}}$,
of the relative momentum, in the CCA. The polarisation states for the external gluon are
identical for the 2 vertices, since polarisations cannot change via soft scattering with the medium constituents. 
Since the polarisation indices can only take two values, $\lambda_{a,b,c}=y$ or $z$, one can distinguish between
8 possible polarised splitting functions, conveniently grouped in 4 cases
(unless otherwise stated, in the equations below, there is no summation over repeated indices):

\texttt{(i)} when all 3 gluons have the same polarisation, $\lambda_a=\lambda_b=\lambda_c$, one finds
\beq
\label{vertex:aaa}
 \Gamma_{a\rightarrow aa}(\bm{P},\zeta ) \Gamma_{a\rightarrow aa}(\bar{\bm{P}},\zeta ) 
 =\frac{ 4  P_a \bar P_a }{\zeta(1-\zeta)} \left[ \frac{1-\zeta}{\zeta} + \frac{\zeta}{1-\zeta} + \zeta(1-\zeta)\right];
\eeq

\texttt{(ii)} when $\lambda_a=\lambda_b\ne \lambda_c$ (e.g. $\lambda_a=\lambda_b=y$, whereas $\lambda_c=z$), one finds
\beq
\label{vertex:aac}
\Gamma_{a\rightarrow ac}(\bm{P},\zeta ) \Gamma_{a\rightarrow ac}(\bar{\bm{P}},\zeta ) 
 =   \frac{4  P_c \bar P_c}{(1-\zeta)^2}\,;
 \eeq
 
\texttt{(iii)} when $\lambda_a=\lambda_c\ne \lambda_b$:
\beq
\label{vertex:aba}
\Gamma_{a\rightarrow ba}(\bm{P},\zeta ) \Gamma_{a\rightarrow ba}(\bar{\bm{P}},\zeta ) 
 =  \frac{4  P_b \bar P_b}{\zeta^2}\,;
 \eeq
 
 \texttt{(iv)} when $\lambda_a\ne \lambda_b=\lambda_c$:
\beq
\label{vertex:abb}
\Gamma_{a\rightarrow bb}(\bm{P},\zeta ) \Gamma_{a\rightarrow bb}(\bar{\bm{P}},\zeta ) 
 = 4  P_a \bar P_a. 
 \eeq

As a final check, let us show that, when summing over polarisation states for all gluons,
we recover the familiar expression of the DGLAP splitting function for unpolarised gluons:
\beq
\frac{1}{2}\sum_{\lambda_{a,b,c}=y,z}
\Gamma_{a\rightarrow bc} (\bm{P},\zeta )\Gamma_{a\rightarrow bc} (\bar{\bm{P}},\zeta )
 =\frac{ 4  \bm{P}\cdot \bar{\bm{P}}}{\zeta(1-\zeta)} \left[ \frac{1-\zeta}{\zeta} + \frac{\zeta}{1-\zeta} + \zeta(1-\zeta)\right]
\eeq
where $\bm{P}\cdot \bar{\bm{P}}= P_y\bar{P}_y+P_z\bar{P}_z$. This is indeed the expected result
in the unpolarised case  \cite{Ellis:1996mzs,Blaizot:2012fh}.

\subsection{BDMPS-Z branching rate in an anisotropic medium}

We shall now construct the medium-induced branching rate for polarised gluons propagating through an
anisotropic plasma by generalising the corresponding results for unpolarised gluons which split
in an isotropic medium. To that aim, we shall use the path-integral
formulation of the BDMPS-Z approach (see e.g. the presentations in \cite{Casalderrey-Solana:2011ule,Blaizot:2012fh}).
More precisely, we follow the analysis in Ref.~\cite{Blaizot:2012fh},
which includes a detailed discussion of the transverse momentum dependence of the branching rate. 
This is indeed important for our purposes, since the polarisation effects depend upon the flow of transverse 
momentum at the splitting vertices, as previously explained.

The medium-induced branching rate for polarised gluons has the same general structure as in the unpolarised
case, except for the fact that it includes the polarised splitting vertices introduced in the previous subsection.
Specifically (see Eq.~(5.1) in Ref.~\cite{Blaizot:2012fh})
\begin{align}
\label{Eq:rate_general}
\frac{d \mathcal{P}_{a\rightarrow bc}}{d\zeta dt}&\, = \frac{g^2 N_c}{8\pi \omega^2 \zeta(1-\zeta)} \,\mathrm{Re}\, \int_0^{L} d\Delta t\int \frac{d^2  \bm{P}}{(2\pi)^2} \int \frac{d^2 \bar{\bm{P}}}{(2\pi)^2} \nonumber\\*[0.2cm]
&\qquad\times
 \Gamma_{a\rightarrow bc} (\bm{P},\zeta )\, \Gamma_{a\rightarrow bc} (\bar{\bm{P}},\zeta )\, \tilde{S}^{(3)}(\Delta t,\bm{P},\bar{\bm{P}}),
\end{align}
where we recall that $\omega\equiv p_a^+$ is the LC longitudinal momentum
of the parent parton, $\zeta\equiv p_b^+/\omega$ is the splitting fraction of gluon $b$, $1-\zeta$ is the
corresponding fraction for gluon $c$,
and the ``time'' variables like $t$ or $\Delta t$ are truly {\it light-cone} times, cf. \eqn{LCdef}.   The integration
variable $\Delta t\equiv t_2-t_1$ is the difference between the splitting times (the temporal locations of
the QCD vertices) in the direct amplitude (DA) and the complex conjugate amplitude (CCA), respectively.
In writing \eqn{Eq:rate_general}, we implicitly assumed that both $t_1$ and $t_2$ lie inside the medium,
that is, we ignored the possibility that the parent gluon splits already before entering the medium, or after exiting
from it. This is indeed legitimate for the typical emissions, which have relatively low energies  $\zeta(1-\zeta)\omega\ll
\omega_c$ and therefore short formation times $t_f\ll L$. In the present context, the formation time $t_f$
is the typical value of $\Delta t$, as fixed by the integrations in  \eqn{Eq:rate_general}.

The branching rate in \eqn{Eq:rate_general} keeps trace of the energies and the polarisation states of the
participating gluons, but not of their transverse momenta which are integrated over. More precisely, the 
integrations run over the {\it relative} transverse momenta at the splitting vertices\footnote{The fact that one
can use relative transverse momenta {\it alone} is a consequence of translational invariance in the transverse
plane $(y,z)$: this is manifest for the QCD vertex \eqref{QCD_vertex}, which is the same as in the vacuum, 
and is also true for the medium effects, since we implicitly assume the medium to be homogeneous
(albeit anisotropic):
e.g. the  transport coefficients $\widehat{q}_z$ and  $\widehat{q}_y$ take the same values at all the points
$(\xi,y,\zeta )$, although these values can be different from each other:
$\widehat{q}_z\ne\widehat{q}_y$.}, 
that is, $\bm{P}$ in the DA (cf. \eqn{Pdef}) and  $\bar{\bm{P}}$  in the CCA. 
As already explained, the difference between $\bm{P}$ and $\bar{\bm{P}}$ reflects the momentum 
transferred by the medium during the formation time $\Delta t$.

The medium effects are encoded in the 3-point function $\tilde{S}^{(3)}(\Delta t,\bm{P},\bar{\bm{P}})$ which
describes the dynamics of the branching system during the time interval $\Delta t$. This is a {\it three}
point function since it refers to the three gluons which ``co-exist'' during $\Delta t$: the two daughter gluons in the DA 
(where the splitting occured at $t_1$) and the parent gluon in the CCA (which splits only later, at $t_2=t_1+\Delta t$).  
This 3-point function controls the values for $\Delta t$, $P_\perp$ and $\bar P_\perp$.
It is most conveniently computed via a Fourier transform from the transverse {\it coordinate} representation, 
that is,
\beq\label{S3FT}
\tilde{S}^{(3)}(\Delta t,\bm{P},\bar{\bm{P}}) =\int d^2\bu\int d^2\bv\,
e^{i \bu \cdot \bm{P} } e^{-i \bv \cdot \bar{\bm{P}}}\, {S}^{(3)}(\Delta t,\bu,\bv).
\eeq
Indeed, as discussed in the previous section and also in Appendix~\ref{app:pt}, 
the transverse coordinate representation is better suited for a study of the propagation of a test particle
which undergoes multiple scattering off the plasma constituents. Each of the three gluons included in $S^{(3)}$
undergoes 2-dimensional quantum diffusion in the presence of the (Gaussian) random potential $A^-$.
The formal solution to this quantum diffusion problem can be given a path integral representation, shown in Appendix~\ref{app:pt}.
The  function ${S}^{(3)}(\Delta t,\bu,\bv)$ is built by multiplying 3 such solutions (for the 3 gluons involved in the branching process)
and then averaging over the random potential according to \eqn{correlmed}. The ensuing path integral can be explicitly computed 
in a  ``harmonic approximation'' which consists in ignoring the logarithmic dependence of the jet quenching parameters
$\widehat{q}_y$ and $\widehat{q}_z$ upon the transverse separations between the 3 partons (recall the discussion after
\eqn{hatqeq}).

The path integral representation for  ${S}^{(3)}(\Delta t,\bu,\bv)$ is well known for the case of an isotropic plasma.
In the harmonic approximation, it reads  
(see e.g. Eq.~(B.26) in Ref.~\cite{Blaizot:2012fh})
 \beq \label{Eq:path_int}
S^{(3)}(\Delta t, \bu,\bv) = \int_{\br(t=0) = \bu}^{\br(t=\Delta t) = \bv} \mathcal{D} \br \exp \left\{ i M 
\int_0^{\Delta t} dt\; \left[\frac{\dot{\br}^2}{2}
+ \frac{\Omega^2  \br^2}{2}\right] \right\},
\eeq
where $\br(t)=\big(r_y(t), r_z(t)\big)$ is a 2-dimensional path in the $(y,z)$ plane and we denoted
\beq
M \equiv \zeta(1-\zeta)\omega,\qquad \Omega^2\equiv i \,[1-\zeta(1-\zeta)]\frac{\hat q}{2 M}\,.\eeq
As suggested by these notations, the path integral \eqref{Eq:path_int} is formally the same as the solution to the Schrödinger 
equation for a non-relativistic  particle with mass $M$ propagating in a harmonic potential with complex frequency $\Omega$.
As generally for a harmonic oscillator, the motions along different directions (here, $y$ and $z$) are independent
from each other. This makes the generalisation of  \eqref{Eq:path_int} to an anisotropic plasma quite straightforward:
it suffices to replace
\beq \hat q\br^2=\hat q(r_y^2+r_z^2) \ \to\ \frac{\widehat{q}_y }{2} \,r_y^2  + \frac{\widehat{q}_z}{2} \,r_z^2\,\eeq
where the factors $1/2$ are related to our normalisation conventions in \eqn{hatqyz}
(namely, the isotropic limit is obtained as $\widehat{q}_z = \widehat{q}_y=\hat q/2$). Then 
the original 2-dimensional path integral factorises into two independent 1-dimensional such integrals,
which are easily evaluated (since Gaussian), to yield
\beq
\label{Eq:S3_result}
\begin{split}
S^{(3)}(\Delta t, \bu,\bv) &= \sqrt{\frac{(1-i) \widehat{k}_y^2}{4\pi \sinh \Omega_y \Delta t}} \exp\left[ \frac{(i-1)\widehat{k}_y^2}
{4 \sinh \Omega_y \Delta t} \left( (u_y^2 + v_y^2) \cosh \Omega_y \Delta t - 2u_y v_y\right)\right] \\
&\times \sqrt{\frac{(1-i) \widehat{k}_z^2}{4\pi \sinh \Omega_z \Delta t}} \exp\left[ \frac{(i-1)\widehat{k}_z^2}{4 \sinh \Omega_z\Delta t} \left( (u_z^2 + v_z^2) \cosh \Omega_z \Delta t - 2u_z v_z\right)\right]
\end{split}
\eeq
where 
\beq
\label{Eq:Omegax}
\begin{split}
\Omega_y \equiv \frac{1+i}{\sqrt{2}} \sqrt{\frac{\widehat{q}_y [1-\zeta(1-\zeta)]}{\zeta(1-\zeta)\omega}} 
\end{split}
\eeq
and 
\beq
\label{Eq:kx2}
\widehat{k}_y^2 \equiv \sqrt{2\zeta(1-\zeta)\omega \,\widehat{q}_y [1-\zeta(1-\zeta)]},
\eeq
and similarly for \(\Omega_z\) and \(\widehat{k}_z^2\). As expected, Eq.~\eqref{Eq:S3_result} reduces to the well-known result
in the literature in the isotropic limit (see Eq.~(B.27) in Ref.~\cite{Blaizot:2012fh}).

As anticipated, the correlation function \eqref{Eq:S3_result} determines the natural values of the time difference
$\Delta t$: when $\Delta t$ is large enough for either $|\Omega_y |\Delta t >1$, or $|\Omega_z |\Delta t >1$,
this correlator decreases exponentially, since e.g. $1/(\sinh\Omega_y \Delta t)
\propto \exp\{-\Delta t/t_{f,y}\}$ for large $\Delta t$; here,
\beq\label{tfy}
t_{f,y}\,\equiv\,\frac{1}{\mathrm{Re} \,\Omega_y}\,=\,\sqrt{\frac{2\zeta(1-\zeta)\omega}{\widehat{q}_y [1-\zeta(1-\zeta)]}}\,,
\eeq
together with a similar expression for $t_{f,z}$. In the isotropic case, $t_{f,y}= t_{f,z}\equiv t_f$
is naturally interpreted as the formation time for medium-induced gluon branching. For an anisotropic medium, the
corresponding role is played by the smallest among  $t_{f,y}$ and $t_{f,z}$ (which corresponds to the largest among
$\widehat{q}_y$ and $\widehat{q}_z$). That is, the formation time is controlled by 
the direction along which the transverse momentum broadening is strongest. After also performing the Fourier transforms
in \eqn{S3FT}, one can infer the typical values for the (anisotropic) transverse momentum broadening during formation:
one thus finds  $\langle P_y^2\rangle =\langle\bar P_y^2\rangle=\widehat{q}_y [1-\zeta(1-\zeta)] t_{f}$, with
$t_f={\rm min}(t_{f,y},\,t_{f,z})$, and similarly for the $z$ components.

\subsection{Medium-induced splitting rates for polarised gluons}

Even though the Fourier transform in \eqn{S3FT} would be straightforward to compute (in the harmonic approximation),
this is actually not needed for computing the branching rate  \eqref{Eq:rate_general}.  Indeed, given the structure
of the QCD vertices in Eqs.~\eqref{vertex:aaa}--\eqref{vertex:abb}, it is enough to compute
\begin{align}\hspace*{-0.6cm}
\label{intS3}
&\,\int \frac{d^2  \bm{P}}{(2\pi)^2} \int \frac{d^2 \bar{\bm{P}}}{(2\pi)^2} \,P_y\bar P_y\,
 \tilde{S}^{(3)}(\Delta t,\bm{P},\bar{\bm{P}}) =\partial_{u_y}\partial_{v_y}
 S^{(3)}(\Delta t, \bu,\bv) \Big |_{\bu=\bv=0}
 \\*[0.2cm]\nonumber
 &\quad=  2\pi \left[\frac{(1-i) \widehat{k}_y^2}{4\pi \sinh \Omega_y\Delta t}\right]^{3/2} \left[\frac{(1-i) \widehat{k}_z^2}{4\pi \sinh \Omega_z \Delta t}\right]^{1/2}\ \ \mathop{\longrightarrow}\limits_{\Delta t\to 0} \ \ 2\pi\left[\frac{1}{2\pi}
 \frac{1-i}{1+i}\,\frac{\zeta(1-\zeta)\omega}{\Delta t}\right]^2\,,
\end{align}
together with a similar double integral with $P_y\bar P_y\to P_z\bar P_z$.

In the second line of \eqn{intS3}, 
we have also shown the behaviour of the result in the limit $\Delta t\to 0$, to emphasise the fact
that the subsequent integration over $\Delta t$ (cf.  \eqn{Eq:rate_general}) would generate a linear divergence
$\int d\Delta t/(\Delta t)^2$ from the inferior limit of the integral at $\Delta t=0$. To understand the origin of this divergence
and also the way to cure it, let us observe that this  limit $\Delta t\to 0$ is in fact equivalent to the vacuum limit
$\widehat{q}_a\to 0$ (with $a=y$ or $z$): indeed, $\Delta t$ enters the previous results only via the products 
$\Omega_y \Delta t$ and $\Omega_z \Delta t$, and $\Omega_a \propto \sqrt{\widehat{q}_a}$.
Hence, the divergence observed when $\Delta t\to 0$ is related to the fact that our results also include 
vacuum-like radiation (bremsstrahlung) and, moreover, they do that in a {\it wrong} way: gluon emissions in the vacuum
can occur at all times $-\infty < t_1, \, t_2 < \infty$, and when all the regions in $t_1$ and $t_2$ are included,
the overall result must vanish (since energy-momentum conservation forbids an on-shell parton 
to decay into two other on-shell partons). Hence, the fact that our results seem to admit a non-trivial (and even divergent)
 ``vacuum-limit'' is unphysical --- this is due to the fact that we restricted the time integrations to $0<t_1,\, t_2<L$, 
which is indeed appropriate for the (relatively soft) medium-induced emissions, but not also for the vacuum-like emissions. 
The way to cure this is however obvious: we are anyway interested in medium-induced emissions alone,
so it is enough the subtract the ``vacuum-limit'' from our results.

As an example, let us exhibit the calculation of the decay rate for the process $y\to z,z$ --- that is,
a parent gluon with linear polarisation along the $y$ axis decays into 2 gluons polarised along the  $z$ axis.
Using the respective splitting function from \eqn{vertex:abb} together with Eqs.~\eqref{Eq:rate_general} and
\eqref{intS3}, and the subtraction of the vacuum contribution as above discussed, we are left with
\begin{align}\hspace*{-1.cm}
\label{intS3}
\frac{d \mathcal{P}_{y\rightarrow zz}}{d\zeta dt}& = \frac{g^2 N_c}{2\pi \omega^2 \zeta(1-\zeta)} 
\,\mathrm{Re} \int_0^{L} d\Delta t
\int \frac{d^2  \bm{P}}{(2\pi)^2} \int \frac{d^2 \bar{\bm{P}}}{(2\pi)^2} \,P_y\bar P_y\,
 \tilde{S}^{(3)}(\Delta t,\bm{P},\bar{\bm{P}})\\*[0.2cm]\nonumber
 &=\frac{g^2 N_c}{\omega^2 \zeta(1-\zeta)} 
\,\mathrm{Re}  \int_0^{L} d\Delta t
\left\{\left[\frac{(1-i) \widehat{k}_y^2}{4\pi \sinh \Omega_y\Delta t}\right]^{3/2} \left[\frac{(1-i) \widehat{k}_z^2}{4\pi \sinh \Omega_z \Delta t}\right]^{1/2}-\left[\frac{1}{2\pi}
 \frac{1-i}{1+i}\,\frac{M}{\Delta t}\right]^2
\right\},
\end{align}
where we recall that $M=\zeta(1-\zeta)\omega$. The remaining integral over $\Delta t$ is now well defined.
Since the integrand  is exponentially suppressed when $\Delta t\gg t_f$ with $t_f\ll L$ (recall our assumption
that $M\ll \omega_c$), one can extend the upper limit to infinity  without loss of accuracy.
Then the above integral is conveniently rewritten as 
\begin{align}\label{intDt}
\mathrm{Re}  \int_0^{\infty} d\Delta t \left\{\cdots\right\}=\,\frac{[\zeta(1-\zeta)\omega]^{3/2}}{2(2\pi)^2}
\,\sqrt{2[1-\zeta(1-\zeta)]}\,\big(\widehat{q}_y \widehat{q}_z\big)^{1/4}
f\left(\sqrt{\widehat{q}_z/\widehat{q}_y} \right) ,
\end{align}
where we defined
\beq\label{deff}
f(a)\equiv \int_0^{\infty} dx\; \left[ \frac{1}{a^{1/2} x^2} - \frac{1}{\sinh ^{1/2} ax\, \sinh^{3/2} x}\right].
\eeq
The isotropic limit $\widehat{q}_z=\widehat{q}_y=\hat q/2$ corresponds to $a=1$ and then one finds $f(1)=1$.
For a medium which is only weakly anisotropic, $\Delta \widehat{q}\ll \hat q$, with
$\Delta \widehat{q}\equiv \widehat{q}_z-\widehat{q}_y$ and $\hat q\equiv \widehat{q}_z+\widehat{q}_y$, 
$a\simeq 1+\Delta \widehat{q}/\hat q$ is close to one, and 
\beq
f(a)\simeq 1 -\frac{a-1}{4}\,\simeq\, 1 -\frac{\Delta \widehat{q}}{4\hat q}
=1 - \frac{1}{4}\,\frac{\widehat{q}_z-\widehat{q}_y}{\widehat{q}_z+\widehat{q}_y}
\qquad\mbox{when}\qquad
\Delta \widehat{q}\ll \hat q\,.\eeq

In the second part of this paper, we shall study the energy distribution of polarised partons, as produced 
via multiple medium-induced branchings. To that aim, we will need the {\it inclusive} rate for the transition
$a\to b$ between 2 given polarisation states $a$ and $b$.
We shall compute this inclusive rate starting with the exclusive
rate for the process $a\to b,c$ and summing over the two possible polarisation states for the gluon $c$.
That is, the daughter gluon that we will explicitly follow is that with splitting fraction $\zeta$. (So long as the
energy of the measured daughter gluon is specified as well, the two daughter gluons cannot be confused
with each other.) We are thus led to define
\begin{align}
\label{Eq:rate_suml}
\frac{d \mathcal{P}_{a\rightarrow b}}{d\zeta dt}\,\equiv\,\sum_{\lambda_c=y,z}
\frac{d \mathcal{P}_{a\rightarrow bc}}{d\zeta dt}\,,
\end{align}
where it is understood that the parent gluon has energy $\omega$ and polarisation state $\lambda_a$,
whereas the measured daughter gluon has energy $\zeta\omega$ and polarisation state $\lambda_b$.

\begin{figure}[t] 
\centerline{
\includegraphics[width=.52\textwidth]{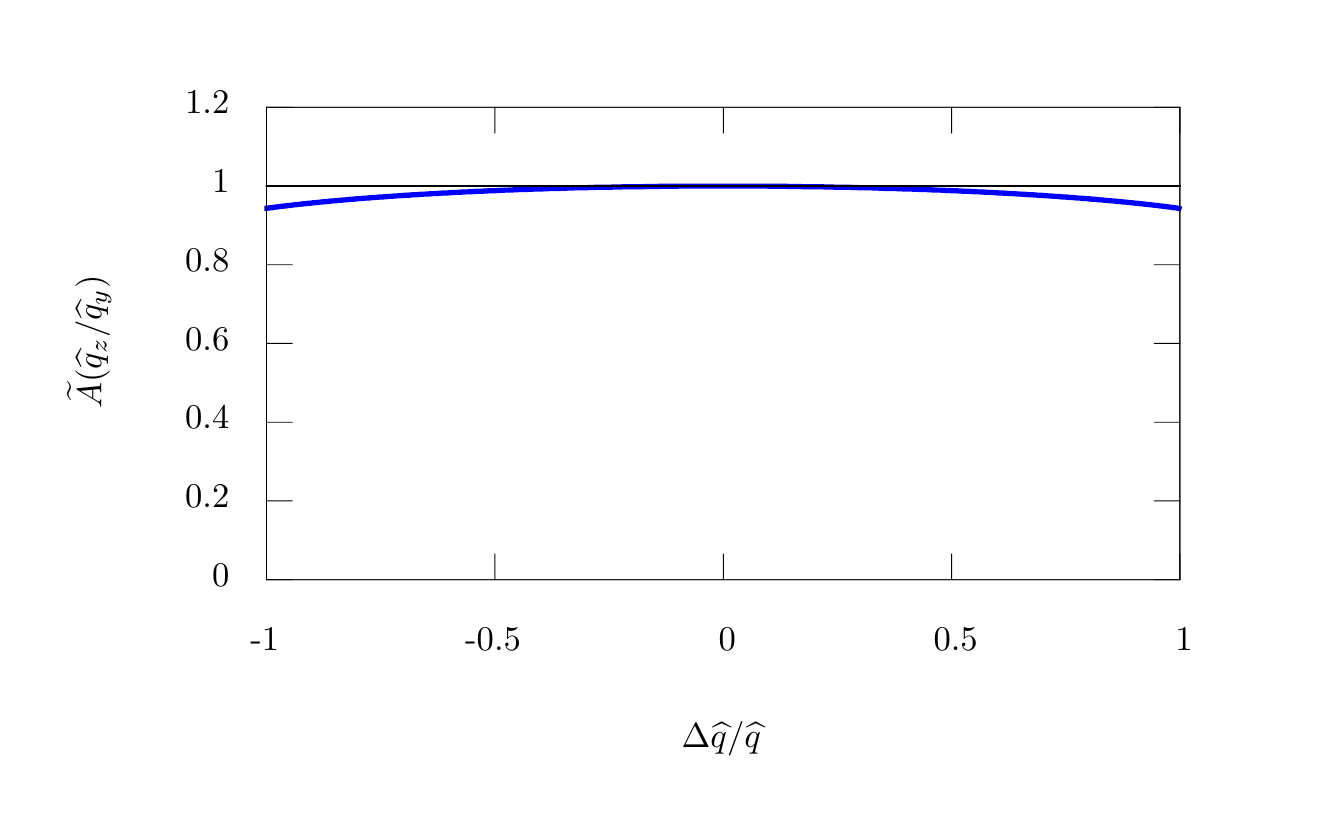}\quad\includegraphics[width=.52\textwidth]{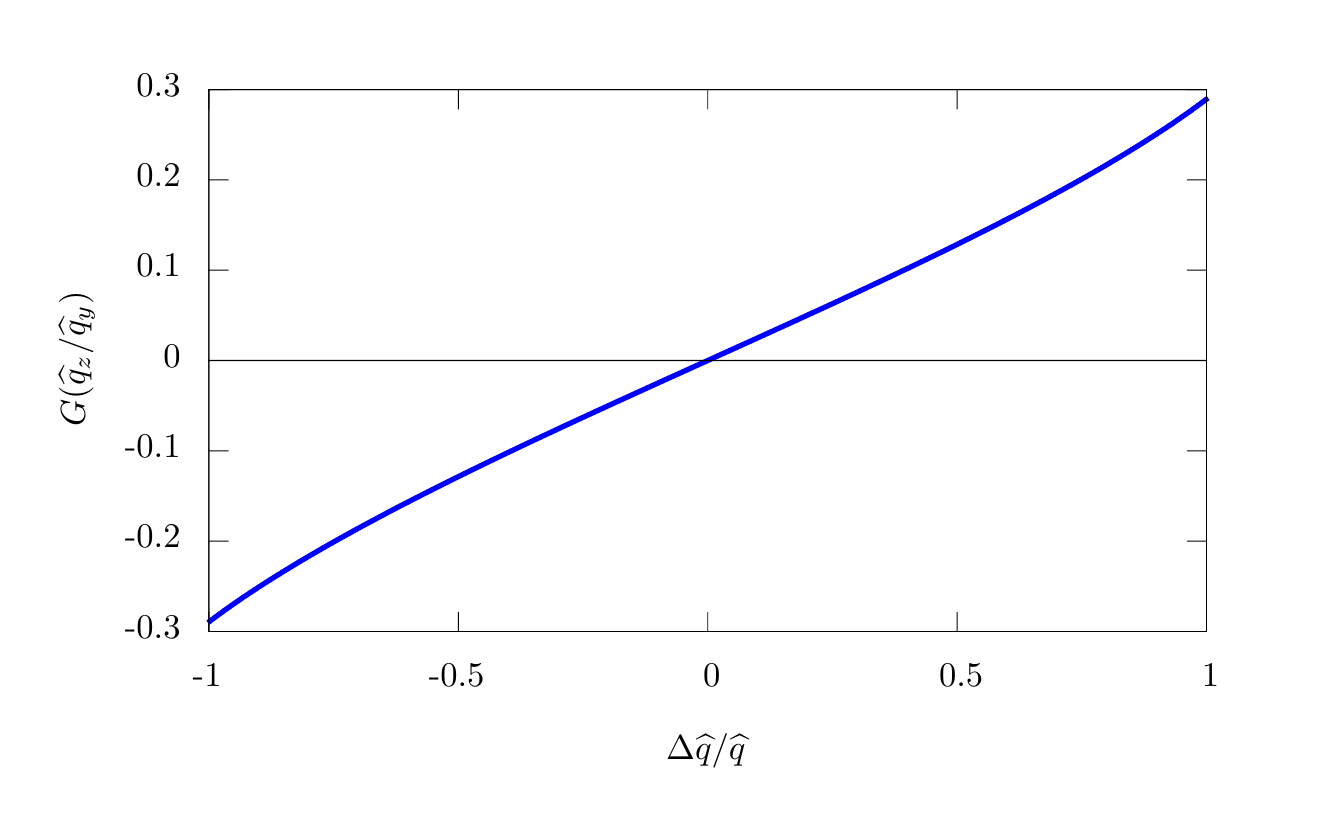}}
\caption{\small The functions \(\widetilde{A}(\widehat{q}_z/\widehat{q}_y)\) and \(G(\widehat{q}_z/\widehat{q}_y)\) which control
the medium-induced branching rates in an anisotropic plasma are plotted as functions of the medium anisotropy
$ \Delta \widehat{q}/\hat q$, where $ \Delta \widehat{q}=\widehat{q}_z-\widehat{q}_y$ and $\hat q=\widehat{q}_z+\widehat{q}_y$.
}
\label{fig:AG}
\end{figure}

It is straightforward although a bit tedious to derive the analog of Eqs.~\eqref{intS3}--\eqref{intDt} for
all the relevant $a\to b$ transition channels. The results are suggestively written as follows 
\beq
\label{Eq:Gammaxtox_full}
\begin{split}
\frac{d \mathcal{P}_{z\rightarrow z}}{d\zeta dt}
= {\abar}\frac{(\widehat{q}_y \widehat{q}_z)^{1/4}}{\sqrt{2\omega}} A(\widehat{q}_z/\widehat{q}_y) \,\gamma(\zeta) \left[ \mathcal{F}^0_{z\rightarrow z}(\zeta) +  G(\widehat{q}_z/\widehat{q}_y) \mathcal{F}^1_{z\rightarrow z}(\zeta)\right].
\end{split}
\eeq
\beq
\begin{split}
\frac{d \mathcal{P}_{z\rightarrow y}}{d\zeta dt}
= {\abar}  \frac{(\widehat{q}_y \widehat{q}_z)^{1/4}}{\sqrt{2\omega}} A(\widehat{q}_z/\widehat{q}_y) \,\gamma(\zeta) \left[ \mathcal{F}^0_{z\rightarrow y}(\zeta) - G(\widehat{q}_z/\widehat{q}_y)\, \mathcal{F}^1_{z\rightarrow y}(\zeta) \right]
\end{split}
\eeq
\beq
\begin{split}
\frac{d \mathcal{P}_{y\rightarrow z}}{d\zeta dt}
= {\abar}\frac{(\widehat{q}_y \widehat{q}_z)^{1/4}}{\sqrt{2\omega}} A(\widehat{q}_z/\widehat{q}_y)\,\gamma(\zeta)
 \left[ \mathcal{F}^0_{y\rightarrow z}(\zeta) + G(\widehat{q}_z/\widehat{q}_y)\, \mathcal{F}^1_{y\rightarrow z}(\zeta)\right]
\end{split}
\eeq
\beq
\label{Eq:Gammaytoy_full}
\begin{split}
\frac{d \mathcal{P}_{y\rightarrow y}}{d\zeta dt}
= {\abar}  \frac{(\widehat{q}_y \widehat{q}_z)^{1/4}}{\sqrt{2\omega}} A(\widehat{q}_z/\widehat{q}_y) \,\gamma(\zeta)  \left[ \mathcal{F}^0_{y\rightarrow y}(\zeta) - G(\widehat{q}_z/\widehat{q}_y)\, \mathcal{F}^1_{y\rightarrow y}(\zeta)\right]
\end{split}
\eeq
where $\abar\equiv \alpha_s N_c/\pi$, $\gamma(\zeta) $ is a splitting kernel which summarises the $\zeta$--dependence
of the medium-induced radiation,
\beq\label{Eq:gamma}
\gamma(\zeta) = \frac{\left[1-\zeta(1-\zeta)\right]^{1/2}}{\zeta^{1/2}(1-\zeta)^{1/2}},
\eeq
and the new functions 
\(A(\widehat{q}_z/\widehat{q}_y)\) and \(G(\widehat{q}_z/\widehat{q}_y)\) are defined as
\beq
\label{Eq:A_def}
A(\widehat{q}_z/\widehat{q}_y) \equiv \frac{1}{2}\left[f(a) + f(1/a)\right]\,,\qquad
G(\widehat{q}_z/\widehat{q}_y) \equiv \frac{f(1/a) - f(a)}{f(a) + f(1/a)}\,,
\eeq
with $a=\sqrt{\widehat{q}_z/\widehat{q}_y} $ and the function $f(a)$ introduced in \eqn{deff}.
In the isotropic limit ($a\to 1$), one has $A\to 1$ and $G\to 0$. For an anisotropic medium,
$G(a)$ is positive when $a>1$ and negative when $a <1$. In particular, in the limit of a small anisotropy, $\Delta \widehat{q}\ll \hat q$,  
one finds $G(\widehat{q}_z/\widehat{q}_y)\simeq \Delta \widehat{q}/4\hat q$. In Fig.~\ref{fig:AG}, we plot \(G\), 
as well as 
\beq
\widetilde{A}(\widehat{q}_z/\widehat{q}_y)\equiv \left[\frac{4\widehat{q}_y \widehat{q}_z}{\widehat{q}^{\,2}}\right]^{1/4}  A(\widehat{q}_z/\widehat{q}_y) 
=\left(1 - \left(\Delta \widehat{q}/{\widehat{q}}\right)^2 \right)^{1/4} A(\widehat{q}_z/\widehat{q}_y)
\eeq
which measures the overall change in the splitting rates in Eqs. \eqref{Eq:Gammaxtox_full}--\eqref{Eq:Gammaytoy_full} due to medium anisotropy.
These functions are plotted in terms of
the dimensionless ratio $ \Delta \widehat{q}/\hat q$, which is the most direct measure of the medium anisotropy. Interestingly, the function $\widetilde{A}$
remains very close to one up to very large asymmetries $ |\Delta \widehat{q}|/\hat q\le 1$, which demonstrates that the anisotropy has only
a tiny effect on the {\it total} branching rate. On the contrary, the function $G$ has a rather pronounced dependence upon the anisotropy,
which is quasi-linear in $ \Delta \widehat{q}$. The quantity \(\Delta \widehat{q}/\widehat{q}\) runs between \(-1\) when \(\widehat{q}_z = 0\) and \(1\) when \(\widehat{q}_y = 0\) such that in a maximally anisotropic plasma \(G\) takes the value \(G(\pm 1) = \pm 0.29.\)

The splitting functions which carry an upper label 0 are the only ones to matter in the case of an isotropic medium.
They have the following expressions:
\begin{align}\label{defF0}
\mathcal{F}^0_{z\rightarrow z}(\zeta)
&\, = \mathcal{F}^0_{y\rightarrow y}(\zeta) =\frac{1}{2} \left( \frac{1-\zeta}{\zeta} + \frac{2 \zeta}{1-\zeta} + \zeta(1-\zeta),
\right)\nn
\mathcal{F}^0_{z\rightarrow y}(\zeta)&\, =\mathcal{F}^0_{y\rightarrow z}(\zeta) =  \frac{1}{2} \left( \zeta(1-\zeta) + \frac{1-\zeta}{\zeta}\right).
\end{align}
For an anisotropic medium, there are additional contributions which involve the splitting functions 
carrying the upper index 1; these read
\begin{align}\label{defF1}
\mathcal{F}^1_{z\rightarrow z}(\zeta) &\,= \mathcal{F}^1_{y\rightarrow y}(\zeta)
 =  \frac{1}{2} \left( \frac{1-\zeta}{\zeta} + \zeta(1-\zeta)\right)
\nn
\mathcal{F}^1_{z\rightarrow y}(\zeta) &\,= \mathcal{F}^1_{y\rightarrow z} (\zeta)= 
\frac{1}{2} \left( \frac{1-\zeta}{\zeta} - \zeta(1-\zeta)\right).
\end{align}
Using the above expressions, it is easily to check that the total branching rate, averaged over the 
polarisations and  summed over the final ones, takes the form
\begin{align}
\label{total-rate}
\frac{d \mathcal{P}}{d\zeta dt}\,\equiv\,\frac{1}{2}\sum_{\lambda_{a,b}=y,z}
\frac{d \mathcal{P}_{a\rightarrow b}}{d\zeta dt}= 
{\abar}  \frac{(\widehat{q}_y \widehat{q}_z)^{1/4}}{\sqrt{2\omega}} A(\widehat{q}_z/\widehat{q}_y) \,\gamma(\zeta) \mathcal{F}(\zeta),
\end{align}
were $ \mathcal{F}(\zeta)$ is the unpolarised DGLAP splitting function:
\beq\label{FDGLAP}
 \mathcal{F}(\zeta) \,=\, \frac{1-\zeta}{\zeta} + \frac{ \zeta}{1-\zeta} + \zeta(1-\zeta).\eeq
Notice that the function \(A(\widehat{q}_z/\widehat{q}_y)\) characterises the strength of the total branching rate, 
whereas the function  \(G(\widehat{q}_z/\widehat{q}_y)\) has disappeared after summing over polarisations 
--- this function only matters for the polarisation flip at the splitting vertices.

As a check, let us verify that, using the above, we recover the well-known result for the
BDMPS-Z branching rate for an unpolarised parton propagating
through an isotropic medium: when $\widehat{q}_z=\widehat{q}_y=\hat q/2$ (and hence $A=1$ and $G=0$), one indeed
finds the expected result
\begin{align}\label{Fiso}
\frac{1}{2}\,\frac{d \mathcal{P}_{z\rightarrow z}+d \mathcal{P}_{y\rightarrow z}}{d\zeta dt}\,=\,\frac{1}{2}\,
\frac{d \mathcal{P}_{y\rightarrow y}+d \mathcal{P}_{z\rightarrow y}}{d\zeta dt}\,=\,\frac{\abar}{4}\,\sqrt\frac{\hat q}{\omega}\,\gamma(\zeta) 
\mathcal{F}(\zeta),
\end{align}
which in particular is independent of the polarisation of the (measured) daughter gluon.

One important aspect of the splitting rates in Eqs.~\eqref{Eq:Gammaxtox_full}--\eqref{Eq:Gammaytoy_full} is the fact that, unlike
the familiar law for bremsstrahlung in the vacuum, they depend not only upon the splitting fraction $\zeta$, but also 
upon the energy $\omega$ of the parent parton: the medium-induced branching probability increases
like $1/\sqrt{\omega}$ with decreasing $\omega$. This feature is of course well known for the case of an isotropic medium,
where it has important consequences (notably, in relation with multiple branching; see e.g. 
\cite{Baier:2000sb,Blaizot:2012fh,Blaizot:2013hx,Blaizot:2013vha,Kurkela:2014tla}),
that we briefly recall now. Its additional implications for the case where the plasma
is anisotropic will be thoroughly discussed in the remaining part of the paper.

So, let us consider the branching rate for an unpolarised parent parton  in an isotropic medium, cf.
Eqs.~\eqref{total-rate} and \eqref{Fiso}.
After integrating this rate over all times $0 < t < L$, with $L$ the medium size available to the parent parton, one finds
\begin{align}\label{Ptf}
\frac{d \mathcal{P}}{d\zeta}\,=\,\abar\frac{L}{t_f(\omega,\zeta)}\,\mathcal{F}(\zeta)\qquad\mbox{with}\qquad
t_f(\omega,\zeta)\equiv\sqrt{\frac{4\zeta(1-\zeta)\omega}{\hat q [1-\zeta(1-\zeta)]}}\,.
\end{align}
The quantity  $\zeta(d \mathcal{P}/d\zeta)$ can be interpreted as the probability for having a single
branching with splitting fraction\footnote{The case $1/2<\zeta<1$ can be similarly treated by using the symmetry of
\eqn{Ptf}  under $\zeta \to 1-\zeta$.} $\zeta<1/2$. When this quantity  becomes of order one or larger, 
the effects of {\it multiple branching} become important and our formalism must be extended to account for them.
According to \eqn{Ptf}, this happens when $\zeta(1-\zeta)\omega\lesssim \ombr\equiv \abar^2 \omega_c$, where we recall
that $\omega_c=\hat q L^2/2$. This condition is satisfied by two types of emissions:
 \texttt{(i)} very asymmetric emissions with $\zeta\ll1$ by a parent parton with
generic energy $\omega$, or \texttt{(ii)} quasi-democratic emissions with generic values of $\zeta$ (say, $\zeta\sim 1/2$), 
but such that the parent parton itself is sufficiently soft: $\omega \lesssim \ombr$. Case  \texttt{(i)} typically
applies to the successive emissions of ``primary'' gluons by the leading parton (whose initial energy $E$ can be very large,
even larger than $\omega_c$).  These primary gluons have typical energies  $\omega \lesssim \ombr$, since
for such energies, their emission probability is of order one.  

After being emitted, the primary gluons are bound to split further,
mostly via democratic branchings, and thus transmit their energy to a myriad of even softer gluons. The typical interval
$\Delta t$  between two such branchings can be estimated by replacing $L\to \Delta t$ in \eqn{Ptf}: the probability 
$\zeta(d \mathcal{P}/d\zeta)$ becomes $\sim\order{1}$ when $\Delta t\sim t_f(\omega)/\abar$, which is 
much smaller than $L$ for low enough  $\omega \ll \ombr$.  These 
``democratic cascades'' are expected to stop when the gluon energies become as small as the typical energies
of the plasma constituents (say, their temperature $T$ for a medium in thermal equilibrium) or as
the Bethe-Heitler energy $\omega_{\rm BH}$, which corresponds to emissions with formation times of the order
of the mean free path.

In Sect.~\ref{sec:evol}, this picture of jet evolution via multiple branching will be generalised to include
anisotropy effects and to keep trace of the parton polarisations.


\subsection{Polarisation distribution after one splitting: a qualitative discussion} 
\label{sec:1split}

To get more intuition for the formalism developed so far, let us compute the transmission of polarisation via a single gluon branching.
Specifically, we shall assume that the parent gluon has probability \(p\) of being polarized in the \(z\) direction and hence probability \(1-p\) 
of being polarized in the \(y\) direction. We are interested in the probility \(p^{\prime}\) for an emitted daughter parton with splitting fraction
$\zeta$ to be polarized in the \(z\) direction. This is given by 
\beq
\label{Eq:pol_prob}
p^{\prime} = \frac{p\, \mathcal{P}_{z\rightarrow z} + (1-p)\mathcal{P}_{y\rightarrow z} }{p \left( \mathcal{P}_{z\rightarrow z} + \mathcal{P}_{z\rightarrow y}\right) + (1-p)\left(\mathcal{P}_{y\rightarrow z} + \mathcal{P}_{y\rightarrow y} \right)}
\eeq
where \(\mathcal{P}_{z\rightarrow z} = d \mathcal{P}_{z\rightarrow z}/d\zeta dt \) etc is a shorthand for the splitting rates introduced
in  Eqs.~\eqref{Eq:Gammaxtox_full}--\eqref{Eq:Gammaytoy_full}. Notice that the dependence upon the energy
$\omega$ of the parent parton disappears in the ratio. Hence, the r.h.s. of \eqn{Eq:pol_prob} merely depends upon the 
splitting fraction $\zeta$.

For more clarity, we first analyze the case of an isotropic medium with \(\widehat{q}_z = \widehat{q}_y=\hat q/2\). In that case,
the dependence upon $\hat q$ drops out in the ratio and we are left with
\beq
\label{Eq:pprime_iso}
p^{\prime}-\frac{1}{2} = h(\zeta) \left(p-\frac{1}{2} \right)
\eeq
where
\beq
\begin{split}
h(\zeta) \equiv \frac{\zeta^2}{(1-\zeta)^2 + \zeta^2 + \zeta^2 (1-\zeta)^2}
\end{split}
\eeq
 We immediately see that for \(p=1/2\), i.e. an unpolarised initial gluon, the emitted gluon is also unpolarised, \(p^{\prime} = 1/2.\)
Importantly, \(0 \leq h(\zeta)\leq 1\) for all \(0\leq \zeta\leq 1\). Therefore, in an isotropic medium, the net polarisation reduces with each splitting. 
After multiple splitting we therefore expect the net polarisation of the initial gluon to have gone away. Furthermore, \(h(\zeta=1) = 1\), so when the daughter parton carries all of the energy fraction of the mother parton, we have \(p^{\prime} = p\). Similarly,  \(h(\zeta=0) = 0\) so that a soft parton has no knowledge of the polarisation of the mother parton and is unpolarised in an isotropic medium. 

We now turn to the case of an anisotropic medium, for which we find
\beq
\label{Eq:pol_prob_full}
p^{\prime}- \frac{1}{2} = \frac{\zeta^2 (p -1/2) + \frac{1}{2}\,G(\widehat{q}_z/\widehat{q}_y) (1-\zeta)^2}{ (1-\zeta)^2 + \zeta^2+ \zeta^2(1-\zeta)^2 +2\,G(\widehat{q}_z/\widehat{q}_y) \,(p-1/2) \zeta^2 (1-\zeta^2)}
\eeq
This equation becomes more intuitive if we assume that the net initial polarization is small, \(p-1/2 \ll 1\),
and/or the medium is only slightly anisotropic,  $\Delta \widehat{q}\ll \hat q$, 
so that we can drop the term proportional to \((p-1/2)G\) in the denominator\footnote{More precisely we are doing a Taylor expansion in \(p-1/2\) and \(\Delta \widehat{q}/\hat q\). In order for corrections to Eq. \eqref{Eq:pol_prob_full} to be subleading we need \((p-1/2)^2 \ll 1\), \((p-1/2)(\Delta\widehat{q}/\hat q) \ll 1\) and \((\Delta \widehat{q}/\hat q)^2 \ll 1\), so that the anisotropy is small and the initial polarization is not too big. Notice that because of the squares in these conditions, the conditions are not very stringent and Eq. \eqref{Eq:pol_prob_appr} is fairly robust.}. Then 
\beq
\label{Eq:pol_prob_appr}
p^{\prime}- \frac{1}{2} = h(\zeta) \left(p-\frac{1}{2}\right) + g(\zeta)\, \frac{1}{2}\,G(\widehat{q}_z/\widehat{q}_y).
\eeq
where 
\beq
\label{Eq:gz_func}
g(\zeta) \equiv \frac{(1-\zeta)^2}{(1-\zeta)^2 + \zeta^2(1-\zeta)^2 + \zeta^2}=h(1-\zeta).
\eeq
The second term on the right hand side of Eq. \eqref{Eq:pol_prob_appr} is a source term that increases polarisation in the \(z\) direction
(assuming that $G>0$). It is independent of the polarisation of the mother parton. 

In Eq. \eqref{Eq:pol_prob_appr} we have two competing effects. The term \(h(\zeta) \left(p-\frac{1}{2}\right)\), which is also present in an isotropic medium, tends to reduce the net polarisation. The other term proportional to \(G(\widehat{q}_z/\widehat{q}_y)\) tends to align the polarisation of
the daughter gluon with the \(z\) direction. When the daughter parton carries nearly all of the energy of its parent, \(\zeta \rightarrow 1\), we find
\beq
p^{\prime} - \frac{1}{2} \simeq p - \frac{1}{2} + (1-\zeta)^2 \frac{1}{2}\,G(\widehat{q}_z/\widehat{q}_y),
\eeq
where the source term $\propto(1-\zeta)^2G$ is very small and should strictly speaking be discarded.
Thus the daughter parton nearly retains the polarisation of the mother parton. In the opposite situation where
the measured daughter parton is soft \(\zeta\rightarrow 0\), we obtain
\beq
p^{\prime} - \frac{1}{2} \simeq \zeta^2 \left(p-\frac{1}{2}\right) + \frac{1}{2}\,G(\widehat{q}_z/\widehat{q}_y).
\eeq
This time, it is the first term $\propto\zeta^2$ which can be discarded, hence the polarisation of the soft daughter gluon
is nearly independent of that of its parent and it is fully driven by the anisotropy of the medium: when $G>0$, its
polarisation is aligned with the \(z\) axis.


\section{Jet evolution in an anisotropic medium}
\label{sec:evol}

So far we have considered a single splitting in an anisotropic medium. Yet, our focus is on the relatively soft emissions with formation times
$t_f\ll L$, for which the effects of multiple branching are expected to be important. 
Hence, in order to understand what a jet looks like after traversing an anisotropic plasma, we need to allow for multiple splittings. We will do this by solving evolution equations for the jet. These are rate equations --- i.e. kinetic equations involving gain and loss terms 
--- with the rates given by the probabilities for one splitting per unit time, as shown in Eqs.~\eqref{Eq:Gammaxtox_full}--\eqref{Eq:Gammaytoy_full}.

We want to track how the polarisation of jet partons at energy fraction \(\xi = \omega /E\) changes with time. Here \(\omega\) is the energy\footnote{More
precisely, $\omega$ and $E$ are the ``plus'' components of the respective LC  momenta, recall the discussion after \eqn{LCdef}; they are referred
to as ``energies'' for brevity.} of a parton and \(E\) is the energy of the initial parton that seeds the jet. 
As before, we assume that the leading parton moves along the $x$ axis (so, it is orthogonal
to the collision axis $z$), that the jet involves only gluons, and that \(\widehat{q}_y\) and \(\widehat{q}_z\) are constant 
throughout the plasma --- albeit generally different from each other. As discussed at the beginning of Sect.~\ref{sec:1br}, our approach is
strictly valid for parton energies within the range $\omega_{\rm BH} \ll \omega \ll \omega_c$, so we shall not track the very soft
gluons with $\omega\lesssim \omega_{\rm BH}$, which propagate at large angles.

A crucial assumption here is that all jet partons move roughly collinearly. This is needed in order for
 the plane transverse to the direction of motion to be the same for all jet partons, including the softest ones. 
 That allows us to describe the polarisation of all partons using fixed directions \(y\) and \(z\). This assumption is 
justified so long as the energies of the jet partons are much larger than the transverse momenta they acquire via collisions in the medium. Roughly
speaking, the condition reads $\omega \gg \sqrt{\hat q L}$ with $\hat q\equiv \widehat{q}_y+\widehat{q}_z$, but this condition gets modified for
the sufficiently soft gluons, whose lifetime (between successive splittings) is smaller than $L$ (see below). 

\subsection{Rate equations for medium-induced emissions of polarised gluons}

Given the structure of Eqs.~\eqref{Eq:Gammaxtox_full}--\eqref{Eq:Gammaytoy_full}, it is convenient to
define a rescaled time variable, which is dimensionless:
\begin{align}\label{deftau}
\tau \equiv \abar\frac{(\widehat{q}_z\widehat{q}_y)^{1/4}}{\sqrt{2E}}  A(\widehat{q}_z/\widehat{q}_y) \,t
\end{align}
We shall  use the notation $\tau_L$ for the maximum value of this variable, corresponding to $t=L$.
To gain more intuition for the maximum value, let us notice that, in the case of an isotropic plasma ($\widehat{q}_z = \widehat{q}_y\equiv \hat q/2$),
\beq
\tau_L\,=\abar\sqrt{\frac{\hat q}{4E}}\,L=\sqrt{\frac{\ombr}{2E}}\qquad\mbox{and}\qquad
\frac{\tau_L}{\sqrt{\xi}} 
=\sqrt{\frac{\ombr}{2\omega}}\,,\eeq
where  $\omega=\xi E$ and we recall that $ \ombr=\abar^2\omega_c=\abar^2\hat q L^2/2$.
From the discussion following \eqn{Ptf}, we recall that multiple branchings
become important when $\omega \lesssim \ombr$. Therefore the most interesting physical regime for us here corresponds
to the case where $1\gg\tau_L\gtrsim\sqrt{\xi}\gg\xi$. The condition $\tau_L\ll 1$ ensures that $E\gg  \ombr$, so the leading parton
survives in the final jet, after crossing the medium\footnote{Note that this condition allows e.g.
$E\sim\omega_c$, in which case $\tau_L\sim\abar$.} (it does not suffer a democratic branching itself). 
The condition $\tau_L\gtrsim\sqrt{\xi}$ means that we concentrate on gluon emissions which are sufficiently soft
 ($\omega=\xi E\lesssim \ombr$) to be sensitive to multiple branching. They can be
either soft ($\zeta\ll 1$) primary emissions by the leading parton, or quasi-democratic branchings
($\zeta\sim 1/2$) of the primary gluons.

In terms of this new time variable, the medium-induced branching rates take a more compact form, e.g.
\beq
\label{Eq:Gammaxtox_2}
\begin{split}
\frac{d \mathcal{P}_{z\rightarrow z}}{d\zeta d\tau}
= \frac{\gamma(\zeta)} {\sqrt{\xi}} \left[ \mathcal{F}^0_{z\rightarrow z}(\zeta) +  G(\widehat{q}_z/\widehat{q}_y) \mathcal{F}^1_{z\rightarrow z}(\zeta)\right].
\end{split}
\eeq
We observe once again that the branching rate depends not only upon the splitting fraction $\zeta$ but also upon the energy fraction $\xi$
of the parent gluon.

These branching rates are the main ingredients of the equations describing the time evolution of the energy distributions of gluons
with a given polarisation state, $\lambda=y$ or  $\lambda=z$. These evolution equations (a.k.a. {\it rate equations}) 
are conveniently written for the respective spectra,
\begin{align}
D_y\equiv \xi\frac{d N_y}{d \xi}\,,\qquad D_z\equiv \xi\frac{d N_z}{d \xi}\,,
\end{align}
where $N_\lambda(\xi)$ is the number of jet constituents with polarisation $\lambda$ and energy fraction $\xi$. 
By following standard techniques (e.g. the method of the
generating functional described in \cite{Blaizot:2013vha}), one deduces the two following coupled equations for $D_y$ and $D_z$:
\beq
\label{Eq:Dyevol}
\begin{split}
\frac{d D_{y}(\xi,\tau)}{d\tau} &= \int_{\xi}^1 d\zeta\; \mathcal{K}_{z\rightarrow y}(\zeta)\, \sqrt{\frac{\zeta}{\xi}}\, D_z\left(\frac{\xi}{\zeta},\tau\right)
  +\int_{\xi}^1 d\zeta\; \mathcal{K}_{y\rightarrow y}(\zeta)\, \sqrt{\frac{\zeta}{\xi}}\, D_{y}\left(\frac{\xi}{\zeta},\tau\right)  \\
& - \frac{1}{2}\int_0^1 d\zeta\;\Big[\mathcal{K}_{y\rightarrow z}(\zeta) + \mathcal{K}_{y\rightarrow y}(\zeta)\Big]\, \frac{1}{\sqrt{\xi}}\, D_{y}(\xi,\tau),
\end{split}
\eeq
and
\beq
\label{Eq:Dxevol}
\begin{split}
\frac{d D_z(\xi,\tau)}{d\tau} &= \int_{\xi}^1 d\zeta\; \mathcal{K}_{z\rightarrow z}(\zeta)\, \sqrt{\frac{\zeta}{\xi}}\, D_z\left(\frac{\xi}{\zeta},\tau\right)  +\int_{\xi}^1 d\zeta\; \mathcal{K}_{y\rightarrow z}(\zeta)\, \sqrt{\frac{\zeta}{\xi}}\, D_y\left(\frac{\xi}{\zeta},\tau\right)  \\
& - \frac{1}{2}\int_0^1 d\zeta\;\Big[\mathcal{K}_{z\rightarrow z}(\zeta) + \mathcal{K}_{z\rightarrow y}(\zeta)\Big]\, \frac{1}{\sqrt{\xi}}\, D_{z}(\xi,\tau).
\end{split}
\eeq
The splitting kernels $ \mathcal{K}_{z\rightarrow z}(\zeta)$ etc. contain the $\zeta$--dependence of the  branching rates like \eqn{Eq:Gammaxtox_2};
they are defined as, e.g.
\beq\label{defKzz}
 \mathcal{K}_{z\rightarrow z}(\zeta)\equiv \,\sqrt{\xi}\,
 \frac{d \mathcal{P}_{z\rightarrow z}}{d\zeta d\tau}\,
= {\gamma(\zeta)}\left[ \mathcal{F}^0_{z\rightarrow z}(\zeta) +  G(\widehat{q}_z/\widehat{q}_y) \mathcal{F}^1_{z\rightarrow z}(\zeta)\right].
\eeq
These equations have a familiar structure, with ``gain terms'' and ``loss terms''. The ``gain terms'', as shown in the first line of each
of the two equations, represent the rate for producing a gluon $(\xi,\,\lambda)$
via the branching of a parent gluon with energy fraction $\xi' =\xi/\zeta>\xi$ and any polarisation state $\lambda'=y$ or $z$.
The  ``loss terms'', as shown in the second line, describe the decay of a gluon $(\xi,\,\lambda)$ into two softer gluons with energy fractions
$\zeta\xi$ and $(1-\zeta)\xi$ and arbitrary polarisation states.  The integrals for the
gain terms are restricted to $\zeta>\xi$ because of the constraint ${\xi}/{\zeta}\le 1$ on the energy fraction of the parent gluon.
The factor \(1/2\) in front of the loss terms is needed to avoid double counting. Its origin is a bit subtle, so it is worth showing a more detailed
argument. 

Recall that a branching rate like $\mathcal{K}_{z\rightarrow z}$ refers to the process where, for a parent gluon with $\lambda=z$,
the daughter gluon with splitting fraction $\zeta$ has polarisation $z$ independently of the polarisation state of the other 
daughter gluon, with splitting fraction $1-\zeta$. Hence,  $\mathcal{K}_{z\rightarrow z}(\zeta)= 
\mathcal{K}_{z\rightarrow zz}(\zeta)+ \mathcal{K}_{z\rightarrow zy}(\zeta)$ 
in obvious notations. Similarly,  $\mathcal{K}_{z\rightarrow y}(\zeta)= \mathcal{K}_{z\rightarrow yz}(\zeta)+ \mathcal{K}_{z\rightarrow yy}(\zeta)$.
Now, in the loss term in \eqn{Eq:Dxevol}, $\zeta$ is a dummy variable which can take any value between 0 and 1. The decays where both
daughter partons are in a same polarisation state, i.e. $\mathcal{K}_{z\rightarrow zz}(\zeta)$ and $\mathcal{K}_{z\rightarrow yy}(\zeta)$,
are symmetric under the exchange of the daughter gluons, since the respective rates are symmetric under $\zeta\to 1-\zeta$ (recall
Eqs.~\eqref{vertex:aaa} and \eqref{vertex:abb}). Clearly, such processes would be counted twice when integrating over all values of $\zeta$.
As for the remaining processes where the daughter gluons have different polarisation states, their rates get interchanged 
when $\zeta\to 1-\zeta$; that is, $\mathcal{K}_{z\rightarrow zy}(1-\zeta)=\mathcal{K}_{z\rightarrow yz}(\zeta)$, cf.
Eqs.~\eqref{vertex:aac} and \eqref{vertex:aba}. So,  these processes too would be counted twice after 
integrating over $\zeta$.

In order to make contact with the known equations for the case of an isotropic plasma and for unpolarised gluons,
it is convenient to use the following linear combinations of $D_y$ and $D_z$:
\beq\label{Dtotpol}
D_{\mathrm{tot}} \equiv D_y + D_z,\qquad\mbox{and}\qquad \widetilde{D} \equiv D_z-D_y.
\eeq
$D_{\mathrm{tot}}$  is the total spectrum for gluons and $\widetilde{D}$ is the jet polarisation. Using
Eqs.~\eqref{Eq:Dyevol} and \eqref{Eq:Dxevol}, it  is straightforward to deduce the corresponding equations for 
$D_{\mathrm{tot}}$ and $\widetilde{D}$. The general equations are not very illuminating (we include them in Appendix~\ref{sec:generaleqs},
where we also check their isotropic limit), but they become more transparent --- and also better suited for constructing (analytic
and numerical) solutions --- in the limit in which we keep only the singular behaviour of the branching rates near $\zeta=0$ and
$\zeta=1$. That is, we neglect the contributions proportional to $\zeta(1-\zeta)$ in Eqs.~\eqref{defF0}--\eqref{defF1} and also
in the structure \eqref{Eq:gamma} of the splitting kernel  $\gamma(\zeta)$.
This approximation, which is similar to that performed in \cite{Blaizot:2013hx} for the case of an isotropic medium, keeps all
the salient features of the branching process, which is indeed driven by its singular points at $\zeta=0$ and $\zeta=1$. 
For an isotropic medium, this has been explicitly checked by comparing the solutions numerically obtained with
the two types of kernel (exact and approximate) \cite{Fister:2014zxa,Blaizot:2015jea}. With this simplification,
the equations obeyed by $D_{\mathrm{tot}}$ and $\tilde{D}$ take particularly suggestive forms:
\beq
\label{Eq:Dtotevol}
\frac{d D_{\mathrm{tot}}(\xi,\tau)}{d\tau} 
 = \int_{\xi}^1 d\zeta\; \mathcal{K}_0 (\zeta)\sqrt{\frac{\zeta}{\xi}}\, D_{\mathrm{tot}}\left(\frac{\xi}{\zeta},\tau\right) 
 - \int_0^1 d\zeta\;\mathcal{K}_0 (\zeta)\, \frac{\zeta}{\sqrt{\xi}}\, D_{\mathrm{tot}}(\xi,\tau) 
\eeq
and
\beq
\label{Eq:Dpolevol}
\begin{split}
\frac{d \widetilde{D}(\xi,\tau)}{d\tau} 
 &= \int_{\xi}^1 d\zeta\; \mathcal{M}_0(\zeta)\, \sqrt{\frac{\zeta}{\xi}}\, \widetilde{D}\left(\frac{\xi}{\zeta},\tau\right) 
  - \int_0^1 d\zeta\;\mathcal{K}_0(\zeta)\, \frac{\zeta}{\sqrt{\xi}}\, \widetilde{D}(\xi,\tau) \\
  &+\int_{\xi}^1 d\zeta\; \mathcal{L}_0(\zeta)\, \sqrt{\frac{\zeta}{\xi}}\, D_{\mathrm{tot}}\left(\frac{\xi}{\zeta},\tau\right).
\end{split}
\eeq
with the new kernels (see Appendix~\ref{sec:generaleqs} for details)
\begin{align}
\label{K0}
\mathcal{K}_0(\zeta) \equiv \frac{1}{\zeta^{3/2} (1-\zeta)^{3/2}},\qquad \mathcal{L}_0 (\zeta)\equiv 
 \frac{(1-\zeta)^{1/2}}{\zeta^{3/2}} \,G(\widehat{q}_z/\widehat{q}_y),\qquad \mathcal{M}_0 (\zeta)\equiv \frac{\zeta^{1/2}}{(1-\zeta)^{3/2}}.
\end{align}

These equations have an intuitive interpretation. The evolution equation for the total number of partons \(D_{\mathrm{tot}}\) takes exactly the same form as for an isotropic medium \cite{Blaizot:2013hx,Blaizot:2013vha}.  This
``coincidence'' is not an exact property --- it only holds within our present approximations, which neglected the non-singular
contributions to the splitting functions. (The general equation, as shown in \eqn{Eq:Dtotevol_first}, is more complicated and
involves contributions proportional to  $\widetilde{D}$.) The exact solution to this equation is known in analytic
form \cite{Blaizot:2013hx} and this will be useful for what follows.

The equation \eqref{Eq:Dpolevol} obeyed by the polarised distribution  $\widetilde{D}$ is new and has an interesting
structure: besides a gain term and a loss term, with kernels $\mathcal{M}_0(\zeta)$ and $\mathcal{K}_0(\zeta)$, respectively,
it also features a {\it source} term, proportional to the total parton distribution  \(D_{\mathrm{tot}}\), which is non-zero only 
in the presence of anisotropy. Indeed, its kernel $\mathcal{L}_0(\zeta)$ is proportional to the function $G(\widehat{q}_z/\widehat{q}_y)$
which  would vanish for an isotropic plasma, cf. \eqn{Eq:A_def}. This demonstrates that,
in an anisotropic medium, non-trivial polarisation can be generated via medium-induced gluon branchings, independently of
the polarisation of the parent gluons. 

This is in agreement with our previous discussion of
a single splitting in Sect.~\ref{sec:1split}: the source term in \eqn{Eq:Dpolevol} is analogous to the second piece, proportional to 
$G(\widehat{q}_z/\widehat{q}_y)$, in the r.h.s. of Eq. \eqref{Eq:pol_prob_appr}. In fact, one can recognise similar properties in both
cases. The kernel  \(g(\zeta)\)  in Eq. \eqref{Eq:pol_prob_appr} vanishes like \( (1-\zeta)^2\) as \(\zeta\rightarrow 1\), whereas the
corresponding kernel $\mathcal{L}_0(\zeta)$  in \eqn{Eq:Dpolevol}  is suppressed by \((1-\zeta)^2\) 
relative to the isotropic functions \(\mathcal{K}_0(\zeta)\) and \(\mathcal{M}_0(\zeta)\). This reflects the fact
that a daughter parton carrying nearly all of the energy of its parent parton does not ``feel'' the anisotropy of the medium 
--- rather,  it retains the polarisation of the parent.

Furthermore, the gain term on the r.h.s. of Eq. \eqref{Eq:Dpolevol}, with kernel $\mathcal{M}_0(\zeta)$, show how 
polarisation gets transmitted from the parent parton to the daughter parton with splitting fraction $\zeta$. It is 
 analogous to the first term on the right hand side in Eq. \eqref{Eq:pol_prob_appr} for a single splitting. Like in that case, the
 respective kernel is suppressed by $\zeta^2$ in the soft limit \(\zeta \rightarrow 0\) (soft gluons do not ``know'' about the
 polarisation of their parents).
 Finally, the loss term on the r.h.s of Eq. \eqref{Eq:Dpolevol} describes the decay of partons carrying polarisation. As their 
 total decay rate is (to our order of approximation) independent of their polarisation, the respective kernel is 
 the unpolarised function \(\mathcal{K}_0(\zeta)\) --- the same as in \eqn{Eq:Dtotevol}.
 
 \subsection{A Green's function for the polarised gluon distribution}
 \label{sec:Green}
 
 In what follows, we shall solve the rate equations \eqref{Eq:Dtotevol} and \eqref{Eq:Dpolevol} with the initial condition that,
 at time $\tau=0$, the jet consists of a single gluon (the ``leading parton'') which is unpolarised:\footnote{Note that in terms
 of the distributions $D_y$ and $D_z$ for gluons with definite polarisations, this initial condition reads $D_y(\xi,\tau=0)=
 D_z(\xi,\tau=0)=   \delta(1-\xi)/2$ (cf. \eqn{Dtotpol}).}
 \beq\label{ICs}
 D_{\mathrm{tot}}(\xi,\tau=0) = \delta(1-\xi)\qquad\mbox{and}\qquad  \widetilde{D}(\xi,\tau=0)=0.
 \eeq 
The corresponding solution for \(D_{\mathrm{tot}}\)  is known in analytic
form and reads  \cite{Blaizot:2013hx,Blaizot:2015jea}
\beq
\label{Eq:Dtot_sol}
D_{\mathrm{tot}}(\xi,\tau) = \frac{\tau }{\sqrt{\xi} (1-\xi)^{3/2}}\, e^{-\pi \tau^2/(1-\xi)}\,.
\eeq
For small $\tau$ and $\xi$ close to 1, this exhibits a pronounced peak at $1-\xi\lesssim \tau^2$ which describes the 
leading parton. The width of this peak reflects the fact that the energy loss by the leading parton, namely
$E(1-\xi)\sim E\tau^2\sim \abar^2\hat q t^2$, is associated with the radiation of soft gluons. For $\xi\ll 1$, the full
spectrum in presence of multiple branching, $D_{\mathrm{tot}}\simeq {\tau }/{\sqrt{\xi}}$, is formally the same as the
spectrum that would be created via a single emission by the leading parton, cf. \eqn{Ptf}. This reflects the turbulent nature
of the democratic gluon cascades  \cite{Blaizot:2013hx}. In particular, the power-law spectrum $1/\sqrt{\xi}$ is a fixed
point of the rate equation \eqref{Eq:Dtotevol}: the gain and loss terms in its r.h.s. exactly cancel each other
for this particular spectrum.

The knowledge of the exact solution \eqref{Eq:Dtot_sol} for the total spectrum allows us to deduce the corresponding
solution $D_{\mathrm{tot}}$ for the polarised distribution  $\widetilde{D}$ without too much effort. To that aim, it is convenient
to first consider the homogeneous version of \eqn{Eq:Dpolevol}, that is, the equation obtained after removing the source term
in the second line of \eqref{Eq:Dpolevol}.
For more clarity, we denote  the respective solution as $\widetilde{D}_0(\xi,\tau)$; this obeys the homogeneous equation
 \beq
\label{Eq:evol_D0}
\begin{split}
\frac{d \widetilde{D}_0(\xi,\tau)}{d\tau} 
 = \int_{\xi}^1 d\zeta\; \mathcal{M}_0(\zeta)\, \sqrt{\frac{\zeta}{\xi}}\, \widetilde{D}_0\left(\frac{\xi}{\zeta},\tau\right) 
  - \int_0^1 d\zeta\;\mathcal{K}_0(\zeta)\, \frac{\zeta}{\sqrt{\xi}}\, \widetilde{D}_0(\xi,\tau),
  \end{split}
\eeq
with the initial condition $\widetilde{D}_0(\xi,\tau=0)=\delta(1-\xi)$. The only difference w.r.t. \eqn{Eq:Dtotevol} 
for $D_{\mathrm{tot}}$ refers to the kernel $\mathcal{M}_0(\zeta)$ for the gain term, which satisfies 
$\mathcal{M}_0 (\zeta)=\zeta^2 \mathcal{K}_0 (\zeta)$ (cf. \eqn{K0}). It is then easy to check that the respective
solutions are related in the same way, that is,
 \begin{align}\label{D0vsDtot}
 \widetilde{D}_0(\xi,\tau)\,=\,\xi^2 D_{\mathrm{tot}}(\xi,\tau)\, = \,\frac{\xi^{3/2} \,\tau}{(1-\xi)^{3/2}}\, e^{-\pi \tau^2/(1-\xi)}\,.
\end{align}
 This  can be demonstrated
by inserting  $ \widetilde{D}_0\equiv \xi^2  \widetilde{D}_1$ within  \eqn{Eq:evol_D0}; one then
 finds that $ \widetilde{D}_1$ obeys the same equation as  $ D_{\mathrm{tot}}$, i.e.  \eqn{Eq:Dtotevol}, 
 with the initial condition $ \widetilde{D}_1(\xi,\tau=0) = \delta(1-\xi)$.
 
\begin{figure}[t] 
\centerline{
\includegraphics[width=.6\textwidth]{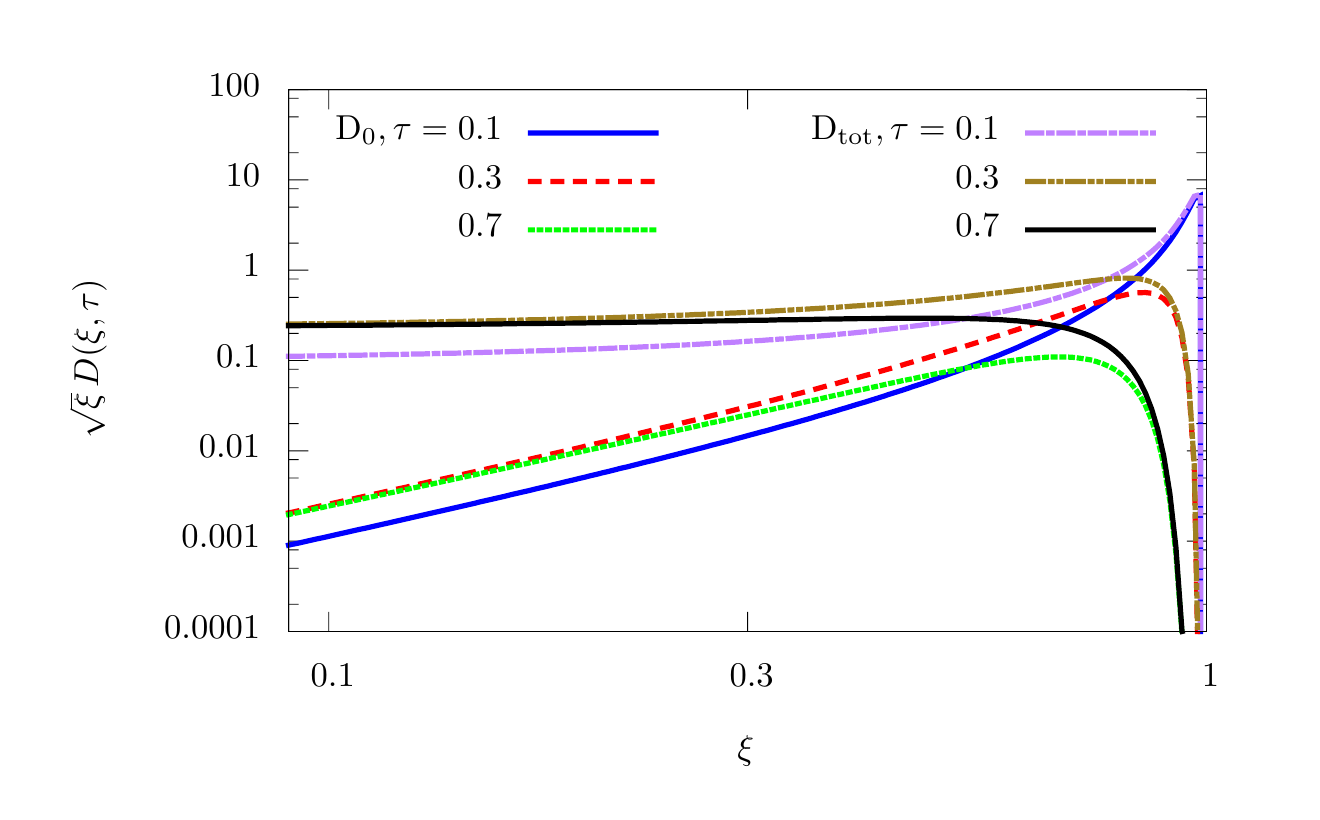}\hspace*{-0.5cm}\includegraphics[width=.6\textwidth]{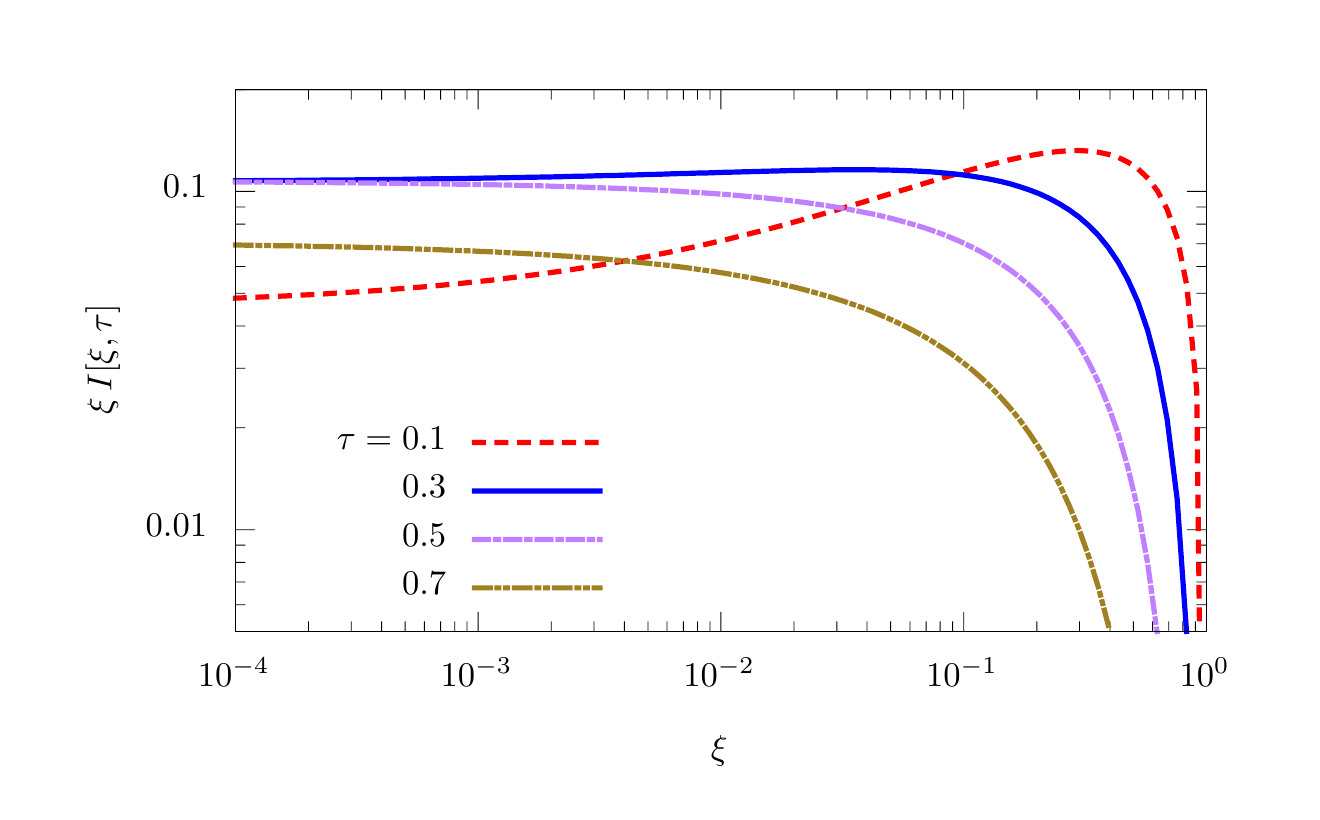}}
\caption{\small {\it Left:} The total (unpolarised) gluon distribution $D_{\mathrm{tot}}(\xi,\tau)$ (cf.  \eqn{Eq:Dtot_sol})
and the polarised distribution $\widetilde{D}_0(\xi,\tau)$ that would be created in an isotropic medium by
a polarised leading gluon (cf.  \eqn{D0vsDtot}). Both functions are multiplied by $\sqrt{\xi}$,
to emphasise the emergence of the turbulent spectrum $D_{\mathrm{tot}}\propto 1/\sqrt{\xi}$
 at small $\xi\lesssim\tau^2$.  {\it Right:} The source term $I(\xi,\tau)$ responsible for inducing polarisation in an
 anisotropic medium,  numerically computed from \eqn{Eq:source_explc} with
 $G(\widehat{q}_z/\widehat{q}_y)=0.3$. The solutions are shown down
 to very small values of $\xi$, to check the quality of the analytic approximation  \eqref{Eq:source_smallx}.}
\label{fig:D0I}
\end{figure}

Here, we merely use \eqn{Eq:evol_D0} as an auxiliary equation, but in fact, this equation 
has an interesting interpretation, which is more transparent in the case of an isotropic plasma. In that case,
\eqn{Eq:evol_D0} describes the transmission of polarisation from the leading parton to the partons in the cascade
via successive branchings. Its solution \eqref{D0vsDtot} represents the polarised gluon distribution 
created by a leading gluon with initial polarisation along the $z$ axis ($D_z(\xi,\tau=0)=   \delta(1-\xi)$ and $D_y(\xi,\tau=0)=0$).
For $\xi\simeq 1$,  \eqn{D0vsDtot} is very similar to the unpolarised distribution  \eqref{Eq:Dtot_sol}; this shows
that the leading parton essentially preserves its initial polarisation so long as it survives
in the medium (i.e. for $\tau\ll 1$). On the other hand, for the much softer modes with $\xi\ll 1$, 
$\widetilde{D}_0(\xi,\tau)$ is suppressed by $\xi^2$ compared to  $D_{\mathrm{tot}}(\xi,\tau)$, 
meaning that the net polarisation is negligible. This is of course consistent with the fact that soft gluons cannot inherit the polarisation
of their parents.  These considerations are illustrated in the left panel of Fig.~\ref{fig:D0I}, where we have plotted  the two functions  
$D_{\mathrm{tot}}(\xi,\tau)$ and $\widetilde{D}_0(\xi,\tau)$.

The above discussion shows that the only way to generate a net polarisation for  the soft gluons
is through the effects of anisotropy, as encoded in the source term in \eqn{Eq:Dpolevol}.
Using the above solution for $\widetilde{D}_0(\xi,\tau)$, one can construct the Green's function  \(G(\xi,\xi_1,\tau)\) which permits
to solve the inhomogeneous equation \eqref{Eq:Dpolevol} for an arbitrary source term.  This Green's function obeys 
the sourceless equation, that is,
 \beq
\label{Eq:evol_greens}
\begin{split}
\frac{d G(\xi,\xi_1,\tau)}{d\tau} 
 = \int_{\xi}^1 d\zeta\; \mathcal{M}_0(\zeta)\, \sqrt{\frac{\zeta}{\xi}}\, G\left(\frac{\xi}{\zeta},\xi_1,\tau\right) 
  - \int_0^1 d\zeta\;\mathcal{K}_0(\zeta)\, \frac{\zeta}{\sqrt{\xi}}\, G(\xi,\xi_1,\tau),
  \end{split}
\eeq
with the initial condition
 \beq\label{ICGreen}
G(\xi,\xi_1,\tau=0)\,=\,\delta(\xi-\xi_1)\,.\eeq
Comparing Eqs.~\eqref{Eq:evol_greens} and \eqref{Eq:evol_D0}, one immediately sees that
 \beq\label{GvsD0}
G(\xi,\xi_1,\tau)\,=\,\frac{\theta(\xi_1-\xi)}{\xi_1}\, \widetilde{D}_0\left(\frac{\xi}{\xi_1},\frac{\tau}{\sqrt{\xi_1}}\right).\eeq
Notice the $\theta$--function in the structure of $G(\xi,\xi_1,\tau)$: it shows that the integral
over $\zeta$ in the gain term in  \eqn{Eq:evol_greens} is truly restricted to $\zeta>\xi/\xi_1$.  
After also using \eqref{D0vsDtot} and\eqref{Eq:Dtot_sol}, one finds 
   \begin{align}\label{Gfinal}
   G(\xi,\xi_1,\tau)&\,=\,\frac{\theta(\xi_1-\xi)}{\xi_1}\left(\frac{\xi}{\xi_1}\right)^2
    D_{\mathrm{tot}}\left(\frac{\xi}{\xi_1},\frac{\tau}{\sqrt{\xi_1}}\right)\nonumber\\*[0.2cm]
    &\,=\,\theta(\xi_1-\xi)\left(\frac{\xi}{\xi_1(\xi_1-\xi)}\right)^{3/2}
   \tau\, e^{-\pi \tau^2/(\xi_1-\xi)}\,.
    \end{align}

We are finally in a position to solve the inhomogeneous \eqn{Eq:Dpolevol} with vanishing initial condition. The solution
can be expressed via the Green's function as 
 \beq
\label{Eq:Dtilde_general}
 \widetilde{D}(\xi,\tau)= \int_{\xi}^1 d\xi_1\; \int_0^{\tau} d\tau_1 \,G(\xi,\xi_1,\tau - \tau_1) I(\xi_1,\tau_1),
\eeq
with $ I(\xi_1,\tau_1)$ the source term in the r.h.s. of  \eqn{Eq:Dpolevol}, that is,
\beq
\label{Eq:source_def}
 I(\xi_1,\tau_1)=\int_{\xi_1}^1 d\zeta\; \mathcal{L}_0(\zeta)\, \sqrt{\frac{\zeta}{\xi_1}}\, D_{\mathrm{tot}}\left(\frac{\xi_1}{\zeta},\tau_1\right).
\eeq
This source term is non-local in energy, since associated with radiation --- polarised gluons with energy fraction $\xi_1$ are 
produced via the decay of gluons with any polarisation and with larger energy fractions ${\xi_1}/{\zeta}$, for any $1>\zeta>\xi_1$ ---,
but is local in time, because of our assumption that medium-induced emissions are quasi-instantaneous.
The ensuing convolution with the Green's function in \eqn{Eq:Dtilde_general} shows how the net polarisation
of gluons with energy fraction $\xi_1$ created at time $\tau_1$ propagates 
(via successive branchings) into the measured bin at $\xi$ at the measurement time $\tau>\tau_1$.

\subsection{The polarised jet distribution}
 \label{sec:sol}
 
 \eqn{Eq:Dtilde_general} with the source term  \eqref{Eq:source_def} represents an exact but formal solution
 to the equation obeyed by the polarised jet distribution $\widetilde{D}(\xi,\tau)$. In order to render this solution more
 explicit and uncover its physical consequences, one needs to (at least, approximately) perform the remaining convolutions
 over $\xi_1$ and $\tau_1$, and also the integral over $\zeta$ in the expression \eqref{Eq:source_def} for the source term.
 As previously explained, we are mainly interested in the production of soft gluons with $\xi\ll 1$ at relatively
 small times $\tau\ll 1$, so we shall adapt our approximations to these conditions.
 
 Let us start by simplifying the source term. Using the expression in \eqn{K0} for the kernel  $\mathcal{L}_0 $ together with 
 \eqn{Eq:Dtot_sol} for  $ D_{\mathrm{tot}}$, one finds
   \begin{align}
\label{Eq:source_explc}
 I(\xi_1,\tau_1)= G(\widehat{q}_z/\widehat{q}_y)\int_{\xi_1}^1 d\zeta\;  (1-\zeta)^{1/2}\,\frac{\zeta}{\xi_1}
  \frac{\tau_1}{ (\zeta-\xi_1)^{3/2}}\, e^{-\pi \tau_1^2 \zeta/(\zeta-\xi_1)}\,.
   \end{align}
We shall later check that, when $\xi\ll 1$, the convolution in   \eqn{Eq:Dtilde_general} is controlled by $\xi_1\ll 1$ as well.
On the other hand, the integral in \eqn{Eq:source_explc} is convergent at its upper limit and hence is dominated by relatively 
large values $\zeta\sim 1$.
Indeed the would-be pole of the integrand at $\zeta=\xi_1$ is regulated by the exponential, which effectively restricts
the support of the integration to $\zeta-\xi_1\gtrsim \zeta\tau_1^2$. We can therefore neglect $\xi_1$ next to $\zeta$ inside
the integrand, to deduce
\beq
\label{Eq:source_smallx}
\begin{split}
I[\xi_1,\tau_1] &\simeq G(\widehat{q}_z/\widehat{q}_y)\,\frac{\tau_1 e^{-\pi \tau_1^2} }{\xi_1}\int_0^1 d\zeta \,\sqrt{\frac{1-\zeta}{\zeta}} 
= \,G(\widehat{q}_z/\widehat{q}_y) \,\frac{\pi \tau_1 e^{-\pi \tau_1^2}}{2\xi_1}.
\end{split}
\eeq
The final integral is indeed controlled by values of $\zeta$ in the bulk, say $\zeta\sim 1/2$, in agreement with our previous discussion.
This argument also shows that the polarisation source at $\xi_1\ll 1$ is generated via the democratic branching of an
unpolarised parent gluon which is itself soft, with energy fraction $\xi'=\xi_1/\zeta\sim 2\xi_1$. This dominance of the
democratic branchings within the source term is not a consequence of the polarised splitting function --- by itself,
the kernel  $\mathcal{L}_0(\zeta) $ in  \eqn{K0} would rather favour very asymmetric splittings with $\zeta\ll 1$ ---, but rather of
the fact that the number of sources $ D_{\mathrm{tot}}(\xi',\tau_1)$ is rapidly increasing with decreasing $\xi'=\xi_1/\zeta$.
Since $\xi_1$ itself is small, this property favours relatively large values of $\zeta$, albeit not {\it too} large --- since the kernel
$\mathcal{L}_0(\zeta) $ vanishes when $\zeta\to 1$. This competition ultimately selects intermediate values $\zeta\sim 1/2$.

In the right panel of Fig.~\ref{fig:D0I}, we display the results for the source term obtained via numerical integration
in \eqn{Eq:source_explc}. Besides confirming the power-law increase $ \propto 1/\xi_1$ at small $\xi_1\ll 1$ as analytically found
in \eqn{Eq:source_smallx}, these numerical results also show an interesting behaviour at larger values $\xi_1 >0.1$: the source
term is rather abruptly generated when decreasing $\xi_1$ below 1. This region at large  $\xi_1$ is not covered by our previous
approximation \eqref{Eq:source_smallx}, yet it can be analytically understood as follows: so long as $\tau_1\ll 1$,
a polarised distribution at $\xi_1\sim\order{1}$ can be generated via primary emissions by the leading parton.
This direct contribution to the source term can be estimated by replacing $D_{\mathrm{tot}}\left({\xi_1}/{\zeta},\tau_1\right)\simeq
\delta({\xi_1}/{\zeta}-1)$ within the integrand of  \eqref{Eq:source_def}, which yields
\beq
\label{Eq:source_LP}
 I(\xi_1,\tau_1)\,\simeq\,\xi_1\mathcal{L}_0(\xi_1)\, =\,
\sqrt{  \frac{1-\xi_1}{\xi_1}} \,G(\widehat{q}_z/\widehat{q}_y).
 \eeq
 This result is independent of time but truly holds only for very small times  $\tau_1\ll 1$. Its dependence upon $\xi_1$
 explains the shape of the curves in Fig.~\ref{fig:D0I} (right) near $\xi_1=1$. As expected, the leading parton contribution \eqref{Eq:source_LP}
 to the source term dominates at $\xi_1> \tau_1^2$, whereas the respective contribution \eqref{Eq:source_smallx} from democratic branchings
 becomes dominant at small $\xi_1\lesssim\tau_1^2$.

Substituting the approximate expression \eqref{Eq:source_smallx}
for the source together with \eqn{Gfinal}  for the  Green's function into  \eqn{Eq:Dtilde_general}, one finds
\beq\label{Dtilde1}
\widetilde{D}(\xi,\tau)
\simeq \frac{\pi}{2} \,
G(\widehat{q}_z/\widehat{q}_y) \int_\xi^1 d\xi_1 \int_0^{\tau} d\tau_1\;\left(\frac{\xi}{\xi_1(\xi_1-\xi)}\right)^{3/2}
(\tau - \tau_1)\; e^{-\pi \frac{(\tau - \tau_1)^2}{\xi_1-\xi}}\;\frac{\tau_1}{\xi_1}\, e^{-\pi \tau_1^2}\,.
\eeq
The integrand involves a competition between the power laws $\xi_1^{-5/2}(\xi_1-\xi)^{-3/2}$, which strongly enhance the contribution
from values of $\xi_1$ as small as possible, meaning  $\xi_1\sim \xi$, and the exponential $\exp{\big\{-\pi (\tau - \tau_1)^2/(\xi_1-\xi)\big\}}$,
which limits the difference $\xi_1-\xi$ to non-zero values $\xi_1-\xi\gtrsim \Delta\tau^2$, with $\Delta\tau\equiv \tau - \tau_1$.
Since however both $\xi_1$ and $\Delta\tau$ are integration variables, it seems quite clear that the maximal contribution to the integral
should come from their minimal values allowed by these two constraints, namely $\Delta\tau^2 \sim \xi_1-\xi\sim \xi $. Indeed, with
these choices, $\xi_1 \sim 2\xi$ is parametrically of order $\xi$, yet the exponent remains of order one, so there is no exponential
suppression. These values for $\xi_1$ and $\Delta\tau$ are in fact correlated with each other:
$\xi_1 \sim 2\xi$ corresponds to a democratic branching and $\Delta\tau \sim \sqrt{\xi}$ is indeed the typical lifetime 
of a gluon with energy fraction $\xi_1 \sim \xi$ before it undergoes such a democratic 
branching\footnote{Indeed, this condition $\Delta\tau \sim \sqrt{\xi}$ is equivalent to $\Delta t\sim t_f(\omega)/\abar$, which as explained
after  \eqn{Ptf} represents the typical interval between two successive democratic branchings for gluons with energies $\omega\sim\xi E$.}, 
as discussed after  \eqn{Ptf}.

\begin{figure}[t] 
\centerline{
\includegraphics[width=.6\textwidth]{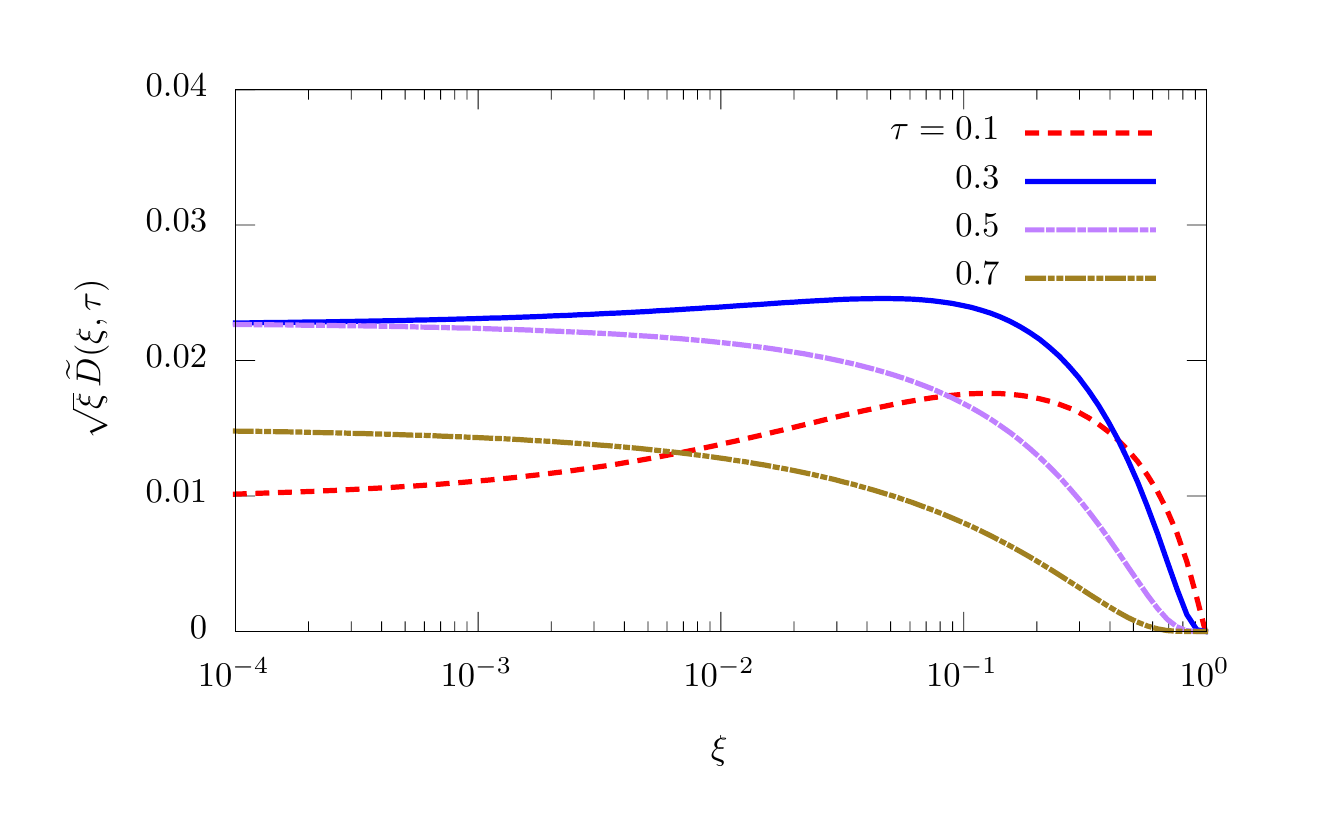}\hspace*{-0.5cm}\includegraphics[width=.6\textwidth]{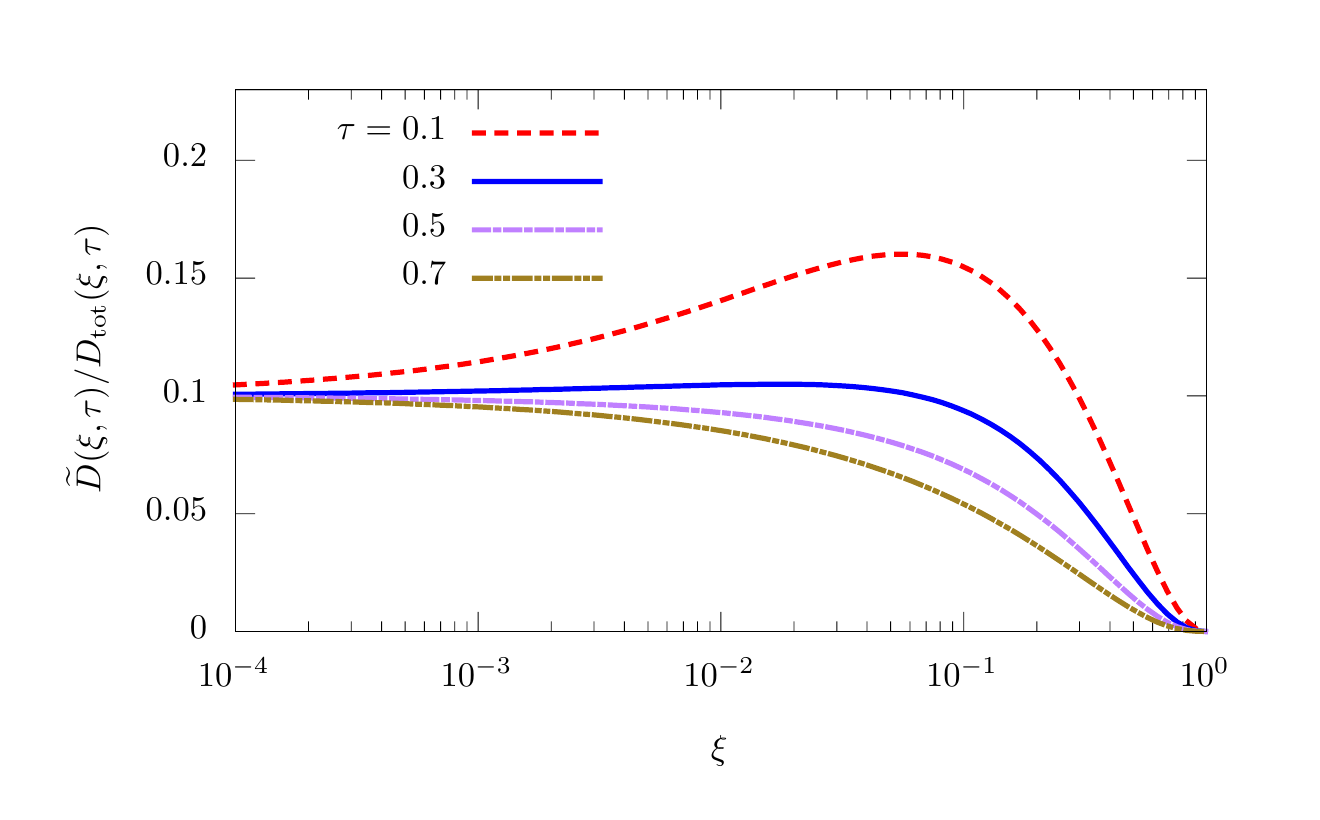}}
\caption{\small  The polarised gluon distribution $\widetilde{D}(\xi,\tau)$ created in an anisotropic medium with 
$G(\widehat{q}_z/\widehat{q}_y)=0.29$ by an unpolarised leading gluon. The results are obtained via numerical integration
in \eqn{Eq:Dtilde_general} with the source term from \eqref{Eq:source_def} and the Green's function \eqref{Gfinal}.
{\it Left:} $\widetilde{D}(\xi,\tau)$ is multiplied by $\sqrt{\xi}$ to emphasise the 
turbulent spectrum at small $\xi\ll\tau^2$ (cf. \eqn{Dtilde3}). {\it Right:} The ratio between the polarised and
unpolarised distributions; one observes their proportionality at small $\xi$, in agreement with \eqref{polD}.}
\label{fig:Dtilde}
\end{figure}

The dominance of democratic branchings in the case of \eqn{Dtilde1} has two well-identified 
physical origins: \texttt{(i)} the fact that
the source term in \eqn{Eq:source_smallx} increases with decreasing $\xi_1$, and  \texttt{(ii)} the fact that the Green's function
\eqref{Gfinal}  is strongly suppressed for very asymmetric splittings with $\zeta\equiv \xi/\xi_1\ll 1$ (which in turn reflects
the fact that soft splittings lose memory of the polarisation of the parent parton).

Although physically appealing, the above argument calls for an explicit mathematical proof, that we now present.
To that aim, it is convenient to perform a change of integration variables, 
from $\tau_1$ and $\xi_1$, to $\Delta\tau= \tau - \tau_1$ and
$v\equiv \sqrt{\xi/(\xi_1-\xi)}$. Then \eqn{Dtilde1} becomes
\beq\label{Dtilde2}
\widetilde{D}(\xi,\tau)
\simeq \frac{\pi}{\xi^{3/2}} \,G(\widehat{q}_z/\widehat{q}_y) \int\limits_{\sqrt{\xi/(1-\xi)}}^\infty dv\left[\frac{v^2}{1+v^2}\right]^{5/2}
 \int_0^{\tau} d\Delta\tau\,\Delta\tau\,\big(\tau-\Delta\tau\big)
\; e^{-\pi \frac{v^2\Delta\tau^2 }{\xi}}\, e^{-\pi (\tau-\Delta\tau)^2}\,.
\eeq
Since $\xi\ll 1$, one can replace ${\sqrt{\xi/(1-\xi)}}\simeq\sqrt{\xi}$ in the lower limit of the integral over $v$.

We will shortly check that this double integral is controlled by values $v\sim 1$ and  $\Delta\tau \sim \sqrt{\xi}$.
In order to analytically perform the integrations, it is convenient to assume that $\tau$ is even larger, namely
$1\gg \tau\gg  \sqrt{\xi}$. 
Physically, this means that we focus on polarised gluons with very low energies
$\omega=\xi E\ll \ombr$  (recall that $ {\xi}\sim \tau^2$ corresponds to $\omega\sim\ombr$). 

Since  $\tau\gg  \sqrt{\xi}\sim\Delta\tau$, we can neglect $\Delta\tau$ next to $\tau$ in the above integrand and then 
the integral over $\Delta\tau$ becomes trivial:
 \begin{align}\label{intDtau}
 \int_0^{\tau} d\Delta\tau\,\Delta\tau\,\big(\tau-\Delta\tau\big)
\; e^{-\pi \frac{v^2\Delta\tau^2 }{\xi}}\, e^{-\pi (\tau-\Delta\tau)^2}\,&\simeq\,
\tau \,e^{-\pi \tau^2}  \int_0^{\tau} d\Delta\tau\,\Delta\tau\,e^{-\pi \frac{v^2\Delta\tau^2 }{\xi}}\nonumber\\*[0.2cm]
&\,=\,\tau e^{-\pi \tau^2} \frac{\xi}{2\pi v^2} \left(1 - e^{-\pi \frac{v^2\tau^2 }{\xi}} \right)
 \end{align}
 By inserting this result into \eqn{Dtilde2}, one finds
 \begin{align}
\label{Dtilde3}
\widetilde{D}(\xi,\tau)&\,
\simeq \,G(\widehat{q}_z/\widehat{q}_y) \,
\frac{\tau e^{-\pi \tau^2} }{2\sqrt{\xi}} 
\int\limits_{\sqrt{\xi}}^\infty dv \frac{v^3}{(1+v^2)^{5/2}} \left(1 - e^{-\pi \frac{v^2\tau^2 }{\xi}} \right)
\nonumber\\*[0.2cm]
&\,\simeq\,G(\widehat{q}_z/\widehat{q}_y) \,
\frac{\tau e^{-\pi \tau^2} }{2\sqrt{\xi}} 
\int\limits_{\sqrt{\xi}/\tau}^\infty dv \frac{v^3}{(1+v^2)^{5/2}}\,=\,
\frac{G(\widehat{q}_z/\widehat{q}_y)}{3} \,
\frac{\tau e^{-\pi \tau^2} }{\sqrt{\xi}}\Big[1 \,+\,\order{\sqrt{\xi}/\tau}\Big]\,.
\end{align}
Notice that the final integral is controlled by values $v\sim 1$, as anticipated. Hence, neither the exponential 
within the integrand in the first line, nor the lower limit $\sqrt{\xi}/\tau$ on the integral over $v$ in the second line
(that was precisely introduced by the exponential) were important for the final result. Indeed, when 
$\tau\gg  \sqrt{\xi}$ and $v\sim 1$, the exponent $v^2\tau^2/\xi$ is large and the exponential is strongly suppressed.
When $v\sim 1$, it is easy to check that the previous integral in  \eqn{intDtau} is indeed controlled by
$\Delta\tau \sim \sqrt{\xi}$.

Remarkably, $\widetilde{D}(\xi,\tau)$ is simply proportional to \(D_{\mathrm{tot}}(\xi,\tau)\) for such small values of $\xi$:
 \begin{align}\label{polD}
\frac{\widetilde{D}(\xi,\tau)}{D_{\mathrm{tot}}(\xi,\tau)}\, \simeq\, \frac{G(\widehat{q}_z/\widehat{q}_y)}{3}\qquad
\mbox{for}\qquad  {\xi}\ll \tau^2.
\end{align}
Although derived here for {\it very} small energies $\omega\ll\ombr$, we expect this relation between the polarised and the total distributions
to remain  qualitatively true for all the gluons affected by multiple branching --- those
with energies $\omega\lesssim\ombr$, or  $\xi\lesssim\tau^2$ (so long as $\tau\ll 1$, of course). This is confirmed
by a numerical calculation of $\widetilde{D}(\xi,\tau)$ based on Eqs.~\eqref{Eq:Dtilde_general}--\eqref{Eq:source_def}
(that is, the numerical evaluation of a series of three convolutions), with the results displayed in Fig.~\ref{fig:Dtilde}.
In this calculation, we have assumed the maximum value $G(\widehat{q}_z/\widehat{q}_y)=0.29$ for the quantity 
which measures the effects of the anisotropy on the branching rates. (As visible in Fig.~\ref{fig:AG}, this value $G=0.29$
corresponds to $ \Delta \widehat{q}/\hat q=1$, i.e. to maximum anisotropy.) Changing the value of $G$ does not
change the trends visible in Fig.~\ref{fig:Dtilde}, but only the normalisation of the curves.
The proportionality relation \eqref{polD} is seen to be numerically verified so long as $  {\xi}\ll \tau^2$, while
small deviations are observed when increasing $\xi$ towards $\tau^2$.

To summarise, we have found that, even for a jet initiated by an unpolarised leading parton, the soft gluon distribution generated via
multiple branching in an anisotropic medium exhibits a net polarisation, which is proportional to the medium
anisotropy (via the function $G(\widehat{q}_z/\widehat{q}_y)$) and also to the total (unpolarised) distribution
--- in particular, it exhibits the same power-law enhancement at small $\xi\ll 1$: $\widetilde{D}\propto 1/ \sqrt{\xi}$.
As our calculation also shows, these remarkable features are the consequence of the fact 
that the polarisation is essentially built in the last few splittings ---  one or two quasi-democratic branchings, which are
close in time and also in energy to the measurement time and the energy bin.

Said differently, the polarisation is continuously created and then washed out at each step of the democratic cascade: unlike
the energy (which is a conserved quantity), it does not propagate between gluons from very different generations. 
Accordingly, the only reason why the polarised distribution appears to
increase with time and with $1/\xi$ is because the number of its sources --- the unpolarised distribution  
\(D_{\mathrm{tot}}(\xi,\tau)\)  --- keeps rising so long as $\tau\ll 1$, i.e. so long as the leading parton
is still ``alive'' and acts as a source of soft primary gluons, which regenerate the cascade.

\section{Conclusions and perspectives}
\label{sec:conc}

In this paper, we have studied the evolution of a jet propagating through an anisotropic quark-gluon plasma, with emphasis on the
consequences of the anisotropy for the polarisation of the jet constituents. The physical problem we had in mind is the production
of jets at central rapidities in ultrarelativistic heavy ion collisions, at RHIC and the LHC. In that context, the anisotropy is naturally
generated by the rapid expansion of the medium along the collision axis, leading to a squeezed momentum distribution for
the partons from the plasma. Our first observation is that the medium anisotropy leads to a difference between the
jet quenching parameters --- the rates for transverse momentum broadening --- along the two directions transverse to the
jet axis: the collision axis and a direction in the plane orthogonal to it. Within the BDMPS-Z mechanism for medium-induced
gluon emissions, such an asymmetry in the transverse momentum broadening naturally leads to a modification in the parton branching
rates. This modification depends upon the partons' polarisation and thus has the potential to dynamically generate net polarisation.
 
 In order to study this, we considered a somewhat simplified scenario, 
in which the plasma is weakly coupled and stationary, and the anisotropy is enforced by merely assuming different values for the 
relevant jet quenching parameters. Also, we considered a jet which is made only of gluons.
Within this set-up, we have generalised the BDMPS-Z mechanism 
to the anisotropic plasma and to the branchings of gluons with definite (linear) polarisation states. Our results for
the splitting rates, as shown in Eqs.~\eqref{Eq:Gammaxtox_full}--\eqref{Eq:Gammaytoy_full}, confirm that, due to the anisotropy, 
the daughter gluons can carry net polarisation even when their parent gluon was unpolarised. 
This effect is particularly remarkable for the emissions of soft gluons
which cannot inherit the polarisation of their parent: in the absence of the anisotropy, such gluons would always be unpolarised, 
irrespective of the polarisation state of their parent (cf. the discussion in Sect.~\ref{sec:1split}).

Using the polarised splitting rates in Eqs.~\eqref{Eq:Gammaxtox_full}--\eqref{Eq:Gammaytoy_full}, we have 
constructed kinetic equations which determine the time evolution of the energy and
polarisation distributions among the jet constituents, Eqs.~\eqref{Eq:Dtotevol} and \eqref{Eq:Dpolevol}. 
They represent natural generalisations of the respective equations for an 
isotropic medium and unpolarised gluons \cite{Blaizot:2013vha}. Like the latter, the polarized kinetic equations exhibit turbulent dynamics: the soft gluons multiply via
quasi-democratic branchings, which now control both the creation of net polarisation (via the decay of unpolarised partons)
and its transmission from one parton generation to the next one. This transmission however is not very efficient: it is
hindered by the fact that, as alluded to above, soft gluons lose memory of the polarisation of their parents. As a result, the
polarisation is both created (by the anisotropy) and washed out (via the subsequent emissions of soft gluons) {\it quasi-locally}
in energy and time --- that is, within just a few, successive democratic branchings. Accordingly, the net polarisation of soft
gluons closely follows the evolution of the unpolarised gluon distribution: the two distributions are simply proportional to each other,
with a proportionality coefficient that would vanish for an isotropic plasma, cf. \eqn{polD}.

These considerations also suggest that for a more realistic scenario, where the plasma anisotropy decreases
with time ---  the momentum distributions of the plasma constituents are expected to evolve towards isotropy, as a result of
elastic collisions which compete with, and eventually take over, the medium expansion \cite{Baier:2000sb} ---, this
mechanism for generating polarisation should become less and less effective with increasing time.  
That said, the numerical solutions to both kinetic theory \cite{Kurkela:2015qoa,Kurkela:2018vqr,Du:2020dvp}
and (viscous) relativistic hydrodynamics \cite{Heinz:2013th,Gale:2013da,Luzum:2013yya,Alqahtani:2017mhy} demonstrate
that, even for relatively large values of the QCD coupling, the approach towards (local) thermal equilibrium is very slow.
So, it is quite plausible that, during the late stages of the jet propagation through the medium --- the most important ones
for its evolution via medium-induced parton branchings --- the anisotropy is still sizeable and only slowly varying with time.
In such a case, our general conclusions in this paper should remain valid (at least qualitatively).


Our present analysis could be improved at several levels, referring both to the description of the medium and to the
treatment of the jet-medium interactions. First, it should be straightforward to include the effects of the 
longitudinal expansion of the medium, by promoting the jet quenching parameters to time-dependent 
quantities\footnote{Notice that,
with these choices, the ratio $\widehat{q}_z(t)/\widehat{q}_y(t)=  \widehat{q}_{z0}/\widehat{q}_{y0}$ remains independent
of time, therefore the same is true for the  functions 
\(A(\widehat{q}_z/\widehat{q}_y)\) and \(G(\widehat{q}_z/\widehat{q}_y)\) which enter the polarised branching rates
\eqref{Eq:Gammaxtox_full}--\eqref{Eq:Gammaytoy_full}. Hence, while the total emission rate becomes time-dependent,
the effects of the anisotropy are still stationary.}, like in Refs.~\cite{Baier:1998yf,Zakharov:1998wq,Baier:2000sb,Arnold:2008iy,Iancu:2018trm,Adhya:2019qse,Caucal:2020uic}:
$\widehat{q}_z(t)=  \widehat{q}_{z0}t_0/t$ and $\widehat{q}_y(t)=  \widehat{q}_{y0}t_0/t$.
A more ambitious approach would involve the microscopic calculation of $\widehat{q}_z(t)$ and $\widehat{q}_y(t)$
for an anisotropic medium, following the lines in Sect.~\ref{sec:ptbroad}.
This would require a theory, or at least a model, for the non-equilibrium (quark and gluon) occupation numbers, 
which enter the gluon polarisation tensor \eqref{PiHTL}. 

Alternatively, the occupation numbers could be taken from numerical solutions to kinetic theory 
\cite{Kurkela:2015qoa,Kurkela:2018vqr,Du:2020dvp}, or (at late stages) from hydrodynamical simulations,
the results of which can be at least approximately matched onto parton occupation numbers
\cite{Heinz:2013th,Gale:2013da,Luzum:2013yya,Alqahtani:2017mhy}.
 In such a case, it could be more convenient to use numerical solutions also
for the rate equations describing the (polarised) parton distributions inside the jet. Alternatively, and perhaps more
efficiently, one could use the Monte-Carlo formulation of the in-medium partonic cascades, as developed in 
\cite{Caucal:2018dla,Caucal:2019uvr,Caucal:2020xad,Caucal:2020uic}; that would also permit to include the
{\it vacuum-like} parton branchings (those resulting from parton virtualities). Yet another approach would be to solve
the linearized kinetic equations describing the evolution of the jet partonic distributions via both elastic
and inelastic collisions \cite{Mehtar-Tani:2022zwf}. In such an approach, the  jet quenching parameters are
dynamically generated. Yet, for the present purposes, the equations in  \cite{Mehtar-Tani:2022zwf} must be
extended to a {\it non-equilibrium} background distribution, which is anisotropic.

Our present conclusions may also have interesting implications for the late stages of the ``bottom-up'' equilibration scenario
 \cite{Baier:2000sb} --- more precisely, for its third stage, where the (semi)hard partons are ``quenched'' via gluon radiation  
 induced by their rescattering off a thermal plasma made with softer gluons.  As a result of this radiation, the ``hard'' partons 
disappear in the medium, while leaving behind ``mini-jets'' which eventually thermalise.
At those early stages, the surrounding plasma is still highly anisotropic, hence the gluons from the mini-jets should carry
significant (net) polarisation. In turn this implies that the plasma created by this early evolution is itself polarised.
That should be the stage at which the hydrodynamics starts to be applicable. It would be therefore interesting to consider
hydrodynamical simulations of a polarised medium.

Having shown that gluons in jets become polarised in an anisotropic medium,  a natural question is how this effect could be measured in experiments.  A natural proposal would be to measure the polarisation of jet hadrons, as angular momentum conservation suggests that polarisation of gluons in the jet 
is transmitted to polarisation of the final-state hadrons.  This proposal is complicated by at least two factors: firstly, polarisation of hadrons is difficult to measure, and secondly, the fragmentation functions specifying the polarisation states for both the initial parton and the final hadron are presently not known.

Yet another strategy to uncover the polarisation of the partons in a jet is based on the fact that, via the hadronisation process,
the polarisation of the initial parton can be correlated with the direction of motion of the outgoing hadron. 
As an example, a quark travelling in the \(x\)-direction with spin aligned along the \(z\)-axis is more likely to emit a hadron 
with momentum component in the \(y\)-direction than in the \(z\)-direction. For quarks this correlation
is captured by the Collins function \cite{Collins:1992kk, Amrath:2005gv} which  has been measured in experiments, see e.g. \cite{BaBar:2013jdt, Anselmino:2007fs}. In principle, one could therefore measure the anisotropic distribution of hadrons inside the jet cone and extract from it the net polarisation of quarks before hadronisation. Our present study only deals with gluons for which the analogue of the Collins function is not known. However, it should be straightforward to generalise our work to include quarks. Another complication is the fact that jet parton polarisation is not the only source of final-state anisotropy of the hadrons inside the jet cone. 
Another obvious source is the anisotropic momentum broadening of the partons prior to hadronisation, which
can be computed and included in a complete phenomenological study. 

Last but not least, we hope that the methodology and the formalism that we have developed on this occasion ---
especially, the generalisation of the BDMPS-Z approach to an anisotropic plasma and to gluons with definite
polarisation states --- will turn out to be useful for other problems as well and will inspire further developments.

\section*{Acknowledgements} 
One of us (S.H.) would like to thank Charles Gale for inspiring discussions during the early stages of this project.

\appendix
\section{Gluon propagation and transverse momentum broadening}
\label{app:pt}

In this Appendix we shall briefly review the calculation of the transverse momentum broadening for an energetic
gluon propagating through a weakly-coupled quark-gluon plasma. Our purpose is to justify \eqn{Pcoord}  for
the relevant probability density. 
The main ingredient of the calculation is the amplitude, $\mathcal{M}^{a}_\lambda(p^+,\bp,t)$ for finding
the gluon with longitudinal momentum $p^+$, transverse momentum $\bp$, polarisation $\lambda$ and colour
$a$ at ``time'' $t\equiv x^+$. This amplitude must be computed for a given configuration of the random gauge field
$A^-$ representing the medium. The average over $A^-$, cf. \eqn{2pA}, should be done after squaring the amplitude.

The time evolution of this amplitude is encompassed in the following recurrence relation
\begin{align}
\mathcal{M}^{a}_\lambda(p^+,\bp, t)
=\int \frac{d^2  \bm{p}_0}{(2\pi)^2} \,\mathcal{G}_{ab}(t, \bp; t_0, \bm{p}_0; p^+)\,\mathcal{M}^{b}_\lambda(p^+,\bm{p}_0, t_0)
\end{align}
where $t_0 < t$ (we assume that the gluon is inside the medium at both $t_0$ and $t$) and 
$\mathcal{G}_{ab}(t, \bp; t_0, \bm{p}_0; p^+)$ is the (relevant component of the) gluon propagator in the
background field $A^-$. This propagator is diagonal in $p^+$ and $\lambda$ (since these
quantities are not modified by the medium) and also independent of $\lambda$ (since both
transverse polarisations propagate in the same way). On the other hand, it describes a non-trivial
change $\Delta \bp=\bp-\bm{p}_0$ in transverse momentum, associated with multiple scattering in the plasma.
To describe this scattering, it is preferable to work in transverse coordinate space, e.g.
\begin{align}
\mathcal{M}^{a}_\lambda(p^+,\bp, t)=\int\rmd^2\bx\,\rme^{i\bp\cdot\bx}\,\mathcal{M}^{a}_\lambda(p^+,\bx, t)\,.
\end{align}
The background-field propagator in mixed Fourier representation  $\mathcal{G}_{ab}(t, \bx; t_0, \bm{\xi}_0; p^+)$ obeys
\beq
\left[ i\partial_t+\frac{\nabla_{\bx}^2}{2p^+} + gA^-(t,\bx)\right]_{ac}
\,\mathcal{G}_{ac}(t, \bx; t_0, \bm{\xi}_0; p^+)=i\delta_{ab}\delta(t-t_0)\delta(\bx-\bm{\xi}_0),
\eeq
which is formally the same as the Schr\" odinger equation
in  two dimensions (the transverse plane $\bx=(y,z)$) for a non relativistic particle of mass $p^+$ 
moving in a time dependent potential $A^-(t,\bx)$. 
As well known, the solution  can be written as  a path integral,
\beq
\mathcal{G}_{ab}(t, \bx; t_0, \bm{\xi}_0; p^+)=\int_{\bu(t_0)=\bm{\xi}_0}^{\bu(t)=\bx}
 {\cal D} \bu\, \exp\left\{i\frac{p^+}{2} \int_{t_0}^t   \rmd t' \, \dot\bu^2 \right\}\, U_{ab}(t,t_0; \bu)
\eeq
where $\bu(t')$ is a 2-dimensional path with the indicated end points and
$U$ is a Wilson line in the adjoint representation, evaluated along this path: 
\beq
 U(t,t_0; \bu)={\rm T}\exp\left\{ ig\int_{t_0}^{t} \rmd t' \, A^-_a(t, \bu(t'))\, T^a   \right\}.
\eeq

To compute the transverse momentum distribution at time $t$, one must take the modulus squared of
the amplitude  $\mathcal{M}^{a}_\lambda(p^+,\bp,t)$ and average over $A^-$. This operation involves
the average of the colour trace of a product of two Wilson lines, one from the direct amplitude (DA), the other one
from the complex conjugate amplitude (CCA). The colour trace appears because the final colour state $a$
must be the same in both the DA and the CCA (since we measure the gluon at time $t$), 
and then the same must be true for the initial colour state $b$ at time $t_0$, by gauge invariance:
\begin{align}\label{Sdef}
\left\langle U_{ab}(t,t_0; \bu) U^*_{ac}(t,t_0; \bv)\right\rangle&\,=\,\delta_{bc} S(t, t_0; \bu, \bv),
\nonumber\\*[0.2cm]
S(t, t_0; \bu, \bv)&\,\equiv\, \frac{1}{N_c^2-1} \left\langle {\rm Tr}\,U(t,t_0; \bu) U^\dagger
(t,t_0; \bv)\right\rangle\,.\end{align}
Here, $\bv(t')$ is the two dimensional path in the CCA, with end points  $\bv(t_0)=\bm{y}_0$ and $\bv(t)=\by$.
Mathematically, the quantity $S(t, t_0; \bu, \bv)$ is the $S$-matrix for the elastic scattering
between a  gluon-gluon colour dipole  --- a pair of gluons in an overall colour singlet state --- and the medium.
A priori, this is a functional of the paths, $\bu(t')$ and $\bv(t')$, of the two gluons. Yet, due to the fact
that the field-field correlator  \eqref{2pA} is local in time and homogeneous in the transverse plane, the 
medium average in \eqn{Sdef} yields a result which only depends upon the {\it difference path} 
$\br(t)\equiv \bu(t)-\bv(t)$:
\beq
S(t, t_0; \bu, \bv)=\exp\left\{-g^2{N_c}\int_{t_0}^t   \rmd t' \, [\gamma(0)-\gamma(\br(t'))]\right\}
\equiv S(t, t_0; \br)\,.\eeq
Then the ensuing path integrals over $\bu(t')$ and $\bv(t')$ greatly simplify: they select only paths which
are such that their difference $\br(t')$ is a linear function of time,  fully determined by its endpoints:
\beq\label{diffpath}
\br(t')=\br(t_0)+[\br(t)-\br(t_0)]\,\frac{t'-t_0}{t-t_0}\,.\eeq
When evaluated along this path, the function $S(t, t_0; \br)$ factorises out from the path integrals over
$\bu(t')$ and $\bv(t')$, which become trivial and provide the respective free propagators. 

To return to momentum space, one must perform Fourier
transforms from the end points $\bx$ and $\by$ to $\bp$ (since the momentum $\bp$ of the measured gluon is
the same in the DA and in the CCA). These two Fourier transforms together with transverse homogeneity 
enforce $\bx-\by=\bm{\xi}_0-\bm{y}_0$, that is, $\br(t)=\br(t_0)$, so the difference path \eqref{diffpath} becomes independent
of time:  $\br(t')=\bx-\by\equiv \br$ for any $t'$. Accordingly, the final momentum distribution is obtained via a 
Fourier transform from $\br$ to $\bp$, as shown in \eqn{Pcoord}.

\section{Rate equations for $D_{\mathrm{tot}}$ and $\tilde{D}$}
\label{sec:generaleqs}

Starting with Eqs.~\eqref{Eq:Dyevol} and \eqref{Eq:Dxevol} for $D_y$ and $D_z$, it  is straightforward to deduce the following equations for 
$D_{\mathrm{tot}}$ and $\tilde{D}$, as defined in \eqn{Dtotpol}:
\beq
\label{Eq:Dtotevol_first}
\begin{split}
\frac{d D_{\mathrm{tot}}(\xi,\tau)}{d\tau} 
 &= \int_ {\xi}^1 d\zeta\; \frac{1}{2}\Big[\mathcal{K}_{ z\rightarrow z}(\zeta) + \mathcal{K}_{ z\rightarrow y}(\zeta) + \mathcal{K}_{y\rightarrow z}(\zeta) + \mathcal{K}_{y\rightarrow y}(\zeta) \Big]\, \sqrt{\frac{\zeta} {\xi}}\, D_{\mathrm{tot}}\left(\frac {\xi}{\zeta},\tau\right) \\
 &+ \int_ {\xi}^1 d\zeta\; \frac{1}{2}\Big[\mathcal{K}_{ z\rightarrow z}(\zeta) + \mathcal{K}_{ z\rightarrow y}(\zeta) - \mathcal{K}_{y\rightarrow z}(\zeta) - \mathcal{K}_{y\rightarrow y}(\zeta) \Big]\, \sqrt{\frac{\zeta} {\xi}}\, \widetilde{D}\left(\frac {\xi}{\zeta},\tau\right) \\
 &- \frac{1}{2} \int_0^1 d\zeta\;\frac{1}{2}\Big[\mathcal{K}_{ z\rightarrow z}(\zeta) + \mathcal{K}_{ z\rightarrow y}(\zeta) + \mathcal{K}_{y\rightarrow z}(\zeta) + \mathcal{K}_{y\rightarrow y}(\zeta) \Big]\, \frac{1}{\sqrt {\xi}}\, D_{\mathrm{tot}}(\xi,\tau) \\
  &- \frac{1}{2}\int_0^1 d\zeta\;\frac{1}{2}\Big[\mathcal{K}_{ z\rightarrow z}(\zeta) + \mathcal{K}_{ z\rightarrow y}(\zeta) - \mathcal{K}_{y\rightarrow z}(\zeta) - \mathcal{K}_{y\rightarrow y}(\zeta)\Big]\, \frac{1}{\sqrt {\xi}}\, \widetilde{D}(\xi,\tau)
\end{split}
\eeq
and
\beq
\label{Eq:Dpolevol_first}
\begin{split}
\frac{d \widetilde{D}(\xi,\tau)}{d\tau} 
 &= \int_ {\xi}^1 d\zeta\; \frac{1}{2}\Big[\mathcal{K}_{ z\rightarrow z}(\zeta) - \mathcal{K}_{ z\rightarrow y}(\zeta) + \mathcal{K}_{y\rightarrow z}(\zeta) - \mathcal{K}_{y\rightarrow y}(\zeta) \Big]\, \sqrt{\frac{\zeta} {\xi}}\, D_{\mathrm{tot}}\left(\frac {\xi}{\zeta},\tau\right) \\
 &+ \int_ {\xi}^1 d\zeta\; \frac{1}{2}\Big[\mathcal{K}_{ z\rightarrow z}(\zeta) - \mathcal{K}_{ z\rightarrow y}(\zeta) - \mathcal{K}_{y\rightarrow z}(\zeta) + \mathcal{K}_{y\rightarrow y}(\zeta) \Big]\, \sqrt{\frac{\zeta} {\xi}}\, \widetilde{D}\left(\frac {\xi}{\zeta},\tau\right) \\
 &- \frac{1}{2}\int_0^1 d\zeta\;\frac{1}{2}\Big[\mathcal{K}_{ z\rightarrow z}(\zeta) + \mathcal{K}_{ z\rightarrow y}(\zeta) - \mathcal{K}_{y\rightarrow z}(\zeta) - \mathcal{K}_{y\rightarrow y}(\zeta) \Big]\, \frac{1}{\sqrt {\xi}}\, D_{\mathrm{tot}}(\xi,\tau) \\
  &- \frac{1}{2}\int_0^1 d\zeta\;\frac{1}{2}\Big[\mathcal{K}_{ z\rightarrow z}(\zeta) + \mathcal{K}_{ z\rightarrow y}(\zeta) + \mathcal{K}_{y\rightarrow z}(\zeta) + \mathcal{K}_{y\rightarrow y}(\zeta)\Big]\, \frac{1}{\sqrt {\xi}}\, \widetilde{D}(\xi,\tau)
\end{split}
\eeq
The new kernels which enter these equations are defined as:
\beq
\label{Eq:K_kernel}
\begin{split}
\mathcal{K}_0(\zeta) &\equiv \frac{1}{2}\Big[\mathcal{K}_{ z \rightarrow z } (\zeta)+ \mathcal{K}_{ z \rightarrow y } (\zeta)+ \mathcal{K}_{ y \rightarrow z } (\zeta)+ \mathcal{K}_{ y \rightarrow y } (\zeta)\Big]\\
&= \gamma(\zeta)\left[ \frac{1-\zeta}{\zeta} + \frac{\zeta}{1-\zeta} + \zeta(1-\zeta)\right]
\simeq \frac{1}{\zeta^{3/2} (1-\zeta)^{3/2}},
\end{split}
\eeq
\beq
\label{Eq:L_kernel}
\begin{split}
\mathcal{L}_0 (\zeta)&\equiv \frac{1}{2}\Big[\mathcal{K}_{ z \rightarrow z } (\zeta)- \mathcal{K}_{ z \rightarrow y } (\zeta)+ \mathcal{K}_{ y \rightarrow z } (\zeta)- \mathcal{K}_{ y \rightarrow y } (\zeta)\Big] \\
&= \gamma (\zeta)\frac{1-\zeta}{\zeta} \, G(\widehat{q}_z/\widehat{q}_y) 
\simeq \frac{(1-\zeta)^{1/2}}{\zeta^{3/2}} \,G(\widehat{q}_z/\widehat{q}_y),
\end{split}
\eeq
and
\beq
\label{Eq:M_kernel}
\begin{split}
\mathcal{M}_0 (\zeta)&\equiv \frac{1}{2}\Big[\mathcal{K}_{ z \rightarrow z } (\zeta)- \mathcal{K}_{ z \rightarrow y } (\zeta)- \mathcal{K}_{ y \rightarrow z } (\zeta)+ \mathcal{K}_{ y \rightarrow y } (\zeta)\Big] \\
&= \gamma(\zeta)\,\frac{\zeta}{1-\zeta} \simeq \frac{\zeta^{1/2}}{(1-\zeta)^{3/2}}.
\end{split}
\eeq
The final approximations have been obtained by ignoring the non-singular contributions multiplied by  \(\zeta(1-\zeta)\). In particular,
the function in Eq. \eqref{Eq:gamma}  has been approximated as
\beq
\gamma (\zeta)\simeq \frac{1}{\zeta^{1/2} (1-\zeta)^{1/2}}.
\eeq
We have furthermore used that the combination
\beq
\begin{split}
&\frac{1}{2}\Big[\mathcal{K}_{ z \rightarrow z } (\zeta)+ \mathcal{K}_{ z \rightarrow y } (\zeta)- \mathcal{K}_{ y \rightarrow z } (\zeta)- \mathcal{K}_{ y \rightarrow y } (\zeta)\Big] \\
=& \gamma (\zeta)\zeta(1-\zeta)\, G(\widehat{q}_x/\widehat{q}_y)
\simeq \sqrt{\zeta(1-\zeta)} \, G(\widehat{q}_x/\widehat{q}_y) 
\simeq 0
\end{split}
\eeq
can be ignored in the same limit. Our approximations are chosen so that they reproduce the first term in a Laurent expansion in \(\zeta \) of the full function, as well as the first term in a Laurent expansion in \(1-\zeta\) of the full function. As an example \(\mathcal{M}_0 (\zeta)= \zeta^{-1/2} + \mathcal{O}(\zeta ^{1/2})\) and \(\mathcal{M}_0 (\zeta)= (1-\zeta)^{-3/2} + \mathcal{O}((1-\zeta)^{1/2})\), both of which are captured by our approximation.

We also  used  the fact that, since \(\mathcal{K}_0(\zeta)\) is symmetric under \(\zeta \rightarrow 1-\zeta\), we can write
\beq
 \int_0^1 d\zeta\;\mathcal{K}_0 (\zeta)=  \int_0^1 d\zeta\;\mathcal{K}_0 (\zeta)\left[ \zeta + (1-\zeta)\right] = 2 \int_0^1 d\zeta\;\mathcal{K}_0 (\zeta)\,\zeta.
\eeq

\bibliographystyle{JHEP}
\bibliography{refs}

\providecommand{\href}[2]{#2}\begingroup\raggedright\begin{thebibliography}{10}

\bibitem{Casalderrey-Solana:2007knd}
J.~Casalderrey-Solana and C.A.~Salgado, \emph{{Introductory lectures on jet
  quenching in heavy ion collisions}}, {\emph{Acta Phys. Polon. B} {\bfseries
  38} (2007) 3731} [\href{https://arxiv.org/abs/0712.3443}{{\ttfamily
  0712.3443}}].

\bibitem{Qin:2015srf}
G.-Y.~Qin and X.-N.~Wang, \emph{{Jet quenching in high-energy heavy-ion
  collisions}}, \href{https://doi.org/10.1142/S0218301315300143,
  10.1142/9789814663717_0007}{\emph{Int. J. Mod. Phys.} {\bfseries E24} (2015)
  1530014} [\href{https://arxiv.org/abs/1511.00790}{{\ttfamily 1511.00790}}].

\bibitem{Blaizot:2015lma}
J.-P.~Blaizot and Y.~Mehtar-Tani, \emph{{Jet Structure in Heavy Ion
  Collisions}}, \href{https://doi.org/10.1142/S021830131530012X}{\emph{Int. J.
  Mod. Phys.} {\bfseries E24} (2015) 1530012}
  [\href{https://arxiv.org/abs/1503.05958}{{\ttfamily 1503.05958}}].

\bibitem{Bjorken:1982qr}
J.D.~Bjorken, \emph{{Highly Relativistic Nucleus-Nucleus Collisions: The
  Central Rapidity Region}},
  \href{https://doi.org/10.1103/PhysRevD.27.140}{\emph{Phys. Rev.} {\bfseries
  D27} (1983) 140}.

\bibitem{Baier:2000sb}
R.~Baier, A.H.~Mueller, D.~Schiff and D.~Son, \emph{{'Bottom up' thermalization
  in heavy ion collisions}},
  \href{https://doi.org/10.1016/S0370-2693(01)00191-5}{\emph{Phys.Lett.}
  {\bfseries B502} (2001) 51}
  [\href{https://arxiv.org/abs/hep-ph/0009237}{{\ttfamily hep-ph/0009237}}].

\bibitem{Dumitru:2009ni}
A.~Dumitru, Y.~Guo, A.~Mocsy and M.~Strickland, \emph{{Quarkonium states in an
  anisotropic QCD plasma}},
  \href{https://doi.org/10.1103/PhysRevD.79.054019}{\emph{Phys. Rev. D}
  {\bfseries 79} (2009) 054019}
  [\href{https://arxiv.org/abs/0901.1998}{{\ttfamily 0901.1998}}].

\bibitem{Burnier:2009yu}
Y.~Burnier, M.~Laine and M.~Vepsalainen, \emph{{Quarkonium dissociation in the
  presence of a small momentum space anisotropy}},
  \href{https://doi.org/10.1016/j.physletb.2009.05.067}{\emph{Phys. Lett. B}
  {\bfseries 678} (2009) 86} [\href{https://arxiv.org/abs/0903.3467}{{\ttfamily
  0903.3467}}].

\bibitem{Thakur:2012eb}
L.~Thakur, N.~Haque, U.~Kakade and B.K.~Patra, \emph{{Dissociation of
  quarkonium in an anisotropic hot QCD medium}},
  \href{https://doi.org/10.1103/PhysRevD.88.054022}{\emph{Phys. Rev. D}
  {\bfseries 88} (2013) 054022}
  [\href{https://arxiv.org/abs/1212.2803}{{\ttfamily 1212.2803}}].

\bibitem{Dong:2022mbo}
L.~Dong, Y.~Guo, A.~Islam, A.~Rothkopf and M.~Strickland, \emph{{The complex
  heavy-quark potential in an anisotropic quark-gluon plasma \textemdash{}
  Statics and dynamics}},
  \href{https://doi.org/10.1007/JHEP09(2022)200}{\emph{JHEP} {\bfseries 09}
  (2022) 200} [\href{https://arxiv.org/abs/2205.10349}{{\ttfamily
  2205.10349}}].

\bibitem{Prakash:2021lwt}
J.~Prakash, M.~Kurian, S.K.~Das and V.~Chandra, \emph{{Heavy quark transport in
  an anisotropic hot QCD medium: Collisional and Radiative processes}},
  \href{https://doi.org/10.1103/PhysRevD.103.094009}{\emph{Phys. Rev. D}
  {\bfseries 103} (2021) 094009}
  [\href{https://arxiv.org/abs/2102.07082}{{\ttfamily 2102.07082}}].

\bibitem{Song:2019cqz}
T.~Song, P.~Moreau, J.~Aichelin and E.~Bratkovskaya, \emph{{Exploring
  non-equilibrium quark-gluon plasma effects on charm transport coefficients}},
  \href{https://doi.org/10.1103/PhysRevC.101.044901}{\emph{Phys. Rev. C}
  {\bfseries 101} (2020) 044901}
  [\href{https://arxiv.org/abs/1910.09889}{{\ttfamily 1910.09889}}].

\bibitem{Romatschke:2004au}
P.~Romatschke and M.~Strickland, \emph{{Collisional energy loss of a heavy
  quark in an anisotropic quark-gluon plasma}},
  \href{https://doi.org/10.1103/PhysRevD.71.125008}{\emph{Phys. Rev. D}
  {\bfseries 71} (2005) 125008}
  [\href{https://arxiv.org/abs/hep-ph/0408275}{{\ttfamily hep-ph/0408275}}].

\bibitem{Ryblewski:2015hea}
R.~Ryblewski and M.~Strickland, \emph{{Dilepton production from the quark-gluon
  plasma using (3+1)-dimensional anisotropic dissipative hydrodynamics}},
  \href{https://doi.org/10.1103/PhysRevD.92.025026}{\emph{Phys. Rev. D}
  {\bfseries 92} (2015) 025026}
  [\href{https://arxiv.org/abs/1501.03418}{{\ttfamily 1501.03418}}].

\bibitem{Churchill:2020uvk}
J.~Churchill, L.~Yan, S.~Jeon and C.~Gale, \emph{{Emission of electromagnetic
  radiation from the early stages of relativistic heavy-ion collisions}},
  \href{https://doi.org/10.1103/PhysRevC.103.024904}{\emph{Phys. Rev. C}
  {\bfseries 103} (2021) 024904}
  [\href{https://arxiv.org/abs/2008.02902}{{\ttfamily 2008.02902}}].

\bibitem{Coquet:2021lca}
M.~Coquet, X.~Du, J.-Y.~Ollitrault, S.~Schlichting and M.~Winn,
  \emph{{Intermediate mass dileptons as pre-equilibrium probes in heavy ion
  collisions}},
  \href{https://doi.org/10.1016/j.physletb.2021.136626}{\emph{Phys. Lett. B}
  {\bfseries 821} (2021) 136626}
  [\href{https://arxiv.org/abs/2104.07622}{{\ttfamily 2104.07622}}].

\bibitem{Lappi:2006fp}
T.~Lappi and L.~McLerran, \emph{{Some features of the glasma}},
  \href{https://doi.org/10.1016/j.nuclphysa.2006.04.001}{\emph{Nucl. Phys. A}
  {\bfseries 772} (2006) 200}
  [\href{https://arxiv.org/abs/hep-ph/0602189}{{\ttfamily hep-ph/0602189}}].

\bibitem{Iancu:2003xm}
E.~Iancu and R.~Venugopalan, \emph{{The color glass condensate and high energy
  scattering in QCD}},  \href{https://arxiv.org/abs/hep-ph/0303204}{{\ttfamily
  hep-ph/0303204}}.

\bibitem{Gelis:2010nm}
F.~Gelis, E.~Iancu, J.~Jalilian-Marian and R.~Venugopalan, \emph{{The Color
  Glass Condensate}},
  \href{https://doi.org/10.1146/annurev.nucl.010909.083629}{\emph{Ann.Rev.Nucl.Part.Sci.}
  {\bfseries 60} (2010) 463} [\href{https://arxiv.org/abs/1002.0333}{{\ttfamily
  1002.0333}}].

\bibitem{Gelis:2012ri}
F.~Gelis, \emph{{Color Glass Condensate and Glasma}},
  \href{https://doi.org/10.1142/S0217751X13300019}{\emph{Int. J. Mod. Phys. A}
  {\bfseries 28} (2013) 1330001}
  [\href{https://arxiv.org/abs/1211.3327}{{\ttfamily 1211.3327}}].

\bibitem{Gelis:2015gza}
F.~Gelis, \emph{{Initial state and thermalization in the Color Glass Condensate
  framework}}, \href{https://doi.org/10.1142/S0218301315300088}{\emph{Int. J.
  Mod. Phys. E} {\bfseries 24} (2015) 1530008}
  [\href{https://arxiv.org/abs/1508.07974}{{\ttfamily 1508.07974}}].

\bibitem{Kumar:2022ylt}
A.~Kumar, B.~M\"uller and D.-L.~Yang, \emph{{Spin polarization and correlation
  of quarks from glasma}},  \href{https://arxiv.org/abs/2212.13354}{{\ttfamily
  2212.13354}}.

\bibitem{Berges:2013eia}
J.~Berges, K.~Boguslavski, S.~Schlichting and R.~Venugopalan, \emph{{Turbulent
  thermalization process in heavy-ion collisions at ultrarelativistic
  energies}},
  \href{https://doi.org/10.1103/PhysRevD.89.074011}{\emph{Phys.Rev.} {\bfseries
  D89} (2014) 074011} [\href{https://arxiv.org/abs/1303.5650}{{\ttfamily
  1303.5650}}].

\bibitem{Berges:2013fga}
J.~Berges, K.~Boguslavski, S.~Schlichting and R.~Venugopalan, \emph{{Universal
  attractor in a highly occupied non-Abelian plasma}},
  \href{https://arxiv.org/abs/1311.3005}{{\ttfamily 1311.3005}}.

\bibitem{Ipp:2020mjc}
A.~Ipp, D.I.~M\"uller and D.~Schuh, \emph{{Anisotropic momentum broadening in
  the 2+1D Glasma: analytic weak field approximation and lattice simulations}},
  \href{https://doi.org/10.1103/PhysRevD.102.074001}{\emph{Phys. Rev. D}
  {\bfseries 102} (2020) 074001}
  [\href{https://arxiv.org/abs/2001.10001}{{\ttfamily 2001.10001}}].

\bibitem{Ipp:2020nfu}
A.~Ipp, D.I.~M\"uller and D.~Schuh, \emph{{Jet momentum broadening in the
  pre-equilibrium Glasma}},
  \href{https://doi.org/10.1016/j.physletb.2020.135810}{\emph{Phys. Lett. B}
  {\bfseries 810} (2020) 135810}
  [\href{https://arxiv.org/abs/2009.14206}{{\ttfamily 2009.14206}}].

\bibitem{Kurkela:2015qoa}
A.~Kurkela and Y.~Zhu, \emph{{Isotropization and hydrodynamization in weakly
  coupled heavy-ion collisions}},
  \href{https://arxiv.org/abs/1506.06647}{{\ttfamily 1506.06647}}.

\bibitem{Kurkela:2018vqr}
A.~Kurkela, A.~Mazeliauskas, J.-F.~Paquet, S.~Schlichting and D.~Teaney,
  \emph{{Effective kinetic description of event-by-event pre-equilibrium
  dynamics in high-energy heavy-ion collisions}},
  \href{https://doi.org/10.1103/PhysRevC.99.034910}{\emph{Phys. Rev. C}
  {\bfseries 99} (2019) 034910}
  [\href{https://arxiv.org/abs/1805.00961}{{\ttfamily 1805.00961}}].

\bibitem{Du:2020dvp}
X.~Du and S.~Schlichting, \emph{{Equilibration of weakly coupled QCD plasmas}},
  \href{https://doi.org/10.1103/PhysRevD.104.054011}{\emph{Phys. Rev. D}
  {\bfseries 104} (2021) 054011}
  [\href{https://arxiv.org/abs/2012.09079}{{\ttfamily 2012.09079}}].

\bibitem{Carrington:2021dvw}
M.E.~Carrington, A.~Czajka and S.~Mrowczynski, \emph{{Jet quenching in
  glasma}}, \href{https://doi.org/10.1016/j.physletb.2022.137464}{\emph{Phys.
  Lett. B} {\bfseries 834} (2022) 137464}
  [\href{https://arxiv.org/abs/2112.06812}{{\ttfamily 2112.06812}}].

\bibitem{Carrington:2022bnv}
M.E.~Carrington, A.~Czajka and S.~Mrowczynski, \emph{{Transport of hard probes
  through glasma}},
  \href{https://doi.org/10.1103/PhysRevC.105.064910}{\emph{Phys. Rev. C}
  {\bfseries 105} (2022) 064910}
  [\href{https://arxiv.org/abs/2202.00357}{{\ttfamily 2202.00357}}].

\bibitem{Mrowczynski:1993qm}
S.~Mrowczynski, \emph{{Plasma instability at the initial stage of
  ultrarelativistic heavy ion collisions}},
  \href{https://doi.org/10.1016/0370-2693(93)91330-P}{\emph{Phys. Lett. B}
  {\bfseries 314} (1993) 118}.

\bibitem{Mrowczynski:2016etf}
S.~Mrowczynski, B.~Schenke and M.~Strickland, \emph{{Color instabilities in the
  quark\textendash{}gluon plasma}},
  \href{https://doi.org/10.1016/j.physrep.2017.03.003}{\emph{Phys. Rept.}
  {\bfseries 682} (2017) 1} [\href{https://arxiv.org/abs/1603.08946}{{\ttfamily
  1603.08946}}].

\bibitem{Hauksson:2020wsm}
S.~Hauksson, S.~Jeon and C.~Gale, \emph{{Probes of the quark-gluon plasma and
  plasma instabilities}},
  \href{https://doi.org/10.1103/PhysRevC.103.064904}{\emph{Phys. Rev. C}
  {\bfseries 103} (2021) 064904}
  [\href{https://arxiv.org/abs/2012.03640}{{\ttfamily 2012.03640}}].

\bibitem{Arnold:2002zm}
P.B.~Arnold, G.D.~Moore and L.G.~Yaffe, \emph{{Effective kinetic theory for
  high temperature gauge theories}},
  \href{https://doi.org/10.1088/1126-6708/2003/01/030}{\emph{JHEP} {\bfseries
  0301} (2003) 030} [\href{https://arxiv.org/abs/hep-ph/0209353}{{\ttfamily
  hep-ph/0209353}}].

\bibitem{Baier:2008js}
R.~Baier and Y.~Mehtar-Tani, \emph{{Jet quenching and broadening: The Transport
  coefficient q-hat in an anisotropic plasma}},
  \href{https://doi.org/10.1103/PhysRevC.78.064906}{\emph{Phys. Rev. C}
  {\bfseries 78} (2008) 064906}
  [\href{https://arxiv.org/abs/0806.0954}{{\ttfamily 0806.0954}}].

\bibitem{Romatschke:2006bb}
P.~Romatschke, \emph{{Momentum broadening in an anisotropic plasma}},
  \href{https://doi.org/10.1103/PhysRevC.75.014901}{\emph{Phys. Rev. C}
  {\bfseries 75} (2007) 014901}
  [\href{https://arxiv.org/abs/hep-ph/0607327}{{\ttfamily hep-ph/0607327}}].

\bibitem{Hauksson:2021okc}
S.~Hauksson, S.~Jeon and C.~Gale, \emph{{Momentum broadening of energetic
  partons in an anisotropic plasma}},
  \href{https://doi.org/10.1103/PhysRevC.105.014914}{\emph{Phys. Rev. C}
  {\bfseries 105} (2022) 014914}
  [\href{https://arxiv.org/abs/2109.04575}{{\ttfamily 2109.04575}}].

\bibitem{Baier:1998yf}
R.~Baier, Y.L.~Dokshitzer, A.H.~Mueller and D.~Schiff, \emph{{Radiative Energy
  Loss of High Energy Partons Traversing an Expanding {QCD} Plasma}},
  \href{https://doi.org/10.1103/PhysRevC.58.1706}{\emph{Phys. Rev.} {\bfseries
  C58} (1998) 1706} [\href{https://arxiv.org/abs/hep-ph/9803473}{{\ttfamily
  hep-ph/9803473}}].

\bibitem{Zakharov:1998wq}
B.G.~Zakharov, \emph{{Quark energy loss in an expanding quark gluon plasma}},
  in \emph{{QCD and high energy hadronic interactions. Proceedings, 33rd
  Rencontres de Moriond, Les Arcs, France, March 21-28, 1998}}, pp.~533--538,
  1998 [\href{https://arxiv.org/abs/hep-ph/9807396}{{\ttfamily
  hep-ph/9807396}}].

\bibitem{Arnold:2008iy}
P.B.~Arnold, \emph{{Simple Formula for High-Energy Gluon Bremsstrahlung in a
  Finite, Expanding Medium}},
  \href{https://doi.org/10.1103/PhysRevD.79.065025}{\emph{Phys. Rev.}
  {\bfseries D79} (2009) 065025}
  [\href{https://arxiv.org/abs/0808.2767}{{\ttfamily 0808.2767}}].

\bibitem{Iancu:2018trm}
E.~Iancu, P.~Taels and B.~Wu, \emph{{Jet quenching parameter in an expanding
  QCD plasma}},
  \href{https://doi.org/10.1016/j.physletb.2018.10.007}{\emph{Phys. Lett.}
  {\bfseries B786} (2018) 288}
  [\href{https://arxiv.org/abs/1806.07177}{{\ttfamily 1806.07177}}].

\bibitem{Adhya:2019qse}
S.P.~Adhya, C.A.~Salgado, M.~Spousta and K.~Tywoniuk, \emph{{Medium-induced
  cascade in expanding media}},
  \href{https://doi.org/10.1007/JHEP07(2020)150}{\emph{JHEP} {\bfseries 07}
  (2020) 150} [\href{https://arxiv.org/abs/1911.12193}{{\ttfamily
  1911.12193}}].

\bibitem{Caucal:2020uic}
P.~Caucal, E.~Iancu and G.~Soyez, \emph{{Jet radiation in a longitudinally
  expanding medium}},
  \href{https://doi.org/10.1007/JHEP04(2021)209}{\emph{JHEP} {\bfseries 04}
  (2021) 209} [\href{https://arxiv.org/abs/2012.01457}{{\ttfamily
  2012.01457}}].

\bibitem{Barata:2022krd}
J.a.~Barata, A.V.~Sadofyev and C.A.~Salgado, \emph{{Jet broadening in dense
  inhomogeneous matter}},
  \href{https://doi.org/10.1103/PhysRevD.105.114010}{\emph{Phys. Rev. D}
  {\bfseries 105} (2022) 114010}
  [\href{https://arxiv.org/abs/2202.08847}{{\ttfamily 2202.08847}}].

\bibitem{Sadofyev:2021ohn}
A.V.~Sadofyev, M.D.~Sievert and I.~Vitev, \emph{{Ab~initio coupling of jets to
  collective flow in the opacity expansion approach}},
  \href{https://doi.org/10.1103/PhysRevD.104.094044}{\emph{Phys. Rev. D}
  {\bfseries 104} (2021) 094044}
  [\href{https://arxiv.org/abs/2104.09513}{{\ttfamily 2104.09513}}].

\bibitem{Andres:2022ndd}
C.~Andres, F.~Dominguez, A.V.~Sadofyev and C.A.~Salgado, \emph{{Jet broadening
  in flowing matter: Resummation}},
  \href{https://doi.org/10.1103/PhysRevD.106.074023}{\emph{Phys. Rev. D}
  {\bfseries 106} (2022) 074023}
  [\href{https://arxiv.org/abs/2207.07141}{{\ttfamily 2207.07141}}].

\bibitem{Romatschke:2007mq}
P.~Romatschke and U.~Romatschke, \emph{{Viscosity Information from Relativistic
  Nuclear Collisions: How Perfect is the Fluid Observed at RHIC?}},
  \href{https://doi.org/10.1103/PhysRevLett.99.172301}{\emph{Phys. Rev. Lett.}
  {\bfseries 99} (2007) 172301}
  [\href{https://arxiv.org/abs/0706.1522}{{\ttfamily 0706.1522}}].

\bibitem{Baier:1996kr}
R.~Baier, Y.L.~Dokshitzer, A.H.~Mueller, S.~Peigne and D.~Schiff,
  \emph{{Radiative Energy Loss of High Energy Quarks and Gluons in a
  Finite-Volume Quark-Gluon Plasma}},
  \href{https://doi.org/10.1016/S0550-3213(96)00553-6}{\emph{Nucl. Phys.}
  {\bfseries B483} (1997) 291}
  [\href{https://arxiv.org/abs/hep-ph/9607355}{{\ttfamily hep-ph/9607355}}].

\bibitem{Baier:1996sk}
R.~Baier, Y.L.~Dokshitzer, A.H.~Mueller, S.~Peigne and D.~Schiff,
  \emph{{Radiative Energy Loss and P(T)-Broadening of High Energy Partons in
  Nuclei}}, \href{https://doi.org/10.1016/S0550-3213(96)00581-0}{\emph{Nucl.
  Phys.} {\bfseries B484} (1997) 265}
  [\href{https://arxiv.org/abs/hep-ph/9608322}{{\ttfamily hep-ph/9608322}}].

\bibitem{Zakharov:1996fv}
B.G.~Zakharov, \emph{{Fully Quantum Treatment of the Landau-Pomeranchuk-Migdal
  Effect in QED and QCD}}, \href{https://doi.org/10.1134/1.567126}{\emph{JETP
  Lett.} {\bfseries 63} (1996) 952}
  [\href{https://arxiv.org/abs/hep-ph/9607440}{{\ttfamily hep-ph/9607440}}].

\bibitem{Zakharov:1997uu}
B.G.~Zakharov, \emph{{Radiative Energy Loss of High Energy Quarks in
  Finite-Size Nuclear Matter and Quark-Gluon Plasma}},
  \href{https://doi.org/10.1134/1.567389}{\emph{JETP Lett.} {\bfseries 65}
  (1997) 615} [\href{https://arxiv.org/abs/hep-ph/9704255}{{\ttfamily
  hep-ph/9704255}}].

\bibitem{Baier:1998kq}
R.~Baier, Y.L.~Dokshitzer, A.H.~Mueller and D.~Schiff, \emph{{Medium-Induced
  Radiative Energy Loss: Equivalence Between the Bdmps and Zakharov
  Formalisms}},
  \href{https://doi.org/10.1016/S0550-3213(98)00546-X}{\emph{Nucl. Phys.}
  {\bfseries B531} (1998) 403}
  [\href{https://arxiv.org/abs/hep-ph/9804212}{{\ttfamily hep-ph/9804212}}].

\bibitem{Wiedemann:1999fq}
U.A.~Wiedemann and M.~Gyulassy, \emph{{Transverse Momentum Dependence of the
  Landau-Pomeranchuk- Migdal Effect}},
  \href{https://doi.org/10.1016/S0550-3213(99)00458-7}{\emph{Nucl. Phys.}
  {\bfseries B560} (1999) 345}
  [\href{https://arxiv.org/abs/hep-ph/9906257}{{\ttfamily hep-ph/9906257}}].

\bibitem{Wiedemann:2000za}
U.A.~Wiedemann, \emph{{Gluon Radiation Off Hard Quarks in a Nuclear
  Environment: Opacity Expansion}},
  \href{https://doi.org/10.1016/S0550-3213(00)00457-0}{\emph{Nucl. Phys.}
  {\bfseries B588} (2000) 303}
  [\href{https://arxiv.org/abs/hep-ph/0005129}{{\ttfamily hep-ph/0005129}}].

\bibitem{Blaizot:2013hx}
J.-P.~Blaizot, E.~Iancu and Y.~Mehtar-Tani, \emph{{Medium-induced QCD cascade:
  democratic branching and wave turbulence}},
  \href{https://doi.org/10.1103/PhysRevLett.111.052001}{\emph{Phys.Rev.Lett.}
  {\bfseries 111} (2013) 052001}
  [\href{https://arxiv.org/abs/1301.6102}{{\ttfamily 1301.6102}}].

\bibitem{Blaizot:2013vha}
J.-P.~Blaizot, F.~Dominguez, E.~Iancu and Y.~Mehtar-Tani, \emph{{Probabilistic
  picture for medium-induced jet evolution}},
  \href{https://doi.org/10.1007/JHEP06(2014)075}{\emph{JHEP} {\bfseries 1406}
  (2014) 075} [\href{https://arxiv.org/abs/1311.5823}{{\ttfamily 1311.5823}}].

\bibitem{Blaizot:2001nr}
J.-P.~Blaizot and E.~Iancu, \emph{{The Quark gluon plasma: Collective dynamics
  and hard thermal loops}},
  \href{https://doi.org/10.1016/S0370-1573(01)00061-8}{\emph{Phys.Rept.}
  {\bfseries 359} (2002) 355}
  [\href{https://arxiv.org/abs/hep-ph/0101103}{{\ttfamily hep-ph/0101103}}].

\bibitem{Kapusta:2006pm}
J.I.~Kapusta and C.~Gale, \emph{{Finite-temperature field theory: Principles
  and applications}}, Cambridge Monographs on Mathematical Physics, Cambridge
  University Press (2011),
  \href{https://doi.org/10.1017/CBO9780511535130}{10.1017/CBO9780511535130}.

\bibitem{Braaten:1989mz}
E.~Braaten and R.D.~Pisarski, \emph{{Soft Amplitudes in Hot Gauge Theories: A
  General Analysis}},
  \href{https://doi.org/10.1016/0550-3213(90)90508-B}{\emph{Nucl. Phys. B}
  {\bfseries 337} (1990) 569}.

\bibitem{Frenkel:1989br}
J.~Frenkel and J.C.~Taylor, \emph{{High Temperature Limit of Thermal QCD}},
  \href{https://doi.org/10.1016/0550-3213(90)90661-V}{\emph{Nucl. Phys. B}
  {\bfseries 334} (1990) 199}.

\bibitem{Blaizot:1993zk}
J.P.~Blaizot and E.~Iancu, \emph{{Kinetic equations for long wavelength
  excitations of the quark - gluon plasma}},
  \href{https://doi.org/10.1103/PhysRevLett.70.3376}{\emph{Phys. Rev. Lett.}
  {\bfseries 70} (1993) 3376}
  [\href{https://arxiv.org/abs/hep-ph/9301236}{{\ttfamily hep-ph/9301236}}].

\bibitem{Mrowczynski:2000ed}
S.~Mrowczynski and M.H.~Thoma, \emph{{Hard loop approach to anisotropic
  systems}}, \href{https://doi.org/10.1103/PhysRevD.62.036011}{\emph{Phys. Rev.
  D} {\bfseries 62} (2000) 036011}
  [\href{https://arxiv.org/abs/hep-ph/0001164}{{\ttfamily hep-ph/0001164}}].

\bibitem{Romatschke:2003ms}
P.~Romatschke and M.~Strickland, \emph{{Collective modes of an anisotropic
  quark gluon plasma}},
  \href{https://doi.org/10.1103/PhysRevD.68.036004}{\emph{Phys. Rev. D}
  {\bfseries 68} (2003) 036004}
  [\href{https://arxiv.org/abs/hep-ph/0304092}{{\ttfamily hep-ph/0304092}}].

\bibitem{Mrowczynski:2004kv}
S.~Mrowczynski, A.~Rebhan and M.~Strickland, \emph{{Hard loop effective action
  for anisotropic plasmas}},
  \href{https://doi.org/10.1103/PhysRevD.70.025004}{\emph{Phys. Rev. D}
  {\bfseries 70} (2004) 025004}
  [\href{https://arxiv.org/abs/hep-ph/0403256}{{\ttfamily hep-ph/0403256}}].

\bibitem{CaronHuot:2008ni}
S.~Caron-Huot, \emph{{O(g) plasma effects in jet quenching}},
  \href{https://doi.org/10.1103/PhysRevD.79.065039}{\emph{Phys.Rev.} {\bfseries
  D79} (2009) 065039} [\href{https://arxiv.org/abs/0811.1603}{{\ttfamily
  0811.1603}}].

\bibitem{Aurenche:2002pd}
P.~Aurenche, F.~Gelis and H.~Zaraket, \emph{{A Simple sum rule for the thermal
  gluon spectral function and applications}},
  \href{https://doi.org/10.1088/1126-6708/2002/05/043}{\emph{JHEP} {\bfseries
  05} (2002) 043} [\href{https://arxiv.org/abs/hep-ph/0204146}{{\ttfamily
  hep-ph/0204146}}].

\bibitem{Salgado:2002cd}
C.A.~Salgado and U.A.~Wiedemann, \emph{{A Dynamical scaling law for jet
  tomography}},
  \href{https://doi.org/10.1103/PhysRevLett.89.092303}{\emph{Phys. Rev. Lett.}
  {\bfseries 89} (2002) 092303}
  [\href{https://arxiv.org/abs/hep-ph/0204221}{{\ttfamily hep-ph/0204221}}].

\bibitem{Salgado:2003gb}
C.A.~Salgado and U.A.~Wiedemann, \emph{{Calculating quenching weights}},
  \href{https://doi.org/10.1103/PhysRevD.68.014008}{\emph{Phys.Rev.} {\bfseries
  D68} (2003) 014008} [\href{https://arxiv.org/abs/hep-ph/0302184}{{\ttfamily
  hep-ph/0302184}}].

\bibitem{Blaizot:2012fh}
J.-P.~Blaizot, F.~Dominguez, E.~Iancu and Y.~Mehtar-Tani, \emph{{Medium-induced
  gluon branching}}, \href{https://doi.org/10.1007/JHEP01(2013)143}{\emph{JHEP}
  {\bfseries 1301} (2013) 143}
  [\href{https://arxiv.org/abs/1209.4585}{{\ttfamily 1209.4585}}].

\bibitem{Ellis:1996mzs}
R.K.~Ellis, W.J.~Stirling and B.R.~Webber, \emph{{QCD and collider physics}},
  vol.~8, Cambridge University Press (2, 2011),
  \href{https://doi.org/10.1017/CBO9780511628788}{10.1017/CBO9780511628788}.

\bibitem{Casalderrey-Solana:2011ule}
J.~Casalderrey-Solana and E.~Iancu, \emph{{Interference effects in
  medium-induced gluon radiation}},
  \href{https://doi.org/10.1007/JHEP08(2011)015}{\emph{JHEP} {\bfseries 08}
  (2011) 015} [\href{https://arxiv.org/abs/1105.1760}{{\ttfamily 1105.1760}}].

\bibitem{Kurkela:2014tla}
A.~Kurkela and U.A.~Wiedemann, \emph{{Picturing perturbative parton cascades in
  QCD matter}},
  \href{https://doi.org/10.1016/j.physletb.2014.11.054}{\emph{Phys.Lett.}
  {\bfseries B740} (2015) 172}
  [\href{https://arxiv.org/abs/1407.0293}{{\ttfamily 1407.0293}}].

\bibitem{Fister:2014zxa}
L.~Fister and E.~Iancu, \emph{{Medium-induced jet evolution: wave turbulence
  and energy loss}}, \href{https://doi.org/10.1007/JHEP03(2015)082}{\emph{JHEP}
  {\bfseries 03} (2015) 082} [\href{https://arxiv.org/abs/1409.2010}{{\ttfamily
  1409.2010}}].

\bibitem{Blaizot:2015jea}
J.-P.~Blaizot and Y.~Mehtar-Tani, \emph{{Energy flow along the medium-induced
  parton cascade}},  \href{https://arxiv.org/abs/1501.03443}{{\ttfamily
  1501.03443}}.

\bibitem{Heinz:2013th}
U.~Heinz and R.~Snellings, \emph{{Collective flow and viscosity in relativistic
  heavy-ion collisions}},
  \href{https://doi.org/10.1146/annurev-nucl-102212-170540}{\emph{Ann. Rev.
  Nucl. Part. Sci.} {\bfseries 63} (2013) 123}
  [\href{https://arxiv.org/abs/1301.2826}{{\ttfamily 1301.2826}}].

\bibitem{Gale:2013da}
C.~Gale, S.~Jeon and B.~Schenke, \emph{{Hydrodynamic Modeling of Heavy-Ion
  Collisions}}, \href{https://doi.org/10.1142/S0217751X13400113}{\emph{Int. J.
  Mod. Phys. A} {\bfseries 28} (2013) 1340011}
  [\href{https://arxiv.org/abs/1301.5893}{{\ttfamily 1301.5893}}].

\bibitem{Luzum:2013yya}
M.~Luzum and H.~Petersen, \emph{{Initial State Fluctuations and Final State
  Correlations in Relativistic Heavy-Ion Collisions}},
  \href{https://doi.org/10.1088/0954-3899/41/6/063102}{\emph{J. Phys. G}
  {\bfseries 41} (2014) 063102}
  [\href{https://arxiv.org/abs/1312.5503}{{\ttfamily 1312.5503}}].

\bibitem{Alqahtani:2017mhy}
M.~Alqahtani, M.~Nopoush and M.~Strickland, \emph{{Relativistic anisotropic
  hydrodynamics}},
  \href{https://doi.org/10.1016/j.ppnp.2018.05.004}{\emph{Prog. Part. Nucl.
  Phys.} {\bfseries 101} (2018) 204}
  [\href{https://arxiv.org/abs/1712.03282}{{\ttfamily 1712.03282}}].

\bibitem{Caucal:2018dla}
P.~Caucal, E.~Iancu, A.H.~Mueller and G.~Soyez, \emph{{Vacuum-like jet
  fragmentation in a dense QCD medium}},
  \href{https://doi.org/10.1103/PhysRevLett.120.232001}{\emph{Phys. Rev. Lett.}
  {\bfseries 120} (2018) 232001}
  [\href{https://arxiv.org/abs/1801.09703}{{\ttfamily 1801.09703}}].

\bibitem{Caucal:2019uvr}
P.~Caucal, E.~Iancu and G.~Soyez, \emph{{Deciphering the $z_g$ distribution in
  ultrarelativistic heavy ion collisions}},
  \href{https://doi.org/10.1007/JHEP10(2019)273}{\emph{JHEP} {\bfseries 10}
  (2019) 273} [\href{https://arxiv.org/abs/1907.04866}{{\ttfamily
  1907.04866}}].

\bibitem{Caucal:2020xad}
P.~Caucal, E.~Iancu, A.~Mueller and G.~Soyez, \emph{{Nuclear modification
  factors for jet fragmentation}},
  \href{https://doi.org/10.1007/JHEP10(2020)204}{\emph{JHEP} {\bfseries 10}
  (2020) 204} [\href{https://arxiv.org/abs/2005.05852}{{\ttfamily
  2005.05852}}].

\bibitem{Mehtar-Tani:2022zwf}
Y.~Mehtar-Tani, S.~Schlichting and I.~Soudi, \emph{{Jet thermalization in QCD
  kinetic theory}},  \href{https://arxiv.org/abs/2209.10569}{{\ttfamily
  2209.10569}}.

\bibitem{Collins:1992kk}
J.C.~Collins, \emph{{Fragmentation of transversely polarized quarks probed in
  transverse momentum distributions}},
  \href{https://doi.org/10.1016/0550-3213(93)90262-N}{\emph{Nucl. Phys. B}
  {\bfseries 396} (1993) 161}
  [\href{https://arxiv.org/abs/hep-ph/9208213}{{\ttfamily hep-ph/9208213}}].

\bibitem{Amrath:2005gv}
D.~Amrath, A.~Bacchetta and A.~Metz, \emph{{Reviewing model calculations of the
  Collins fragmentation function}},
  \href{https://doi.org/10.1103/PhysRevD.71.114018}{\emph{Phys. Rev. D}
  {\bfseries 71} (2005) 114018}
  [\href{https://arxiv.org/abs/hep-ph/0504124}{{\ttfamily hep-ph/0504124}}].

\bibitem{BaBar:2013jdt}
{\scshape BaBar} collaboration, \emph{{Measurement of Collins asymmetries in
  inclusive production of charged pion pairs in $e^+e^-$ annihilation at
  BABAR}}, \href{https://doi.org/10.1103/PhysRevD.90.052003}{\emph{Phys. Rev.
  D} {\bfseries 90} (2014) 052003}
  [\href{https://arxiv.org/abs/1309.5278}{{\ttfamily 1309.5278}}].

\bibitem{Anselmino:2007fs}
M.~Anselmino, M.~Boglione, U.~D'Alesio, A.~Kotzinian, F.~Murgia, A.~Prokudin
  et~al., \emph{{Transversity and Collins functions from SIDIS and e+ e-
  data}}, \href{https://doi.org/10.1103/PhysRevD.75.054032}{\emph{Phys. Rev. D}
  {\bfseries 75} (2007) 054032}
  [\href{https://arxiv.org/abs/hep-ph/0701006}{{\ttfamily hep-ph/0701006}}].

\end{thebibliography}\endgroup

\end{document}